\pdfoutput=1
\documentclass[12pt]{iopart}

\usepackage{iopams}
\usepackage{graphicx}
\usepackage{rangecite}
\begin{document}

\title[Nanoantennas for visible and infrared radiation]{Nanoantennas for visible and infrared radiation}

\author{Paolo Biagioni}
\address{CNISM-Dipartimento di Fisica, Politecnico di Milano, Piazza Leonardo da Vinci 32, I-20133 Milano, Italy}

\author{Jer-Shing Huang}
\address{Department of Chemistry and Frontier Research Center on Fundamental and Applied Science of Matters, National Tsing Hua University, Hsinchu 30013, Taiwan}

\author{Bert Hecht}
\address{Nano-Optics \& Biophotonics Group, Department of Experimental Physics 5, R\"ontgen Research Center for Complex Material Research (RCCM), Physics Institute, University of W\"urzburg, Am Hubland, D-97074 W\"urzburg, Germany}
\ead{hecht@physik.uni-wuerzburg.de}

\begin{abstract}
Nanoantennas for visible and infrared radiation can strongly enhance the interaction of light with nanoscale matter by their ability to efficiently link propagating
and spatially localized optical fields. This ability unlocks an enormous potential for applications ranging from nanoscale optical microscopy and spectroscopy over solar
energy conversion, integrated optical nanocircuitry, opto-electronics and density-of-states engineering to ultra-sensing as well as enhancement of optical nonlinearities. Here we review the current understanding of metallic optical antennas based on the background of both well-developed radiowave antenna engineering and plasmonics. In particular, we discuss the role of plasmonic resonances on the performance of nanoantennas and address the influence of geometrical parameters imposed by nanofabrication. Finally, we give a brief account of the current status of the field and the major established and emerging lines of investigation in this vivid area of research.
\end{abstract}

\tableofcontents

\submitto{\RPP}
\maketitle

\section{Introduction}
In 1959, when nanoscience as we know it today was still far from being a reality, Richard Feynman gave a talk at the annual meeting of the American Physical Society, entitled ``There's plenty of room at the bottom'' \cite{Feynman}. In this talk Feynman anticipated most of the experimental fields and issues of concern which, more than twenty years later, would become key issues in the understanding of phenomena at the nanometer scale. While talking about the possibility of building nanoscale electric circuits, he also posed the question: ``...is it possible, for example, to emit light from a whole set of antennas, like we emit radio waves from an organized set of antennas to beam the radio programs to Europe? The same thing would be to beam the light out in a definite direction with very high intensity...''. Today, we can safely state that Feynman's suggestion has already become reality and the research on nanoantennas that work at optical frequencies has developed into a strong branch of nanoscience - nano-optics in particular - with many exciting perspectives. It is the goal of this Report to summarize and explain the current understanding of optical antennas on the background of both, the highly developed field of antenna engineering \cite{lee,balanis} and plasmonics \cite{maier,Ozbay06,Gramotnev10}.

Although Feynman's work was right before the eyes of everybody for a long time, it took the solid development of near-field optics \cite{novotny07b} to acquire enough proficiency in using nanostructures to influence the flow of light at deep subwavelength scales with the required precision. Although it is not the intention of this Report to provide a detailed account on the chronological development of the field, we nevertheless would like to mention a few selected publications that inspired the authors to enter into the field of nanoantennas. First of all there is the visionary book chapter by Dieter W.~Pohl  \cite{pohl99}, in which he points out the similarities between ``fluorescing molecules, small scattering particles etc.~and telecommunication antennas'' and suggests to ``inspect antenna theory for concepts applicable and useful to near-field optics''. Another eye opener was the paper by Grober \textit{et al.}~\cite{Grober97}, in which the authors explicitly discuss the use of nanoantennas for scanning near-field optical microscopy and provide proof-of-principle experiments using microwave radiation - an idea later on brought close to realization by Oesterschulze \textit{et al.}~\cite{Oesterschulze01}. Many other efforts dealing with antenna-like structures date back into the Eighties and even before, mostly driven by the need for efficient infrared (IR) detectors. A good account is given in recent reviews \cite{Bharadwaj09,Novotny11}.

\subsection{Antenna basics: Radiation and near field of a time-dependent charge distribution}
Antennas are used either to create electromagnetic (e.m.) waves with a well-defined radiation pattern, which can then travel over large distances, or to receive e.m.~waves from a remote source in order to extract some encoded information, to measure changes in their intensity, or to exploit the transmitted power \cite{balanis}. Today the importance of antennas is dominated by their ability to provide an interface between localized information processing using electrical signals and the free-space wireless transmission of information encoded in various parameters of e.m.~waves, such as e.g.~amplitude, phase, and frequency. Due to these properties, antennas and e.m.~radiation have become indispensable assets to science and technology as well as to our everyday life.

The function of an antenna is based on the fact that free charge carriers are constricted into certain well-defined regions of space.  These charges may start to oscillate if an ac-voltage is applied or an e.m.~wave is reaching such a region. Examples for such systems are the conduction electrons in pieces of metal \cite{lee,balanis} as well as electrons and ions in a gas discharge tube \cite{Borg99}. An ac-voltage applied to a piece of metal changes the spatial distribution of charges as a function of time, which in turn will eventually affect the electric field of the charge distribution at any distance from the source. Due to the finite speed of light $c$, any change in the charge distribution that occurs at time $t_{\rm o}$ results in a change in the electric field at a remote point at a distance $R$ only after a time $t_{\rm o} + \frac{n R}{c}$, where $n$ is the refractive index of the medium. A well-known fundamental source of such e.m.~disturbances is a harmonically oscillating dipole which may be pictured as two metallic spheres connected by a thin wire as it was realized in H.~Hertz's pioneering experiments \cite{Hertz}. If such a system is prepared in an initial state where some negative charge is on one sphere and the corresponding positive charge on the other one, the system - when left alone - will start to perform an exponentially damped harmonic oscillation at a frequency $\omega_{\rm o} =\frac{1}{\sqrt{LC}}$ (where we assume small damping),  in which $L$ and $C$ are the inductance and the capacitance of the system, respectively. The fact that the system is exponentially damped, i.e.~energy loss is proportional to the energy still stored within the system, has two reasons: (i) a finite (Ohmic) resistance felt by the charge carriers in the metal wire and (ii) loss of energy due to radiation of e.m.~waves. This so-called radiation loss occurs due to the fact that the oscillation eventually creates time-dependent electric fields at remote distances, which must then be accompanied by magnetic fields that vary according to Maxwell's equations. At large enough distance these fields transform into plane waves which are free-space solutions of the wave equation. If the dipole oscillation would be suddenly switched off, those far-away fields, or simply far fields, would continue to propagate since they carry energy that is stored in the fields themselves and has been removed from the energy originally stored in the charge distribution we have been starting out with. On the  contrary, the so-called near-field zone corresponds to the instantaneous electrostatic fields of the dipole, which do not contribute to radiation but return their energy to the source after each oscillation cycle or when the source is turned off (reactive power).

\subsection{Towards optical antennas: From perfect metals to plasmonic materials}
In order to tune an antenna in such a way that it is resonant at optical frequencies one needs to adjust both $L$ and $C$ to bring the resonance into the optical regime. As R.~Feynman already pointed out in 1959, in order to achieve a resonance in the optical wavelength regime one would have to make both, $L$ and $C$, very small \cite{Feynman}. This can be achieved by shrinking the dimensions of the antenna to the scale of the wavelength \cite{Hecht07}. However, if we are moving to higher and higher frequencies in order to eventually end up with IR and visible light, metals no longer behave as perfect conductors. The main difference between the interaction of low-frequency and very-high-frequency e.m. waves with the conduction electrons in metals stems from a finite effective mass of electrons. Such effective mass causes the electrons to react with increasing phase lag to the oscillating e.m. field as the frequency increases. This behavior is in perfect analogy to a mass on a spring excited by an oscillating external force. In the case of electrons in a metal, the restoring force is the Coulomb interaction with the stationary metal ions. For low frequencies, the electrons follow the excitation without phase lag. For increasing frequency of the excitation, they exhibit an increasing oscillation amplitude as well as an increasing phase lag. As soon as the phase lag approaches 90$^\circ$ the amplitude of the charge oscillation goes through a maximum and is only limited by the internal (Ohmic and radiation) damping of the system. In metallic nanoparticles, this resonance corresponds to the localized plasmon resonance which for certain materials (such as gold, silver, aluminum, and copper) happens to appear in or close to the visible spectral range. Plasmon resonances do not appear in ``perfect'' conductors (metals at low enough frequencies) since in those materials by definition no phase lag exists between excitation and charge response. The presence of localized plasmon resonances  is therefore characteristic for optical frequencies and can be exploited to balance drawbacks of antenna systems in this frequency range, such as e.g.~enhanced Ohmic losses compared to the radio-frequency (RF) regime. It should be noted here for completeness that the metallic character of doped semiconductors at low frequencies makes it possible to excite surface plasmons resonant at mid infrared, THz, and microwave frequencies \cite{Allen77,Boltasseva11}, while even perfect metals if periodically structured can support excitations which behave very similar to surface plasmon polaritons, so-called spoof plasmons \cite{Pendry04}.
\begin{figure}[htbp]
    \centering
     \includegraphics[width=0.5\textwidth]{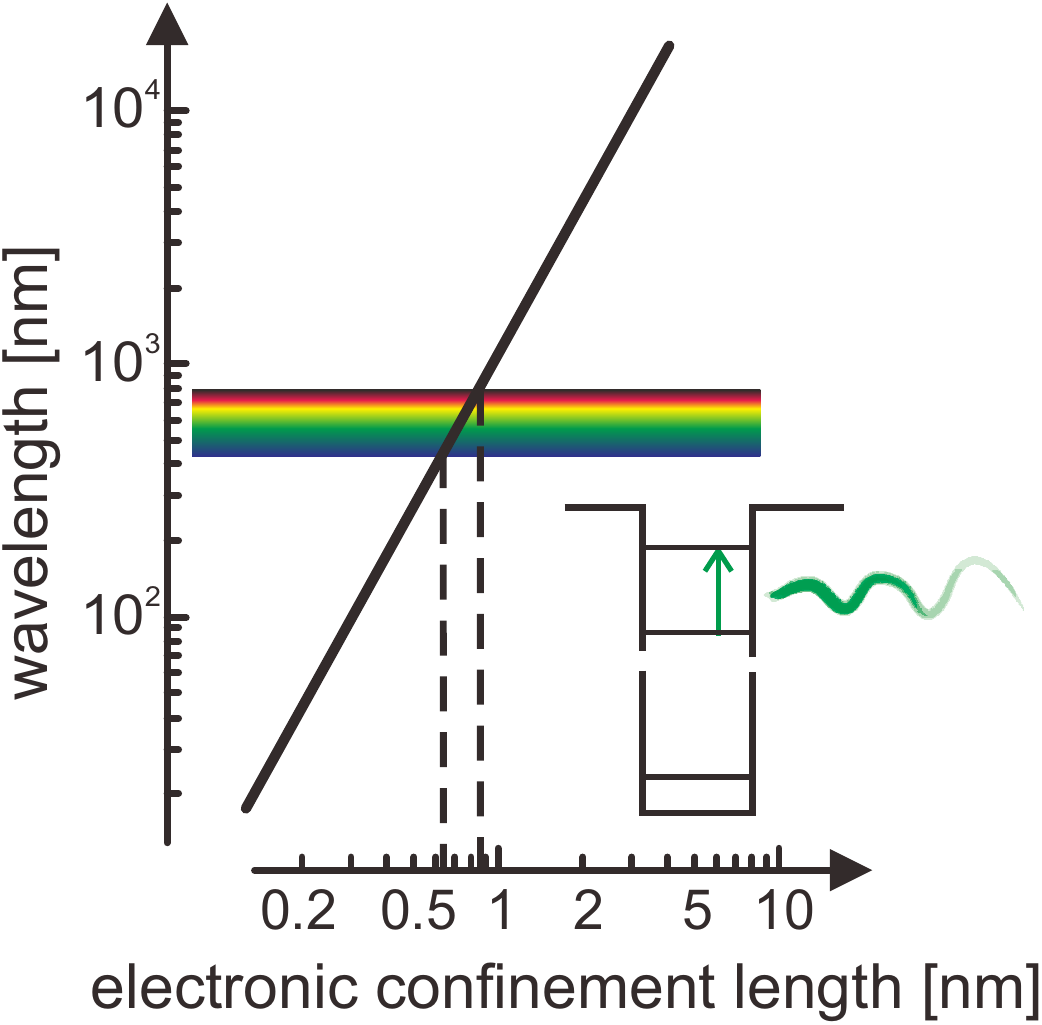}
     \caption{Electronic confinement vs. wavelength of associated radiation due to a HOMO-LUMO transition. Compared to the electron confinement length, the corresponding wavelength of the emission in the visible regime is typically 2 to 3 orders of magnitude larger. Such a mismatch leads to very inefficient
absorption and emission of photons by the quantum emitter.}
     \label{fig: confinement-wavelength}
\end{figure}

\subsection{Potential of nanoantennas at optical frequencies}
\label{potentialofopticalantennas}
Why would anybody be interested in antennas at optical frequencies? What would be the advantage of using  an antenna over standard means, such as lenses and mirrors, to manipulate e.m.~waves at optical frequencies? The wavelength of visible light in vacuum in the green spectral range is about 500~nm, corresponding to an energy of about 2.5~eV. Photons with such an energy can interact with matter through transitions between electronic states of spatially confined electrons.

Using the simplest quantum mechanical approach to describe such a system, the particle-in-a-box
model, it is easy to show that the length scale of electron confinement, i.e.~the length of the box, must be on the order of 1~nm if we require the lowest energy transition to occur in the visible spectral range. Fig.~\ref{fig: confinement-wavelength} illustrates the quadratic relation between electronic confinement and wavelength of the related e.m.~wave obtained within such a model. Electrons that show such a spatial confinement are typically encountered in larger organic molecules and artificial quantum confined systems, e.g. quantum dots and the like, which we call ``quantum emitters'' for simplicity. Note the strong mismatch between the electronic confinement length, which determines the spectroscopic properties and the local interactions of a quantum emitter on a nanometer scale, and the wavelength of the e.m.~radiation.

Since the wavelength governs effects of diffraction, e.g. in the focusing of light, this mismatch prevents propagating photons from being confined to the same spatial extension as the electrons of a quantum emitter. This leads, for example, to a typical behavior of single molecules at ambient conditions, which is that they absorb only very little light even when illuminated with a tightly focused laser beam \cite{Kukura10,Chong10,Celebrano11}. Similar arguments explain the small cross-section for the generation of excitons in a semiconductor material - the fundamental process for solar energy conversion. A further important consequence of the length-scale mismatch is a rather long lifetime of the excited state of a typical quantum emitter. Since the size of the molecule is so much smaller than the free-space wavelength of light, the birth of a photon from a quantum emitter is a highly inefficient process \cite{Keller00}. This is nicely illustrated considering the total power emitted by a time-harmonic line current element in a homogeneous space with a length $\Delta l$ much shorter than the wavelength $\lambda_0$,
\begin{equation}\label{pointdipoletotalpower}
P_{\rm o} =\frac{I^2}{3}\pi\eta\left(\frac{\Delta l}{\lambda_0}\right)^2,
\end{equation}
\noindent where $I$ is the current amplitude and $\eta = \sqrt{\mu_{\rm o}/\varepsilon_{\rm o}} \simeq 377$ $\Omega$ the wave impedance of free space \cite{lee}. Classically, such a current element can be considered a model for the oscillatory motion of electrons in a molecule. Obviously, the radiated power is proportional to the square of the length-to-wavelength ratio. For the typical extension of a molecule of 1~nm this expression well reproduces the experimentally found relatively low excited state decay rates on the order of $10^{9}~\rm s^{-1}$.
Due to this - on a molecular timescale - very long excited state lifetime, there is plenty of time for the excited-state energy to be dissipated through alternative nonradiative channels or for the molecule to become destroyed by photochemical processes. Furthermore, the maximum number of photons that can be emitted per unit time is relatively small. It is the low photon emission rate that limits the usability of single quantum emitters as sources of single photons \cite{Lounis05} and their detectability in sensing and spectroscopic applications. Finally, as a third consequence of the mentioned size mismatch, we note that in a typical far-field experiment spatially-resolved spectroscopic analysis of photons emitted simultaneously by an ensemble of closely packed quantum emitters is hindered by diffraction, which limits spatial resolution to about half of the emission wavelength \cite{novotny}.

Since optical antennas are able to (i) confine e.m. radiation to very small dimensions and (ii) very efficiently release radiation from localized sources into the far field, they provide the possibility to tailor the interaction of light with nano-matter in such a way that the three mentioned fundamental shortcomings can be lifted to a large extent. Therefore the idea of an ``optical antenna'' is a fundamental concept in the general field of light-(nano)matter interaction.

Potential applications of optical antennas are therefore closely related to their ability to strongly localize and enhance optical fields upon illumination into the feed point - e.g.~the gap between two antenna arms. Both field confinement and enhancement trigger strong interest related to nonlinear optical effects, ultrasensing, imaging, as well as solar energy conversion and opto-electronics, all of which will be discussed in more detail later on in this Report.

Moreover, and as a further illustration, let us consider the field of optical communication: Since the higher the frequency, the more information can be encoded, the visible and infrared wavelength band is widely used in today's high-speed data communication networks. As an interesting side effect, when entering the optical regime, the frequency becomes large enough such that detection of single radiation quanta is readily achievable and that quantum jumps in single molecules and atoms can be induced and observed \cite{Hwang09}. Therefore, in the optical regime, quantum aspects of the interaction of radiation and matter can be exploited in the context of long distance communication \cite{Duan01}. In this picture photons are considered as ``flying qubits'', while the atoms and molecules act as immobilized qubits \cite{Lvovsky09}. Due to the fact that antenna emission patterns can be tuned in such a way that radiation is emitted in specific directions with sharp angular characteristics, the use of single quantum emitters in combination with optical antennas opens up fascinating new perspectives for quantum communication and data processing \cite{Brien07}.

\subsection{Outline}
To provide a solid background, we will begin with a brief account on the theory of classical RF antennas and introduce important antenna parameters. In contrast to perfectly conducting antennas, antennas at optical frequencies consist of nanometer-sized metal particles.  Their interaction with light is determined by the frequency-dependent complex dielectric function, which we will introduce first. We then start out to investigate the resonant behavior of single wires which are the basic constituents of optical antennas. We then move to isolated structures that consist of at least two strongly interacting nanoparticles and then to more complex structures and to optical antennas interacting with a ``driving circuit''. While analyzing the resonances of these systems we will pinpoint the similarities and differences in the behavior of optical antennas as compared to their RF counterparts. What follows is then a brief account on fabrication methods that can be used to create optical antennas as well as an overview of the most important nanoantenna geometries that have been investigated so far. Once optical antennas have been fabricated, it is important to be able to verify the expected performance. Here we provide an account of the currently used optical characterization techniques and their respective strengths. We conclude our discussion with a brief review of current applications and fields of intense study in the context of nanoantennas.

\section{Elements of classical antenna theory}
Classical antenna theory uses Maxwell's equations to describe the interaction of time-dependent currents with electromagnetic waves. Most characteristic features of classical antennas are related to the two facts that (i) antenna wires are represented by a perfect conductor and (ii) critical dimensions, such as the antenna feed-gap and wire thickness, can be considered to be negligibly small compared to the wavelength.
\subsection{Introduction to ``antenna language''}
\label{antennalanguage}
\begin{figure}[htbp]
        \centering
        \includegraphics[width=0.5\textwidth]{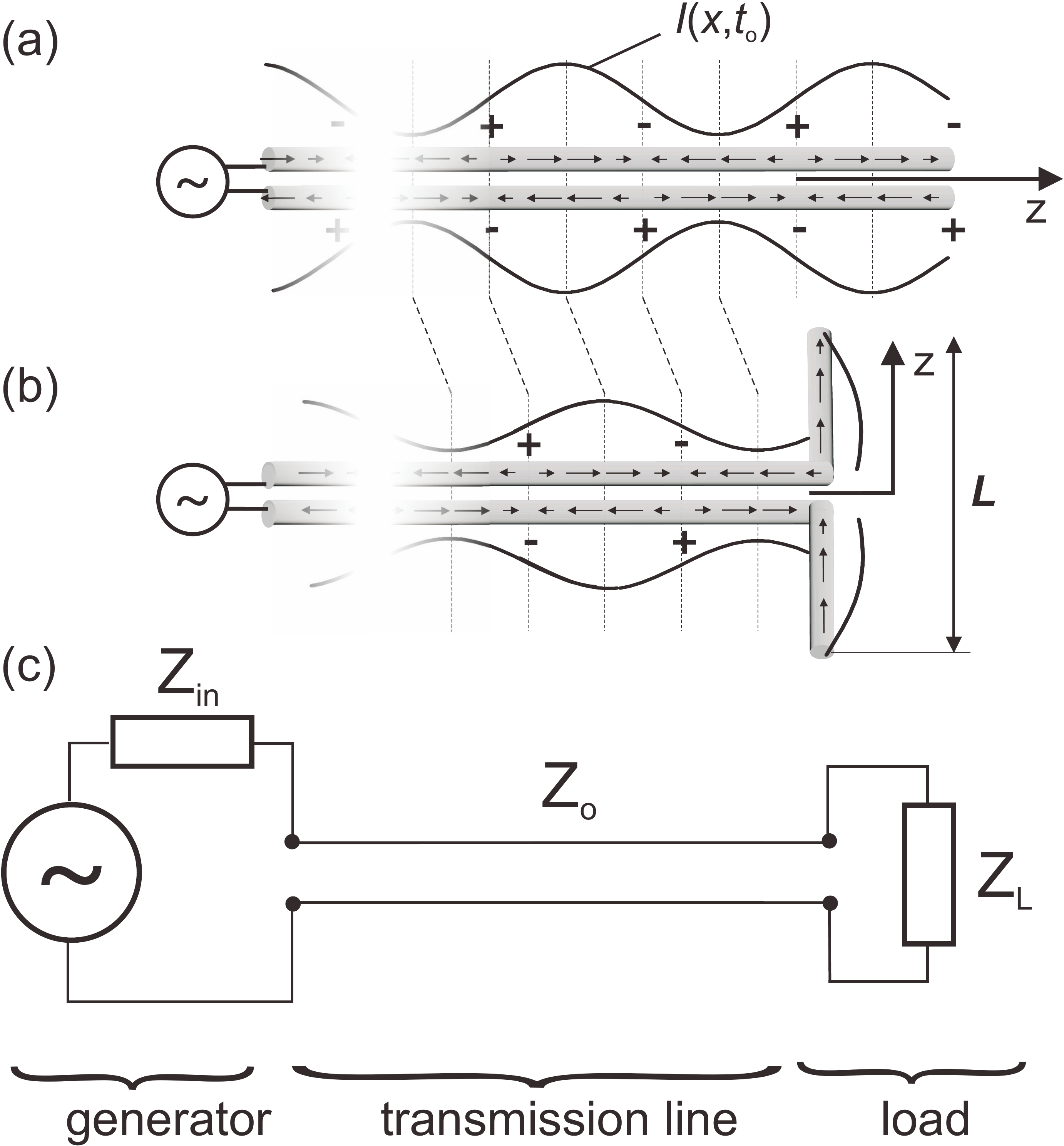}
        \caption{\label{fig:transmissionline} Harmonically driven two-wire transmission line terminated by (a) an open end and (b) a finite-length antenna. For a given instant of time, the arrows indicate magnitude and direction of the current, plus and minus signs indicate local charge accumulation, and the solid line indicates the standing current wave; (c) equivalent circuit of the system including the internal impedance of the generator, $Z_{\rm in}$, the characteristic impedance of the transmission line, $Z_{\rm o}$, as well as the impedance of the antenna, $Z_{\rm L}$, acting as a load.}
\end{figure}

We assume time-harmonic fields throughout this Report. The e.m.~field emitted by an antenna is completely determined as soon as the time-harmonic current density ${\bf j}({\bf r})$ along the antenna wires is known, from which the charge density $\rho({\bf r})$ then follows according to the continuity relation $\nabla \cdot {\bf j}({\bf r})=-\frac{\rm \partial \rho({\bf r})}{\rm \partial t}=i \omega \rho({\bf r})$. The reason for this is that in the Lorenz gauge, $\nabla \cdot {\bf A}({\bf r}) = i\omega \mu_o\mu\varepsilon_o\varepsilon \Phi({\bf r})$, the vector potential ${\bf A}({\bf r})$ and the scalar potential $\Phi({\bf r})$ satisfy a set of four inhomogeneous scalar Helmholtz equations
\begin{eqnarray}
\left[\nabla^2 + k^2\right]{\bf A}({\bf r}) & = & -\mu_o\mu {\bf j}({\bf r}) \\
\left[\nabla^2 + k^2\right]\Phi({\bf r}) & = & -\frac{1}{\varepsilon_o\varepsilon}\rho({\bf r}) \label{2}
\end{eqnarray}
where $k=2\pi/\lambda_0$, with $\lambda_0$ being the free-space wavelength. The field distribution, radiation pattern and total power radiated by the antenna are then found e.g.~by calculating a spatial convolution of the respective scalar Green's function $G_{\rm o}({\bf r},{\bf r'})$  for the given problem with the current density and the charge density present on the antenna as
\begin{eqnarray}\label{solution of wave equation with greensfunction}
{\bf A}({\bf r}) & = & \mu_o\mu \int_V {\bf j}({\bf r'}) G_{\rm o}({\bf r},{\bf r'})dV'    \\
\Phi({\bf r}) & = & \frac{1}{\varepsilon_o\varepsilon}\int_V \rho({\bf r'}) G_{\rm o}({\bf r},{\bf r'})dV'   \; .
\end{eqnarray}
The scalar Green's function is the solution of Eq.~(\ref{2}) for a Dirac delta source distribution \cite{Jackson}. Once the potentials are known, the fields can be determined by straightforward differentiation according to the definitions
 \begin{eqnarray}
 {\bf B}({\bf r}) & = &\nabla\times{\bf A}({\bf r})\\
 {\bf E}({\bf r}) & = &-\nabla \Phi({\bf r})- \frac{\partial {\bf A}({\bf r})}{\partial t}\; .
 \end{eqnarray}
 It turns out, however, that the current distribution on the antenna is quite difficult to determine exactly. For center-fed antennas with small feed-gaps and thin wires, an approximate current distribution can be found by solving an integral equation (see e.g.~\cite{King} for details). Here, for reasons of simplicity, we discuss important antenna parameters under the assumption that the current distribution on a dipole antenna has a sinusoidal shape inherited  from the standing wave pattern that builds up in a two-wire transmission line terminated by an open end, driven by a high-frequency voltage source. This is the kind of circuitry that is often used to drive an antenna. The configuration is sketched in Fig.~\ref{fig:transmissionline} (a).
The transmission line itself, although it sustains time-harmonic currents with a spatially varying amplitude, does not radiate into the far field if the gap between the wires is small, since each current element in one wire has its antiparallel counterpart in the other wire oscillating $180^\circ$ out of phase and therefore radiation largely cancels in the far field albeit a strong near field that is localized between the wires. Since good conductors are considered in RF circuits, the wavelength of the standing wave is practically the same as the wavelength in free space.  For an infinitely-long transmission line the local ratio of the voltage between the wires and the current through a wire is a constant called \emph{ characteristic impedance}, $Z_{\rm o} = U(z)/I(z)$, independent of the position $z$ along the line. It depends solely on the materials used and on the geometry of the transmission line \cite{cheng}.

In the lowest approximation one may assume that this sinusoidal current distribution is not significantly changed when we start to bend the wires at a certain distance $L/2$ from the open end - one upwards and one downwards. The strongest radiation from such a system of total length $\sim L$ is obtained for a bending angle of 90$^{\circ}$ [see Fig.~\ref{fig:transmissionline} (b)]. It can be shown that for antennas made of thin wires compared to the wavelength the current is indeed very well described by a sinusoidal distribution
\begin{equation}
\label{current_eq}
I(z) = I_{\rm max} \sin\left[k\left(\frac{1}{2}L-|z|)\right)\right]\textrm{,}
\end{equation}
in which the amplitude becomes $I_{\rm max}=I(0)/\sin(\frac{1}{2}kL)$ as expected from the simplistic standing-wave model \cite{lee,balanis}. The actual current amplitude, however, differs from that found in the unbend transmission line. The reason for this behavior lies in the fact that the antenna itself can now be thought of as a resonant circuit with a total complex impedance $Z_{\rm L} \neq Z_{\rm o}$, leading in general to a reflection at the bending point and a shift in the standing-wave pattern as sketched in Fig.~\ref{fig:transmissionline} (b). It is then natural to define the input impedance of an antenna by the ratio of the voltage measured over the input terminals and the current flowing into each antenna arm, $Z_{\rm L} = U(0)/I(0) = R_{\rm L} + iX_{\rm L}$. As for any frequency-dependent complex impedance, the equivalent circuit of the antenna shows a resonance for the driving frequency for which ${\rm Im}(Z_{\rm L})= X_{\rm L} = 0$, which also leads to a maximum in the current amplitude. We will refer to such a resonance as an ``antenna resonance''.

The absorbed power is determined by the real part of the antenna impedance $R_{\rm L}$, which includes Ohmic losses, $R_{\rm nr}$,  as well as losses due to radiation, $R_{\rm r}$,  and accordingly
\begin{equation}
R_{\rm L} = R_{\rm r} + R_{\rm nr} \textrm{.}
\end{equation}
Once the radiation resistance is known, the radiated power can be calculated as $P_{\rm r} = \frac{1}{2}R_{\rm r} I(0)^2$. A corresponding relation holds for the nonradiative power dissipated into heat. The \emph{radiation efficiency} of an antenna can therefore be defined as \cite{lee,balanis}
\begin{equation}
\eta = \frac{R_{\rm r}}{R_{\rm r} + R_{\rm nr}}\textrm{,}
\label{efficiency}
\end{equation}
describing the ratio of the radiated power to the total power absorbed by the antenna, in analogy to the quantum yield of a fluorescent molecule \cite{novotny}. Since Ohmic losses for RF antennas are very small, radiation efficiencies are typically larger than 99\%.

Together with a simple model for the high-frequency generator driving the antenna via the transmission line, which is described by a lossless ac-voltage source and a complex internal impedance $Z_{\rm in}$, we can come up with the equivalent circuit model for the whole system depicted in Fig.~\ref{fig:transmissionline} (c). The equivalent circuit model allows one to describe all relevant parameters of the circuit.

\begin{figure}[htbp]
        \centering
        \includegraphics[width=0.5\textwidth]{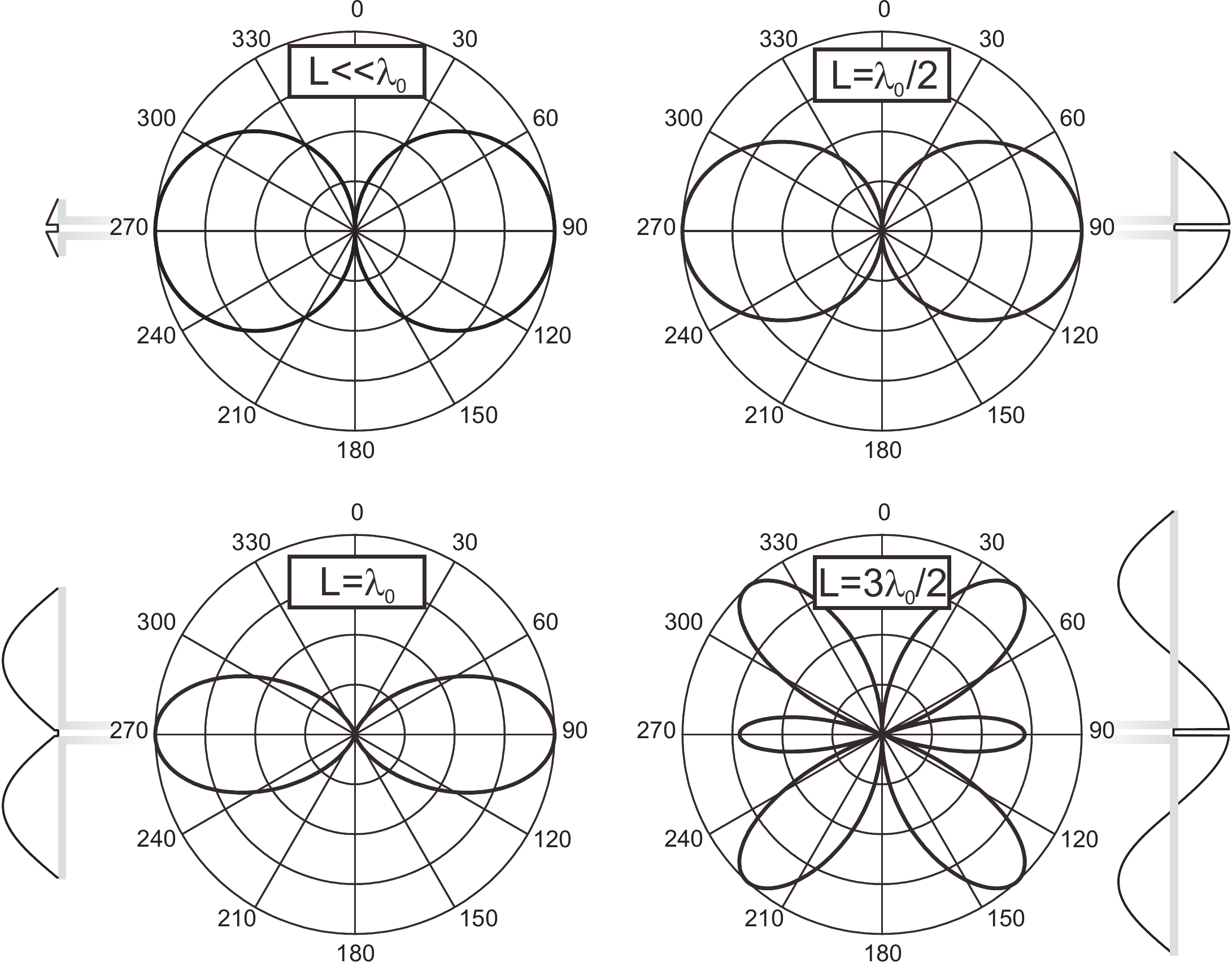}
        \caption{Normalized emission patterns for a point-like dipole ($L\ll\lambda_0$) and for perfectly conducting thin-wire antennas of length $L=\lambda_0/2$, $\lambda_0$, and $3\lambda_0/2$ \cite{lee}. The gap antenna attached to an impedance-matched waveguide effectively behaves as a single-wire antenna. A sketch of the current standing wave is provided beside each emission pattern.}
        \label{pattern}
\end{figure}

Before we get involved with this in more detail we would like to discuss the \emph{radiation pattern} $p(\theta,\phi)$ of a linear antenna with sinusoidal current distribution (Eq. \ref{current_eq}), which is described by \cite{lee}
\begin{equation}
p(\theta,\phi) \sim \left|\frac{\cos \left(\frac{1}{2}kL\cos \theta\right) - \cos\left(\frac{1}{2}kL\right)}{\sin \theta}\right|^2 \rm{,}
\end{equation}
where the angle $\theta$ is measured from the direction of the antenna wires and $\phi$ is the azimuthal angle. As one might expect, the emission pattern for antenna lengths up to $\lambda_0$ is very similar to the pattern of a Hertzian dipole ($L\ll\lambda_0$) only that its angular dependence becomes narrower. Only when the antenna length increases beyond $\lambda_0$, current elements are introduced on the same wire that oscillate $180^\circ$ out of phase, causing strong interference effects which lead to the development of a multi-lobed pattern (see Fig.~\ref{pattern}). The radiation pattern can be further influenced by deviating from the linear shape of the antenna or by adding additional wires as passive elements at well-chosen positions as it is done in the famous Yagi-Uda antenna design \cite{lee} to be discussed later on.

In order to quantify and compare the ability of different antennas to radiate power preferentially into a certain direction, antenna engineers introduce \textit{directivity} \cite{balanis}
\begin{equation}
D(\theta,\phi) =\frac{p(\theta,\phi)}{P_{\textrm{r}}/4\pi} \textrm{,}
\end{equation}
which is defined as the ratio of the radiation intensity $p(\theta,\phi)$ to the total radiated power $P_{\textrm{r}}=\int p(\theta,\phi) {\rm sin} \theta d\phi d\theta$ per unit solid angle (corresponding to an ideal isotropic radiator). An equally important figure of merit is the antenna \textit{gain}, which is defined as the ratio of $p(\theta,\phi)$ to the total input power $(P_{\textrm{r}}+P_{\textrm{nr}})$ that is to be re-radiated per unit solid angle (corresponding to the power that would be radiated by an antenna with no losses). Obviously, gain $G$ and directivity $D$ are related by the radiation efficiency of the antenna:
\begin{equation}
G(\theta,\phi)=\frac{p(\theta,\phi)}{(P_{\textrm{r}}+P_{\textrm{nr}})/4\pi}=\eta D(\theta,\phi) \textrm{.}
\end{equation}
These and other relevant figures of merit are of course strongly frequency-dependent. Therefore, in antenna design it is important to specify the \textit{bandwidth} over which a certain performance is achieved.

\subsection{Reciprocity theorem}
\label{reciprocitytheorem}
So far we have considered antennas mostly as devices which create e.m.~waves. However, naturally, antennas can also be used to collect e.m.~waves. One may ask whether there is a relation between the ability of an antenna to emit e.m.~waves and its ability to collect them. Indeed, such relations exist and are typically discussed by calling upon different forms of reciprocity theorems. We will not give any derivation here, but only state the most important reciprocity relation for antenna-like e.m.~systems and mention the conclusions that can be drawn. Assuming time-harmonic fields in linear media in which the tensors $\epsilon$ and $\mu$ are symmetric, the reciprocity theorem, sometimes referred to as the Rayleigh-Carson reciprocity theorem, reads as \cite{LandauLifshitz}
\begin{equation}
\label{reciprocity}
\int\!{\bf j}_1\cdot{\bf E}_2\;dV_1 =\int\!{\bf j}_2\cdot{\bf E}_1\;dV_2\textrm{,}
\end{equation}
where ${\bf j}_i$ $(i=1,2)$ are time-harmonic source currents which may run through antenna wires and ${\bf E}_i$ $(i=1,2)$ are the corresponding fields that originate from the respective currents. Note that Eq.~(\ref{reciprocity}) describes a situation with two independent currents and the resulting fields, i.e.~two antennas. The integrals in Eq.~(\ref{reciprocity}) only run over the volume of the respective source currents because the integrands vanish everywhere else. Eq.~(\ref{reciprocity}) can be used to proof (i) that the shape of the angular receiving pattern of an antenna equals that of its angular emission pattern \cite{lee,balanis} and (ii) that the ratio of the power delivered from the first antenna to the second antenna and the power supplied to the first antenna is equal to the ratio of the power delivered from the second to the first antenna and the power supplied to the second antenna \cite{balanis}. These two reciprocity relations for antennas are very useful in antenna engineering and remain valid also at optical frequencies where they are more and more frequently used \cite{Dorfmuller10}.

In the case of an optical antenna that is coupled to a quantum emitter in its vicinity, Eq.~(\ref{reciprocity}) can be used to derive a reciprocity relation that links the polarization- and angle-of-incidence-dependent rate for excitation $\gamma_{exc,i}(\theta,\phi)$ and the radiative decay rate $\gamma_{\rm rad}$ via the polarization-dependent directivity $D_{i}(\theta,\phi)$ \cite{Bharadwaj09,novotny}

\begin{equation}
\frac{\gamma_{exc,i}(\theta,\phi)}{\gamma^o_{exc,i}(\theta,\phi)}=\frac{\gamma_{\rm rad}}{\gamma^o_{\rm rad}}\frac{D_{i}(\theta,\phi)}{D^o_{i}(\theta,\phi)} {\rm ,}
\end{equation}
where $i \in \{\theta,\phi\}$ denotes the two polarization directions of the transverse radiated far fields and $^o$ denotes quantities in absence of the antenna. In the derivation a second dipolar emitter is assumed in the far field of the antenna. The excitation rate $\gamma_{exc,i}(\theta,\phi)$ is therefore the rate at which the antenna-coupled emitter is excited by a plane wave with polarization $i$ and direction $(\theta,\phi)$. Note that this reciprocity relation cannot make any predictions about the nonradiative decay rate, i.e. Ohmic losses that occur in the antenna, since nonradiative antenna modes are near-field effects and therefore are independent of the reciprocity consideration.

\subsection{What RF-antenna engineers may be concerned with}
The question which antenna performs best for a given application is often not easy to answer since contradicting requirements, e.g.~a strongly directed emission pattern and a large bandwidth or a small overall size and a large radiation resistance, need to be combined. Radiowave antenna engineering strongly benefits from the fact that metals at radio frequencies can be considered to be nearly lossless. This makes it possible to screen a very large variety of antenna shapes to achieve a certain performance without having to pay much attention to the radiation efficiency. Things start to change as we move towards shorter wavelengths, and already for the microwave regime (the THz domain) losses become a constraint that antenna and circuit engineers have to deal with.

The simplest antenna circuit we can think of has already been drawn in Fig.~\ref{fig:transmissionline}(c). In the general case, the impedance discontinuity at the load will result in reflection of the forward-traveling voltage wave, with a reflection coefficient given by \cite{cheng}
\begin{equation}
\label{reflection}
\Gamma=\frac{Z_{\rm L}-Z_0}{Z_{\rm L}+Z_0}\rm{.}
\end{equation}
Since impedances in general also possess an imaginary part related to their reactive properties, the reflection coefficient is a complex quantity, describing both the amplitude of the reflected back-traveling signal and its phase relation with the forward wave. From Eq.~(\ref{reflection}), it is clear that reflectionless coupling can be achieved when the characteristic impedance of the transmission line matches the antenna impedance, i.e.~when $Z_{\rm L}=Z_0$.
In an unmatched situation it is possible that the antenna is on resonance but very little power is delivered to it via the transmission line because of a large impedance mismatch. This is a situation that occurs for example for an antenna with $L=\lambda_0$ in which the current vanishes in the gap according to Eq.~\ref{current_eq}. Although it has favorable properties, such an antenna cannot be fed by connecting wires at the feed-gap since the related antenna impedance diverges leading to a strong impedance mismatch.

In order to be able to efficiently deliver energy to an antenna, RF antenna engineers have developed strategies to achieve efficient impedance matching between the generator and the antenna even for exotic antennas. This is often achieved using external circuits, e.g.~passive stubs, which consist of short pieces of transmission lines connected in series or parallel close to the antenna feed point \cite{lee}. Such matching circuits act as resonators, storing a considerable amount of power, and modify the phase and amplitude of the reflected voltage wave. For perfect conductors, only a small amount of power is consumed by such passive matching elements. Therefore the overall radiation efficiency of the antenna including the matching circuit is only slightly smaller than that of the antenna alone. However, for higher frequencies, this is not true anymore, and standing-wave stub currents need to be properly minimized to keep losses low. Therefore, for antennas at optical frequencies, such strategies cannot be copied without careful consideration, since stub-like resonator structures as well as antenna circuits that exhibit a rather large number of passive elements may have rather strong losses and consequently a strongly reduced overall radiation efficiency. Moreover, while everything can be intuitively understood in terms of voltage-wave reflection, things can become more subtle when power reflection is considered \cite{Kurokawa65,Huang09b}. The question of how to feed an antenna in an optimal way will be of particular importance for nanoantennas at optical frequencies as we will discuss later on.

\section{Properties of metals at optical frequencies}
The constricted electron gas needed to build an antenna is very often provided by metals. However, at optical frequencies metals no longer behave as perfect conductors. Their optical response is described by a complex frequency-dependent dielectric function $\varepsilon(\omega)=\varepsilon_1(\omega)+i\varepsilon_2(\omega)$, relating the electric field $ E(\omega)$ and the induced polarization density as $P(\omega)=\varepsilon_0 [\varepsilon(\omega)-1]E(\omega)$ \cite{Jackson}.
In order to qualify for use in optical antennas, Ohmic losses in the chosen metal should be as low as possible. Ohmic absorption is proportional to the material conductivity $\sigma(\omega)$, which in turn is related to $\varepsilon(\omega)$ by $\varepsilon_2(\omega)=\frac{\sigma(\omega)}{\varepsilon_0 \omega}$, and Ohmic losses take place in close proximity to the surface within the so-called penetration depth \cite{maier}. Typical penetration depths for metals at visible frequencies are of the order of several tens of nm, e.g. about 13 and 31~nm for aluminum and gold at 620~nm wavelength, respectively \cite{Kreibig95}.
Material losses in metal nanostructures can be kept low either by choosing a metal with large (negative) real part of $\varepsilon(\omega)$, in order to reduce the penetration depth, or by selecting a low imaginary part of $\varepsilon(\omega)$, in order to intrinsically keep the Ohmic losses low.

Moreover the dielectric properties of a metal, as we will see, can cause a particle plasmon resonance in the visible spectrum, which is connected to large local fields and enhanced scattering. For small particles in vacuum, on resonance, the real part of $\varepsilon(\omega)$ takes on the value $\varepsilon_1(\omega)=-2$ for a wavelength in the blue-green region, so that elongated particles and dimers show a resonance in the red or near-IR region, as will be discussed. Since $\varepsilon_1(\omega)=-2$ is not very large, the imaginary part of $\varepsilon(\omega)$ cannot be neglected when discussing plasmonic resonances.

\subsection {Drude-Sommerfeld model}
The optical response in metals is dominated by the collective behavior of the free electron gas. To a first approximation, the conduction electrons in the metal can be treated as an ideal electron gas moving in the background of the positive metal ions. Using the Drude-Sommerfeld model, the dielectric function of the metal can be expressed as
\begin{equation}
\varepsilon_{\rm{Drude}}(\omega)=1-\frac{\omega_{\rm{p}}^2}{\omega^2+i\gamma\omega}\rm{,}
\end{equation}
where $\omega_{\rm{p}}$ is the volume plasma frequency, which increases with increasing carrier density, and $\gamma$ a damping constant \cite{maier,novotny}. For noble metals at optical frequencies, typically $\omega<\omega_{\rm{p}}$ (e.g. for Au $\omega_{\rm{p}} \simeq 13.8\cdot 10^{15}$ s$^{-1}$ and $\gamma \simeq 1.07 \cdot 10^{14}$ s$^{-1}$ \cite{novotny}) and therefore this model accounts for (i) a negative real part, meaning that the conduction electrons do not oscillate in phase with the external field, which is - by the way - the reason for the high reflectivity of metal surfaces, and (ii) a significant imaginary part.

\subsection{Interband transitions}
\begin{figure}[htbp]
        \centering
        \includegraphics[width=0.5\textwidth]{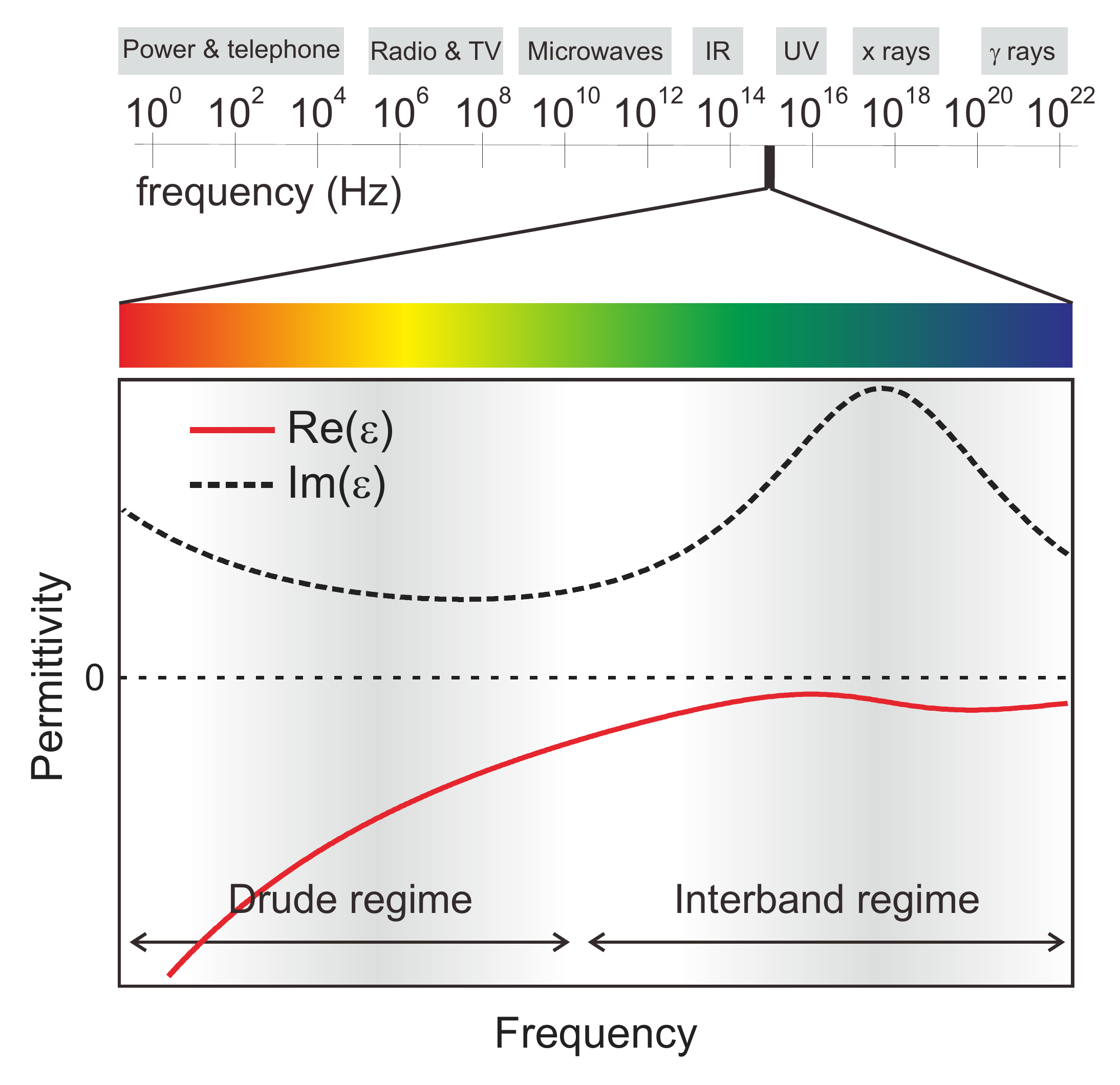}
        \caption{Sketch of a typical dielectric function of a metal at optical frequencies which represent a small part of the whole spectrum of e.m.~waves (top). An interband-transition peak, visible in the imaginary part of $\varepsilon(\omega)$, is superimposed to the monotonic Drude-like behavior of the free electron gas.}
        \label{fig:drude}
\end{figure}

The Drude-Sommerfeld model does not account for the possibility that photons with high-enough energy cause interband transitions by promoting electrons from lower-lying valence bands to higher-energy conduction bands. This further degree of freedom is related to bound electrons and can classically be described by a collection of damped harmonic oscillators with well-defined resonance frequencies $\omega_0$, yielding contributions to the dielectric response of the type
\begin{equation}
\varepsilon_{\rm{Lorentz}}(\omega)=1+\frac{\tilde\omega_{\rm{p}}^2}{(\omega_0^2-\omega^2)-i\tilde\gamma\omega}\rm{,}
\end{equation}
where $\tilde\omega_{\rm{p}}$ depends on the density of bound electrons involved in the absorption process and $\tilde\gamma$ is a damping constant for the bound electrons. This absorption channel leads to a strong deviation from the free electron gas model near $\omega_0$, leading to a maximum in the imaginary part of $\varepsilon(\omega)$ and therefore to strongly increased damping. In Fig.~\ref{fig:drude} we schematically show the combined contribution of free and bound electrons to the complex dielectric constant of a typical metal in the visible.

\subsection{Comparison of relevant metals}
\label{comparison}
The choice of the best plasmonic material for a given application is still a subject of discussion and research \cite{Boltasseva11,West10}. Mainly Au, Ag, Al, and Cu have been considered so far to be used as materials for metallic optical antennas. The respective material-dependent spectral behavior is discussed in \cite{Mohammadi09}. Here we would like to mention the most relevant properties. Gold and copper have very similar dielectric constants, with a Drude-like response below 2.1~eV (wavelength $> 600$~nm) and an onset of interband transitions occurring around 2.3~eV (530-550~nm). This makes them excellent candidates to build antennas for the red and near-IR spectral region. Other materials should be preferred for the blue-green part of the visible spectrum. For silver the first interband transition is above 3.1~eV (wavelength $<400$~nm), which makes it superior to gold for wavelengths around 450-550~nm. Finally, aluminum has a larger (negative) real part of the dielectric function, and can therefore be considered the material which among these four metals best approximates an ideal metal, especially in the 400-600~nm spectral region. However, it has an interband absorption peak located around 800~nm wavelength, so that its use in the near-IR region is problematic. Apart from the spectral properties of the dielectric function, also the chemical stability of antenna materials is an experimentally relevant issue. Ag and Cu are known to quickly corrode in ambient conditions (formation of oxides and sulfides), while Al is known to form thin passivation layers of Al$_2$O$_3$. Au is the material that is mostly used experimentally, since it combines a favorable dielectric function in the red and near IR with excellent chemical stability.

\section{Properties of isolated optical antennas}
\begin{figure}[htbp]
        \centering
        \includegraphics[width=0.5\textwidth]{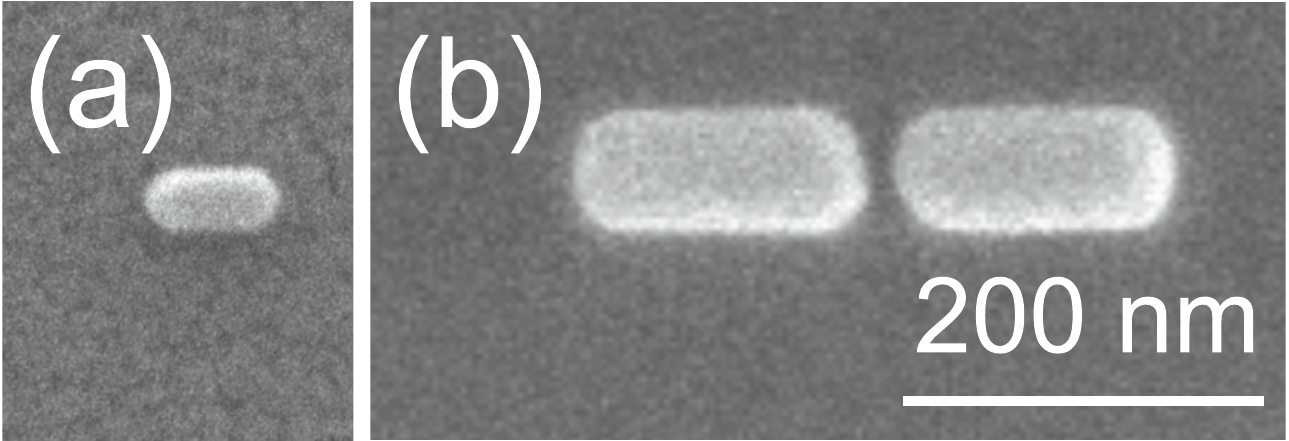}
        \caption{SEM images of (a) a single-wire antenna (colloidal Au nanorod) \cite{Nanopartz}  and (b) a two-wire optical antenna (produced by focused-ion beam milling, see Section \ref{fabrication}) \cite{Huang10b}. Scale bar is the same for both panels.}
        \label{dipole_antenna}
\end{figure}

We now introduce important models for the description of {\em optical} antennas. In contrast to RF antennas that always appear as circuit elements connected to a feeding circuit, optical antennas often appear as {\em isolated} structures whose resonant properties will be discussed now. The most basic optical antenna geometries are single- and two-wire antennas, consisting of a single nanorod and of two end-to-end aligned rods separated by a small gap, respectively. Scanning electron microscopy (SEM) images of respective prototypes (a single-crystalline colloidal rod \cite{Nanopartz} and a nanostructured two-wire antenna \cite{Huang10b}) are shown in Fig.~\ref{dipole_antenna}. A single wire can be viewed as the fundamental building block of more complex antennas. We therefore start with a  discussion of single-particle optical resonances first. We will introduce the Mie description of optical resonances which is often used but is not very intuitive. Much more physical insight can be obtained by introducing a mass-and-spring model as well as a Fabry-P\'{e}rot model for the optical resonances of elongated particles. The Fabry-P\'{e}rot model connects to the RF theory but takes into account the strongly reduced wavelength of plasmonic wire waves.
Towards more complex structures, the fundamental two-wire nanoantenna is of particular interest due to the strongly enhanced and deeply-subwavelength-confined fields that occur in its feed-gap upon external illumination. Using the mass-and-spring model we will discuss the more complex spectra of two-wire nanoantennas which arise due to mode hybridization caused by the strong e.m.~interaction between the particles. In order to illustrate these concepts we will present finite-difference time-domain (FDTD) simulations \cite{fdtd} of fundamental prototype structures. Finally, we will discuss the differences in the emission patterns between RF and optical antennas.

\subsection{Single-particle plasmon resonances}
\subsubsection{Mie description}
Localized plasmon resonances are resonant collective oscillations of the electron cloud in a metal nanoparticle, originating from the characteristic dielectric response of metals at optical frequencies. They are accompanied by resonantly enhanced polarizabilities and accordingly enhanced scattering and absorption as well as enhanced near-field intensities. The response of a spheroidal object to plane-wave illumination is analytically described in the frame of Mie theory. When applied to sub-wavelength particles, only the first-order dipolar term needs to be considered \cite{bohren}. To illustrate this point of view, let us consider a sphere of polarizable material with radius $r$ and  dielectric constant $\varepsilon$, embedded in a medium with dielectric constant $\varepsilon_{\rm{env}}$, under the influence of a static electric field $\mathbf{E}_0$. The dipole moment induced in the sphere by the external field can be written as \cite{bohren}
\begin{equation}
\mathbf{\mu}_{\rm{p}}=4 \pi r^3 \varepsilon_0 \frac{\varepsilon-\varepsilon_{\rm{env}}}{\varepsilon+2\varepsilon_{\rm{env}}}\mathbf{E}_0 \; .
\label{dipole1}
\end{equation}
For $\varepsilon_{\rm{env}}=1$ (vacuum) this expression exhibits a resonance when the real part of $\varepsilon$ approaches $-2$. When we move to optical frequencies and consider now a plane wave with wavelength $\lambda_0$ illuminating a very small metal sphere (radius $r\ll \lambda_0$), the external field can be considered constant over the particle. In this quasistatic approximation, the phase is also constant over the particle and retardation effects can be neglected. Therefore Eq.~(\ref{dipole1}) is still valid, however, the static dielectric constant $\varepsilon$ is replaced with $\varepsilon(\omega)=\varepsilon_1(\omega)+i\varepsilon_2(\omega)$ \cite{novotny}. The resonance condition $\varepsilon_1(\omega)\rightarrow -2$ is met for particles consisting of Au, Ag, and Cu around the visible range, and for Al in the near-ultraviolet range. On resonance, the vanishing real part of the denominator of Eq.~(\ref{dipole1}) leads to a strongly increased induced dipole moment and therefore to enhanced local fields and scattering. However, due to the spectral position of the interband transitions, particles consisting of different materials still show different optical properties, as discussed in Section \ref{comparison}, and are therefore suitable for applications in different spectral regimes.

Let us now move from the plasmonic resonances of nanospheres to those of elongated prolate particles. We are interested in charge oscillations along the main axis (length $d$) of such an ellipsoid. Quantitatively, by means of a simple transformation of Eq.~(\ref{dipole1}) one can obtain the induced dipole moment of a prolate ellipsoid of volume $V$, being the prototype of an elongated particle, which reads as \cite{bohren}
\begin{equation}
\mathbf{\mu}_{\rm{p}}=V \varepsilon_0 \frac{\varepsilon-\varepsilon_{\rm{env}}}{P_j\varepsilon+(1-P_j)\varepsilon_{\rm{env}}}\mathbf{E}_0 \; .
\label{dipole2}
\end{equation}
Here,  $P_j$ is a function of the {\em aspect ratio R}, i.e. the ratio of the long axis radius $d/2$ to the short axis radius $r$ of the particle. Detailed analysis of the resonance positions as a function of the aspect ratio using Eq.~(\ref{dipole2}) shows that the resonance position depends approximately linearly on the aspect ratio such that an increased aspect ratio leads to a red-shift of the resonance. Roughly speaking, for gold, visible wavelengths are covered by sub-wavelength particles with aspect ratios ranging from 1 to 3 \cite{Sonnichsen}.

\subsubsection{Mass-and-spring model}
\label{mass_spring}
\begin{figure}[htbp]
        \centering
      \includegraphics[width=0.5\textwidth]{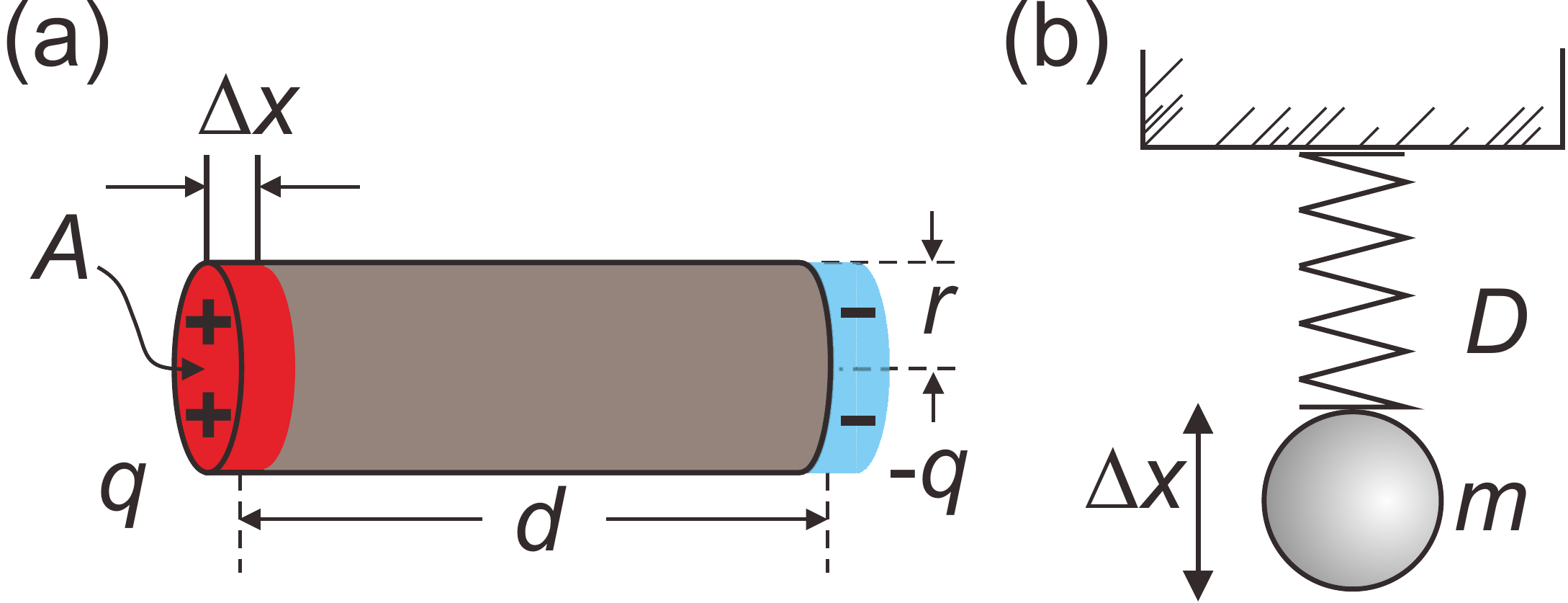}
        \caption{Mass-and-spring model for plasmonic resonances: (a) Sketch of a plasmonic particle whose electron cloud has been displaced by $\Delta x$. The resulting positive and negative charge at the ends are treated as point-like charges that possess potential energy due to Coulomb interaction. (b) The resulting oscillation can be modeled by an effective spring constant $D$ and the effective mass $m$ of the moving electron cloud.}
        \label{massandspringmodel}
\end{figure}

Although the quasistatic approximation of Mie theory provides a good prediction of the resonances of prolate particles within its range of validity, it provides little physical insight into what is the cause of the linear scaling of the resonance energy with the aspect ratio. To capture the physics of a non-spherical plasmonic particle, we aim at representing the plasmon resonance by a simple mass-and-spring model with a resonance frequency $\omega_{\rm res} = \sqrt{D/m}$, where $D$ and $m$ are the elastic constant of the restoring force and the total effective mass of the electron system, respectively. To estimate the restoring force we consider an elongated particle with dimensions given in Fig.~\ref{massandspringmodel}(a).
We assume that the particle has cylindrical shape. When the electron cloud in the particle is displaced by $\Delta x$, opposite point charges $\pm q$ build up at both ends whose magnitude depends on the charge carrier density $n$ and the cross-sectional area of the cylinder $A$ as $q=n e A \Delta x$, where $e$ is the elementary charge. The Coulomb potential energy of the two charges is then
\begin{equation}
W(\Delta x) = \frac{1}{4\pi \varepsilon_{\rm o}}\frac{q^2}{d}= \frac{1}{4\pi \varepsilon_{\rm o}}\frac{(n e  A)^2}{d} \Delta x^2 \; .
\label{potential}
\end{equation}
The restoring force can now be determined as
\begin{equation}
F(\Delta x) =-\frac{\partial W(\Delta x)}{\partial \Delta x} = -\frac{1}{2\pi \varepsilon_{\rm o}} (n e)^2 \frac{A^2}{d}  \Delta x = - D \Delta x \textrm{,}
\end{equation}
from which the spring constant $D$ is obtained. The linear relation between displacement and the resulting force leads to harmonic oscillations of the system, which allows drawing an analogy with a simple mass-and-spring model. The relevant mass that is involved is the mass of the whole electron cloud which is given by $m = n  m_{\rm e}  A  d$, where $m_{\rm e}$ is the effective electron mass. We therefore obtain the approximate resonance frequency $\omega_{\rm res}$ of the particle plasmon of an elongated particle
\begin{equation}
\omega_{\rm res} = \frac{\omega_{\rm p}}{2\sqrt{2}} \frac{1}{R} \rm{,}
\label{resonance-eleongated-particle}
\end{equation}
where we substituted the plasma frequency $\omega_{\rm p}^2=ne^2/(\varepsilon_{\rm o} m_{\rm e})$ as well as $A=\pi r^2$.  Due to the fact that in this simple model we assume that the charges at the end are more or less point-like, we cannot expect the result to reproduce the exact resonance frequencies for shorter and thicker particles, however, the trend that the resonance frequency is inversely proportional to the aspect ratio $R$ is nicely reproduced.

The physical reason for the aspect ratio scaling behavior lies in the fact that the electric field of the charge distribution has a dipolar character. Here, it is worth noting that such a linear dependence on the aspect ratio no longer exists as the radius of the rod gets larger compared to the wavelength \cite{Prescott06,Bryant08}.
For a homogeneous field, like in a plate capacitor, which would be a good description for a sufficiently extended system, a resonance in the plasma oscillations occurs at the bulk plasma frequency $\omega_{\rm p}$ independent of the geometry \cite{maier}. The possibility to describe the resonances and resonance shifts of plasmonic particles by using simple mechanical models will be picked up again for the discussion of coupling effects in more complex antenna systems. Note that the mass-and-spring model also accounts for a shift in the resonance peak position between a resonance spectrum measured in the near field vs.~a far-field spectrum. This shift is related to the Ohmic damping of the resonance \cite{Zuloga11}.

\subsubsection{Fabry-P\'{e}rot model}
\label{fabryperot}
In order to connect again to RF theory and to get a better idea about the nature of the eigenmodes of plasmonic structures, we introduce a further point of view. To understand the resonances of single-wire antennas, one may also start by considering the fundamental guided modes of thin metal wires with a complex dielectric function. Modes that are propagating along such wires can be qualitatively classified as having mostly either ``surface'' or ``bulk'' character \cite{Novotny94}. Due to the large imaginary part of the dielectric function and a skin depth that, in the optical regime,  is often comparable to the wire diameter, all modes suffer from exponential damping and possess a finite propagation length. Surface-like modes result from collective surface oscillations of electrons propagating along the wire. They are associated with near fields that evanescently decay into both the metal and the surrounding dielectric. As a consequence of the evanescent decay in the transverse direction, these guided modes must have a shorter effective wavelength and therefore a reduced propagation speed compared to light in vacuum \cite{Novotny94}.

\begin{figure}[htbp]
        \centering
        \includegraphics[width=0.5\textwidth]{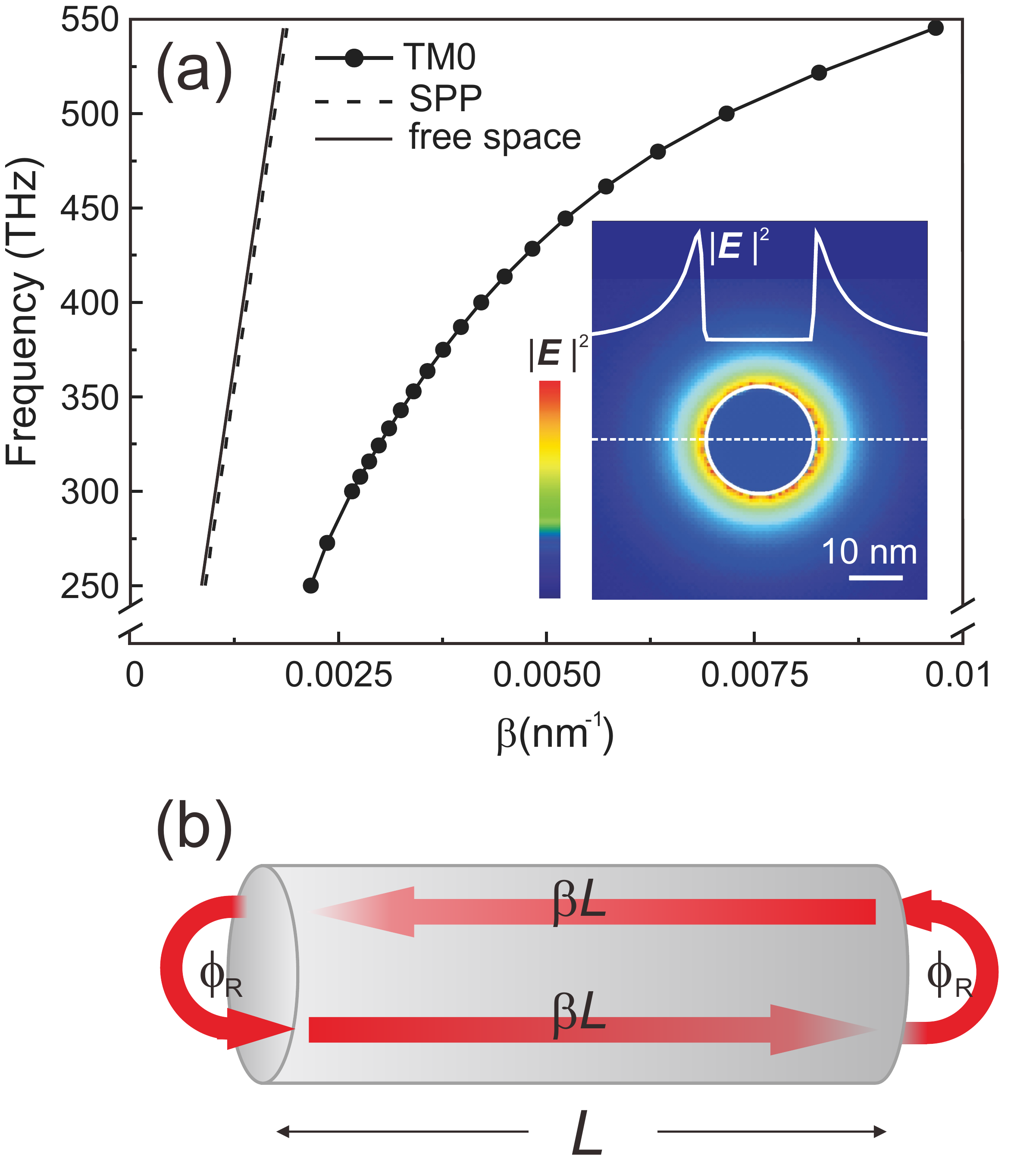}
        \caption{Resonances in a single-wire antenna: (a) Simulated dispersion relation for the fundamental TM0 mode on a Au wire (radius 10~nm) in vacuum, by the finite-difference frequency-domain method \cite{fdfd}, compared to that of free-space propagation and to that of a surface plasmon polariton (SPP) propagating at the Au/air interface (which in this energy range basically coincides with the free-space dispersion relation). The inset shows the mode profile of the guided mode; (b) Sketch of accumulated phase contributions upon propagation and reflection in a truncated wire, leading to Fabry-P\'{e}rot resonances.}
        \label{tm0}
\end{figure}

Consider, for example, an infinitely long metal wire with a circular cross-section situated in vacuum. Its fundamental TM0 mode is rotationally symmetric with respect to the wire axis, as shown in the inset of Fig.~\ref{tm0}(a). The field amplitude of such a mode along the wire can be expressed as ${\bf E}(x,y,z)={\bf E}(x,y,0)e^{-\gamma z}$, where the $z$-axis coincides with the wire direction and $\gamma=\alpha+i\beta$ is the complex propagation constant. As the wire radius is decreased, $\beta$ increases and diverges, thus resulting in an almost ideal, one-dimensional waveguide \cite{Takahara97}. This is an effect also exploited in the adiabatic focusing of plasmons in a tapered wire \cite{Stockman04,Janunts05}. Although the mode has no cut-off, squeezing
the radius results in a larger confinement of fields inside the metal wire and therefore increases the losses due to Ohmic damping \cite{Takahara97}. For a fixed wire radius, the guided mode exhibits a dispersion relation as illustrated in Fig.~\ref{tm0}(a). The deviation from the free-space light line is an important  feature of surface plasmon modes on noble-metal nanowires, and the corresponding reduced wavelength plays a crucial role in optical antenna design \cite{Dorfmuller10,Novotny07,taminiau09}, as we will discuss below.

A single-wire antenna of length $L$ can be pictured as a finite piece of such a wire waveguide. While mode propagation along an infinitely long metal wire is not accompanied by radiation, the single-wire antenna can radiate significantly due to the broken translational symmetry. The two open ends represent mirror-like discontinuities with a near unity reflection coefficient for the fundamental TM0 mode. In such a one-dimensional cavity, a standing wave builds up once the accumulated phase per round trip equals an integer multiple of $2 \pi$. In other words, a Fabry-P\'{e}rot resonance builds up if the correct resonance length is chosen for the truncated wire \cite{Bryant08,Novotny07,taminiau09,Ditlbacher05,Douillard08,Kolesov09,Barnard08,Bozhevolnyi07,Aizpurua05,Dorfmuller09}. For a given wire cross-section and material, the resonance condition, when perfect reflection at the ends is considered, satisfies the simple relation $\beta L_{\rm{res}}=n \pi$, where $n = 1,2, ...$ is the resonance order. However, for plasmonic single-arm antennas, since the two open ends possess a strongly reactive character at optical frequencies \cite{Feigenbaum08}, the fields extend outside the physical boundaries of the metal structure, which results in a phase shift $\phi_{\rm{R}}$ of the fundamental TM mode upon reflection. Such a phase shift has the same effect as some additional length of propagation \cite{cheng}. As a result, the effective length experienced by the mode bouncing back and forth along the wire is different from the actual rod length, and an offset must be added to take such a phase shift into account \cite{Novotny07,taminiau09,Dorfmuller09}. A simple relation between the antenna length $L_{\rm{res}}$ and the mode wavelength $\lambda=2\pi / \beta$ for the $n$-th order resonance therefore reads as:
\begin{equation}
\beta L_{\rm{res}}+\phi_{\rm{R}}=n \pi \textrm{,}
\label{lambda_eff}
\end{equation}
where $\phi_{\rm{R}}$ is strongly dependent on the actual end-cap geometry. This description, which is also sketched in Fig.~\ref{tm0}(b), retains its validity for arbitrary arm cross-section, provided the proper mode constant $\beta$ is considered. Interestingly, it has been noticed that if the phase shift upon reflection can be engineered to become negative, plasmonic systems can support a so-called zero-order mode resonance for which $n=0$ \cite{Feigenbaum08,Maier06b}.

Eq.~(\ref{lambda_eff}) clearly shows a linear relation between the resonance length $L_{\rm{res}}$ of the rod and the wavelength $\lambda=2\pi / \beta$ of the propagating mode. An effective wavelength scaling rule, relating the mode wavelength $\lambda$ with the free-space wavelength $\lambda_0$, has also been analytically derived for a given wire radius \cite{Novotny07}. Notably, for the case of Drude-like dielectric properties, it has been shown that a linear relation of the form
\begin{equation}
\lambda=a+b \lambda_0
\label{novotny_scaling_law}
\end{equation}
holds, where $a$ and $b$ are wavelength-independent coefficients \cite{Novotny07}. Deviations from this ideal linear scaling law arise at frequencies where the optical response of the metal is dominated by interband transitions.

From the two scaling laws that have just been discussed (Eqs. \ref{lambda_eff} and \ref{novotny_scaling_law}), it is then clear that for a Drude-like material a linear relationship is also expected between the resonance length $L_{\rm{res}}$ of a single wire and the  free-space wavelength $\lambda_0$ used for illumination. This fact is well established and has been discussed before in section \ref{mass_spring} in terms of the linear dependence of the resonance position of an elongated particle on its aspect ratio \cite{Sonnichsen,Bryant08,Link99}.

\subsection{Resonances of two-wire antennas}
\label{section:coupling}
The end-to-end coupling between two wires over a narrow gap can create highly localized and strongly enhanced optical near fields inside the gap. It is this effect which makes such an arrangement a highly efficient antenna for light. To understand this coupling it is useful to consider again the mechanical analogue of a plasmon resonance based on harmonic oscillators \cite{Rechberger03}. Since the external field creates oscillating surface charges on the nanoparticle, each rod can be thought of as a spring with a respective effective mass attached to it. If two end-to-end-aligned particles come close to each other, an additional spring needs to be introduced which accounts for the interaction between the surface charges on the ends of both particles that are facing each other. This effect becomes significant for distances comparable with the wire diameters.
\begin{figure}[htbp]
\centering
\includegraphics[width=0.5\textwidth]{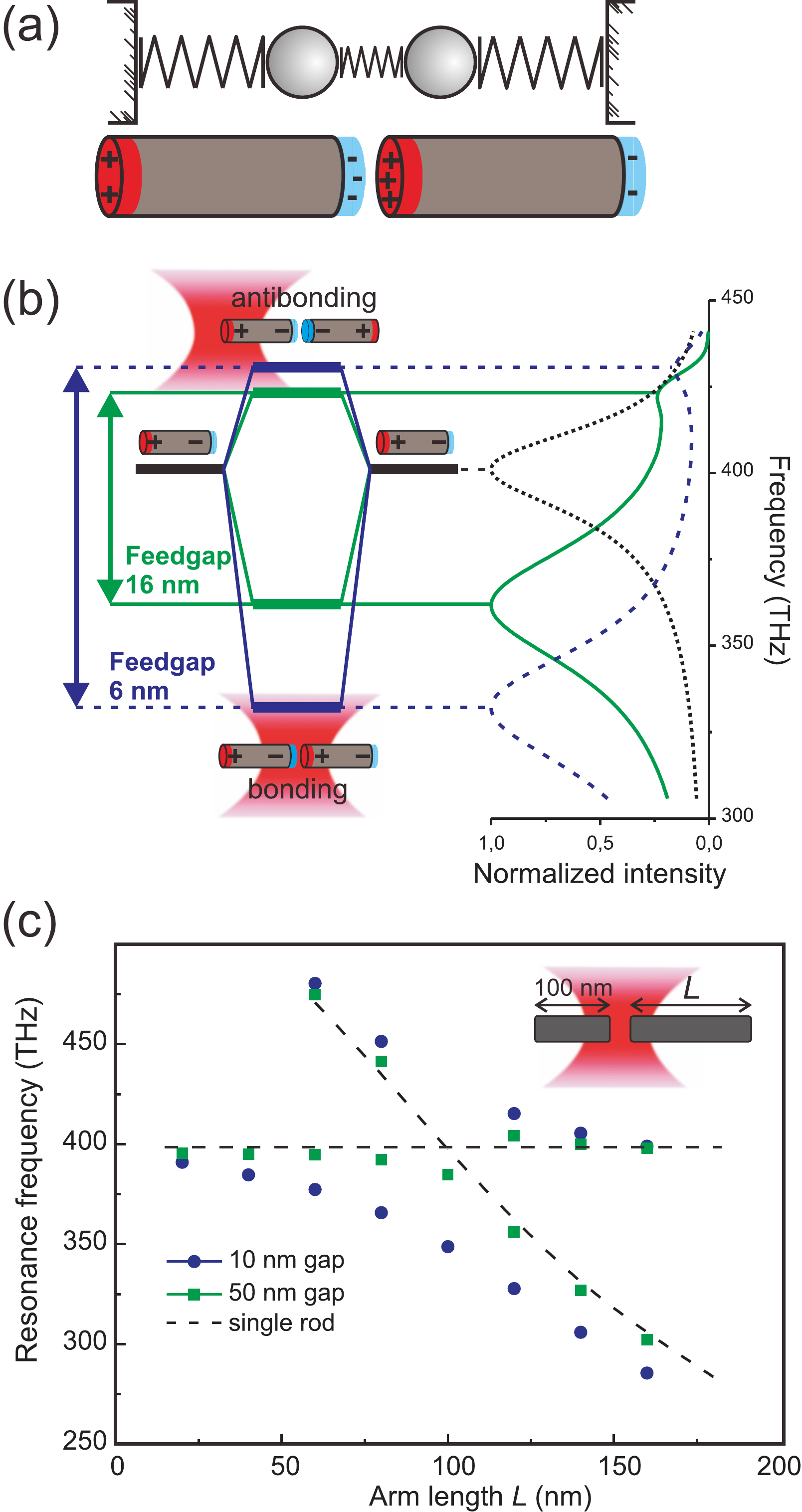}
\caption{Interparticle coupling and mode splitting in two-wire antennas: (a) Sketch of the mass-and-spring model for the coupling between two plasmonic oscillators; (b) energy-level diagram and simulated near-field intensity spectra for 30~nm high, 50~nm wide, and 110~nm long (arm length) symmetric two-wire antennas with 6~nm (blue dashed) and 16~nm (green solid) gap, as well as single-wire antennas with same dimension (black dotted); (c) avoided-crossing behavior for the new eigenmode frequencies in a system of two coupled rods observed as the resonance of one rod is tuned via its length to cross the resonance of the second rod. Note that data points for the antibonding mode are missing for the case of a symmetric dipole antenna since the excitation used in the simulation is fully symmetric for this case. The position of the illuminating beam in the simulations is also sketched in (b) and (c). Panel (b) adapted with permission from Huang \textit{et al.} \cite{Huang10}. Copyright 2010, American Chemical Society.}
\label{coupling}
\end{figure}
Fig.~\ref{coupling}(a) illustrates the basic idea of such a coupled-harmonic-oscillator system. The coupling of the two spring-mass systems (antenna wires) through a third spring (antenna feed-gap) in this picture results in the appearance of two new eigenmodes. One eigenmode exhibits in-phase oscillation of the two springs, for which the interaction spring has fixed length and therefore does not exert any additional force on the masses. The other eigenmode is characterized by a respective antiphase oscillation in which the interaction spring shifts the resonance to higher frequencies. This very simple and intuitive classical model already contains the most characteristic features of strongly coupled systems \cite{Novotny10}.

The interparticle coupling can also be described in terms of a hybridization model \cite{Huang10,Prodan03,Nordlander04,Funston09,Halas11} which strongly relates to molecular orbital theory, where the overlap of wave functions is taken into account to derive the energy splitting and symmetry character of ``bonding'' and ``antibonding'' orbitals. In general, for ensembles of plasmonic particles one may state that whenever modes are spectrally and spatially overlapping, their coupling generates new resonances with an energy splitting, analogous to atomic orbital hybridization. We conclude that in a linear dipole nanoantenna consisting of two identical nanorods separated by a subwavelength gap, the two individual (degenerate) fundamental single-wire resonances split into two resonances. This behavior is sketched in Fig.~\ref{coupling}(b) \cite{Huang10}.

The bonding resonance mode, which is red-shifted compared to the original single-wire resonance, is characterized by dipole-like charge oscillations. Due to its dipolar character, this oscillation mode can be excited by plane-wave illumination polarized along the antenna axis. It couples strongly with the radiation field via dipole radiation in addition to some Ohmic damping. As a result, lower-energy bonding antenna modes appear ``bright'' in spectroscopic measurements under plane-wave excitation \cite{Dorfmuller09,Funston09}. It is noteworthy that the red shift of this mode is a feature that is not reproduced by the simple mass-spring model described above, where the in-phase oscillation frequencies of the two coupled systems are degenerate with those of each isolated system. This observation can still be interpreted in the frame of such a model if one allows for a reduced restoring force of the two single-particle springs. Phenomenologically, such a weakened spring constant is related to the mutual induction of charges,  which in the coupled system are displaced towards the gap, thus reducing the effective restoring force [see also Fig.~\ref{sim}(h) and (i) later on].

The higher-energy, antibonding mode is characterized by a charge distribution which exhibits mirror symmetry with respect to the feed-gap [see Fig.~\ref{coupling}(b)]. As a consequence, the antibonding mode does not efficiently emit into the far field since the two individual dipoles oscillate out of phase and therefore cancel each other to a large extent in the far field. Furthermore, the antibonding mode cannot be excited in far-field spectroscopy as long as the illumination path remains fully symmetric \cite{Dorfmuller09,Funston09,Ghenuche08}. The antibonding mode therefore is designated as a ``dark'' mode. For these two reasons, most of the experiments so far reported mainly on the red-shifted bonding modes (see e.g. \cite{Funston09,Ghenuche08,Muhlschlegel05}). To excite the dark antibonding mode, the symmetry of the system needs to be broken. For a symmetric antenna this can be achieved by using asymmetric excitation conditions, which is the case for excitation by a localized point-like source \cite{Liu09}, total-internal-reflection \cite{Yang10}, tilted plane wave excitation\cite{Dorfmuller09}, or a displaced focused excitation beam \cite{Huang10}. Due to reduced radiation damping, dark resonances are also expected to have a higher quality factor \cite{Huang10}.

The coupling between the wires of a two-wire antenna strongly increases with decreasing gap width. The smaller the feed-gap size, the larger the energy splitting. The energy splitting between the antibonding and bonding antenna resonances is the analogue of the energy splitting in the avoided-crossing behavior seen in the adiabatic strong-coupling case of interacting quantum systems \cite{Novotny10}. Since the resonance frequency of each antenna wire can be tuned by its length independently, two-wire antennas provide the opportunity to visualize the strong coupling between nanoparticles by tracing the shifts of the maxima of the involved resonances. Fig.~\ref{coupling}(c) shows results from FDTD simulations of asymmetric two-wire antennas where we tune the resonance frequency of one wire from well above to well below the resonance frequency of the other one. The gap is set to be either 10 or 50~nm in order to visualize the gap-dependent coupling strength. The latter can be inferred from the energy splitting ``on resonance'' in analogy to the strong coupling between an atom and a cavity. A clear avoided-crossing behavior is observed. From an experimental point of view, provided a proper excitation geometry is used to excite both modes, such a clear avoided-crossing curve can only be obtained for small enough gaps because otherwise the splitting may not be sufficient to spectrally separate both rather broad resonances.

The coupling between nanoparticles plays a very important role with respect to their spectral properties. In fact, by its engineering, for example by controlling the assembly geometry of the nanoparticles, one may, in principle, shape the spectral response of a more complex system and  obtain desired optical properties \cite{Prodan03,Halas11,Fan10}.  Furthermore, coupling of bright and dark modes has found applications in plasmon-induced transparency in metamaterials \cite{Zhang08,Liu09b} and Fano-like resonances \cite{Fan10,Verellen09,Lukyanchuk10,Woo11}, where such coupling produces sharp dips in the broad resonance peak which may be useful in sensing applications.

\subsection{A case study of single- and two-wire antennas by simulations}
\label{somesimulations}
\begin{figure}[htbp]
        \centering
        \includegraphics[width=0.5\textwidth]{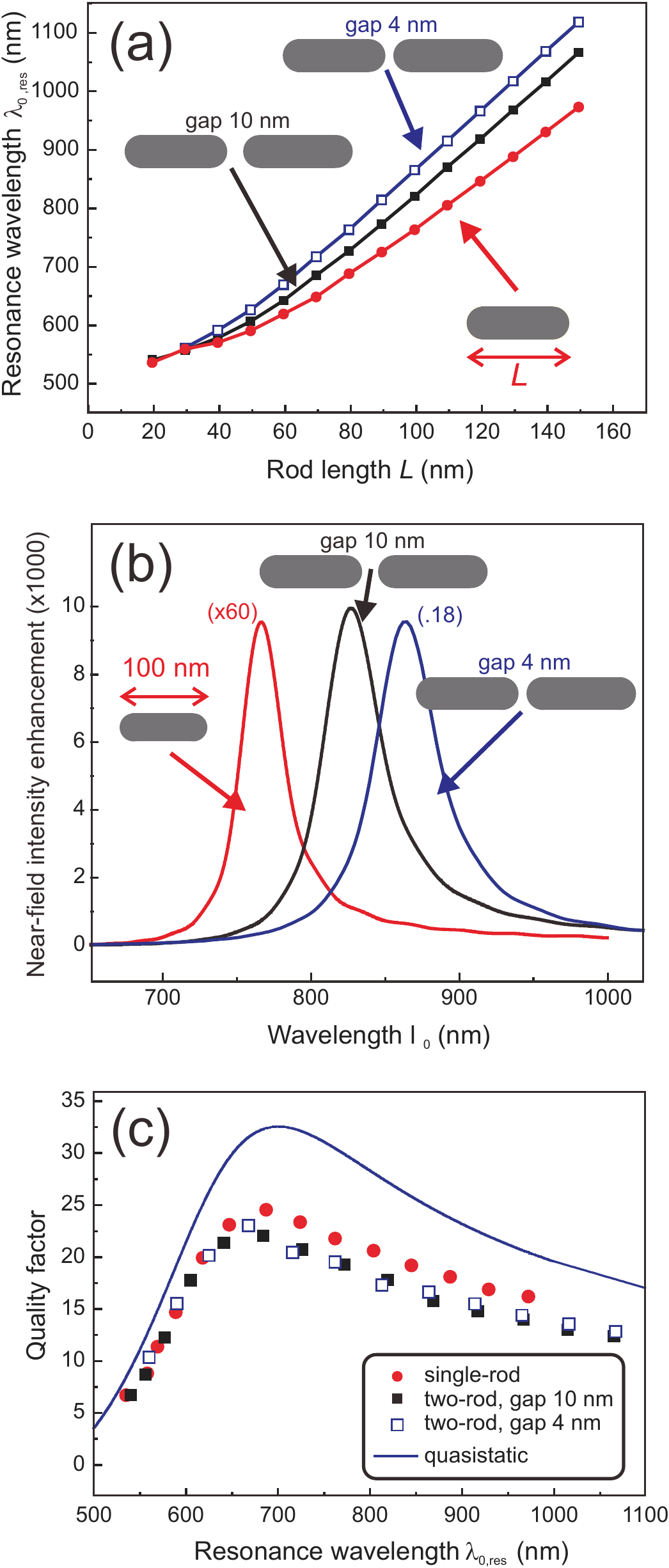}
        \caption{FDTD simulation results for single- and two-wire Au nanoantennas: (a) Resonance wavelength as a function of the rod length (red circles: single-wire antenna; black solid squares: 10-nm gap two-wire antenna; blue empty squares: 4-nm gap two-wire antenna); (b) representative near-field spectra for antennas constituted of 100-nm-long wires; (c) quality factor as a function of the resonance wavelength.}
        \label{sim}
\end{figure}

In order to illustrate what was introduced so far, we are now discussing the results of a computer experiment using a set of FDTD simulations \cite{taflove} of single- and two-wire optical antennas concentrating only on the bonding mode resonance. Each antenna wire is modeled as a Au cylinder with hemispherical end caps and 10~nm radius, in vacuum. The system is symmetrically illuminated with a centered Gaussian beam (0.6 numerical aperture), linearly-polarized along the wire axis, and near-field intensity spectra are recorded 5~nm away from the single-wire apex or in the middle of the gap for the two-wire antenna. The gap is set to be either 10 or 4~nm. From the simulated near-field spectra, the resonance wavelength and the quality factor for the antennas are determined. Simulation results are shown in Fig.~\ref{sim}. In panel (a) we plot the free-space wavelength $\lambda_{\rm 0, res}$ at which the rod is resonant as a function of the rod length $L$ for single-wire and two-wire antennas. The linear scaling behavior, resulting from the combination of Eq.~(\ref{lambda_eff}) and Eq.~(\ref{novotny_scaling_law}), is apparent in the red and near-IR portion of the spectrum, while deviations from linearity appear towards the green, where interband transitions become effective, as already discussed in Section \ref{fabryperot}.
Notably, the slopes of the linear segments are different for the single- and two-wire antennas. In the frame of the Fabry-P\'{e}rot description of resonances in linear antennas, this observation can be attributed to a modified reflection coefficient at the gap ends in each wire of the two-wire antennas, due to the close proximity of the other wire.

Representative near-field intensity spectra are shown in panel (b) for 100~nm-long wires. A redshift due to inter-wire coupling can clearly be observed. The resonance wavelength redshifts from about 770~nm to about 830~nm when going from a single-wire structure to a 10-nm-gap two-wire structure and further shifts to about 845~nm when the gap is reduced to 4~nm. Only the bonding mode resonance of the wires can be observed in these spectra due to the symmetric illumination. Panel (c) displays the quality factor Q of the resonances (dots) as a function of the resonance wavelength $\lambda_{\rm 0, res}$. It can be approximately calculated as the ratio $Q \simeq \frac{\lambda_{\rm 0, res}}{\Delta\lambda_0}$, where $\Delta\lambda_0$ is the full width at half maximum of the resonance. The higher the quality factor is, the longer the energy can be stored inside a resonator. Here the simulated quality factors are compared to the results of analytic calculations [solid line in panel (c)] in the quasistatic limit \cite{Wang06b}, where only Ohmic losses are considered for a point-like dipole plasmonic oscillator.
The low Q values obtained for the fundamental bonding modes in wire antennas can be attributed to the combined effect of Ohmic losses and radiation losses. The fact that the simulated Q factors qualitatively follow the trend obtained within the quasistatic approximation points to the fact that Ohmic losses dominate in these structures. The quasistatic approximation represents a fundamental barrier only for small particles which can be described in the quasistatic limit. For larger systems retardation effects are expected to lead to deviations from this idealized model. It has been shown, for example, that a very small cavity with a zero-order resonance can be designed by accurately engineering the reflection phase shift, and that in this way quality factors far beyond the quasistatic limit can be achieved \cite{Feigenbaum08}. Similar effects may appear for particular antenna systems. Antibonding modes, in this frame, are predicted to possess lower radiation losses and therefore larger quality factors \cite{Huang10}.

\begin{figure}[htbp]
        \centering
        \includegraphics[width=1\textwidth]{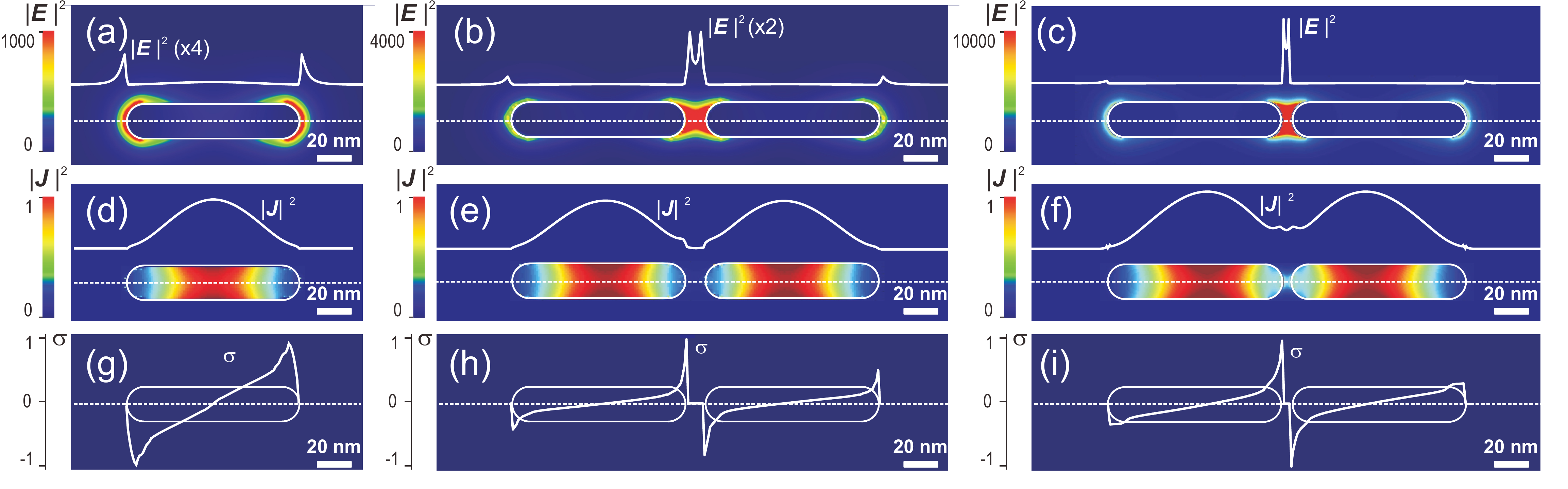}
        \caption{FDTD simulation results for Au nanoantennas composed of 100-nm-long wires in vacuum. Panels (a), (b), and (c): near-field intensity enhancement maps in a plane cutting through the antenna for a single-wire antenna (a) and a two-wire antenna with either 10-nm (b) or 4-nm (c) gap. Panels (d), (e), and (f): normalized total current density maps for the same antenna structures. Panels (g), (h), and (i): normalized surface-charge density profiles.}
        \label{maps}
\end{figure}

Now that we have discussed the spectral properties of the fundamental antenna resonance, let us have a look at field, current, and charge distribution maps. Figs.~\ref{maps}(a),(b), and (c) show on-resonance near-field intensity enhancement maps of the nanostructures already considered in Fig.~\ref{sim}(b). All intensities are normalized to the source. For the two-wire antenna in panel (b), a larger near-field intensity enhancement exists in the gap which exceeds the cumulative effect of the individual wires.
At the same time, the extent of the enhancement volume - if defined by an iso-intensity surface using the single wire peak value - also increases compared to the single wire. This provides better conditions for the coupling of nanomatter to the two-wire antenna. By comparing panels (b) and (c), a clear increase in the enhancement as the gap width is reduced is also apparent.
It is important to stress that the enhanced fields of a resonant linear antenna, because of the symmetry and boundary conditions, will be strongly polarized along the antenna axis. Light-matter interaction via the antenna near fields is therefore restricted to well-defined dipolar transitions \cite{Taminiau08a}, unless special geometries are used \cite{Biagioni09b}.

The total current distribution (Ohmic and displacement currents) for the fundamental resonance of a single-wire antenna [panel (d)] strongly resembles that of a RF $\lambda/2$ antenna, as expected (see Fig.~\ref{pattern} for a qualitative comparison). In the RF regime, when a very small feed-gap is added to a thin-wire antenna which is then connected to an impedance-matched waveguide, only small changes in the current distribution are expected for the two-wire antenna. In the plasmonic regime, however, the air gap is a strongly mismatched insertion for the rod, and causes strong deviations from this idealized situation. Indeed, if we look at the total current distribution for the fundamental resonance of the two-wire nanoantenna, as depicted in panels (e) and (f), two current maxima appear instead of one, making its behavior more similar to that of a $\lambda$ antenna than to that of a $\lambda/2$ antenna.
Notably, by comparing the 10~nm gap [panel (e)] with the 4~nm gap [panel (f)], it can be seen that for the smaller gap the two current maxima are slightly shifted towards the gap, thus providing larger charge accumulation and therefore larger fields as discussed for panel (c).
Finally, in panels (g) to (i) normalized surface charge density profiles are shown, where the dipolar charge separation, already discussed in the previous section \ref{mass_spring} [Fig.~\ref{massandspringmodel}(a)] and \ref{section:coupling} [Fig.~\ref{coupling}(a)], is clearly visible. Again, the effect of shrinking the gap results in increased charge accumulation at the gap and concurrent suppression of charge accumulation at the outer antenna ends.

\subsection{Radiation patterns of plasmonic linear antennas}
\label{radiation_section}
\begin{figure}[htbp]
\centering
\includegraphics[width=0.5\textwidth]{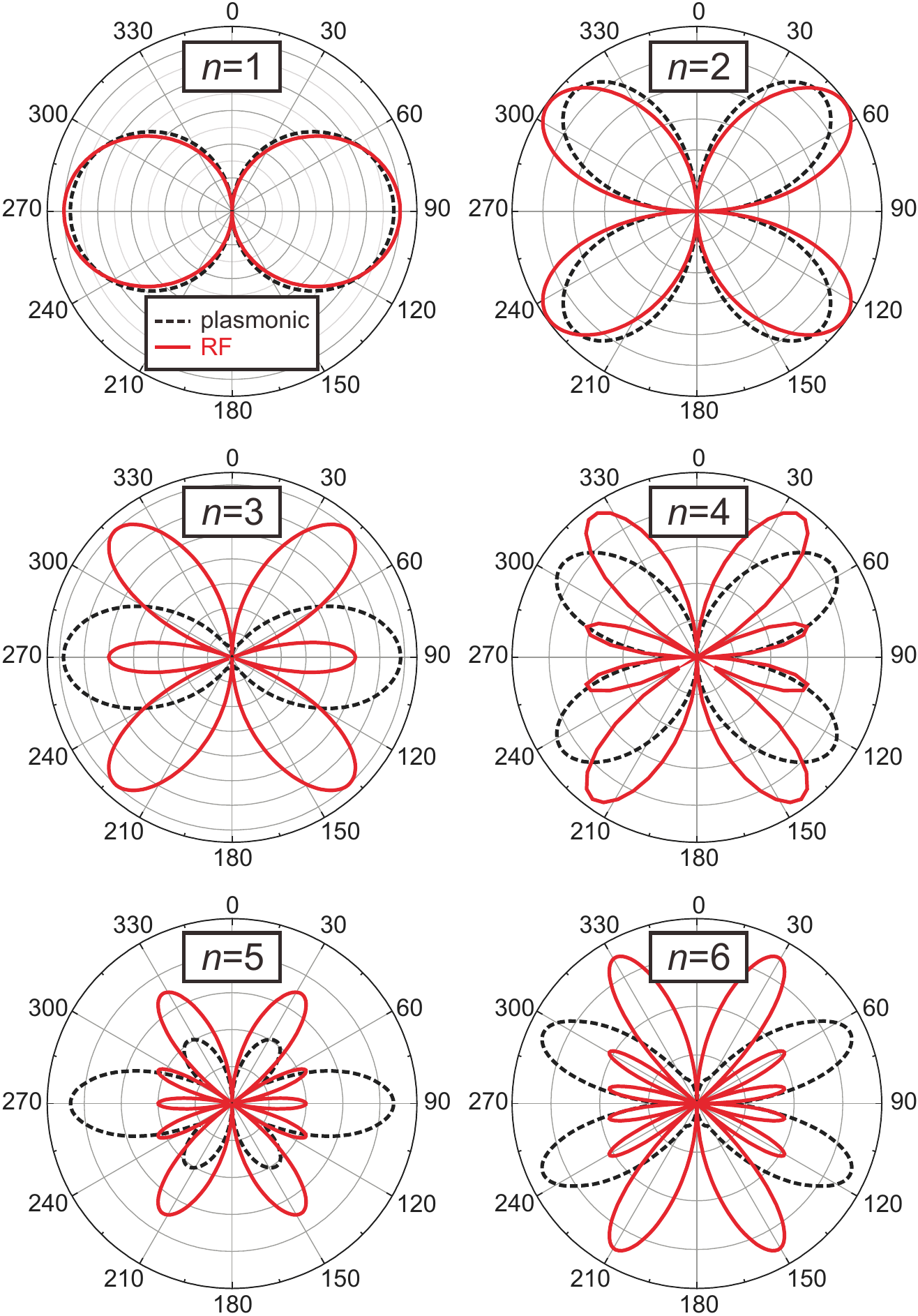}
\caption{Comparison between the emission patterns of plasmonic (black dashed line, courtesy of Jens Dorfm\"{u}ller \cite{Dorfmuller10}) and RF (red solid line \cite{lee}) single-wire linear antennas for the first 6 eigenmodes.}
\label{emission_pattern}
\end{figure}

As we have just discussed, a strongly reduced wavelength is typical for propagating e.m.~fields in plasmonic materials \cite{Novotny07}. This large mismatch between the wavelength of free-space plane waves and that of antenna modes is something which has no counterpart in the radio-frequency regime. As a consequence, as shown in Fig.~\ref{emission_pattern}, antenna radiation patterns are strongly modified \cite{Dorfmuller10,taminiau09}. In particular, all odd optical (plasmonic) modes present a strong emission in the direction perpendicular to the wire, something which is only weakly present in higher-order radio-frequency odd modes. The differences in the emission patterns - and by reciprocity, in the excitation patterns - have consequences for the coupling of optical antennas to quantum emitters placed in their proximity.

\section{Elements of optical antenna theory}
\label{antenna+circuits}
The aim of this section is to discuss to which extent circuit theory, which is a standard for RF antennas, can be applied to optical antennas. In the previous section we have discussed solitary optical antennas driven by propagating fields. Alternatively, one may also consider to drive a nanoantenna by a non-propagating localized field. To this end, a local source may be placed in the near field of the nanoantenna. The local source can be considered a nanometer-sized ac-voltage source, i.e.~a nanometer-sized frequency generator in the quasistatic limit, which drives the nanoantenna via capacitive coupling. Such a local source, realized experimentally e.g.~by a single quantum emitter (or a cluster of them), would only emit weakly in absence of the nanoantenna [see Eq.~(\ref{pointdipoletotalpower})]. However, when placed in close proximity to the antenna feed point it would drive a current in the optical antenna by means of its strong near field. The radiation (or the Ohmic losses for the case of coupling to a bad antenna mode) due to this current then dominates the overall behavior of the coupled system. Another possibility to drive an optical antenna is to rely on a nanoscopic two-wire transmission line as a feeding structure, as in the RF case discussed in Section \ref{antennalanguage}, which consists of wires that have roughly the same cross-section as those used in the nanoantenna itself \cite{Huang09b,Wen09}. The two-wire transmission line supports a localized, nonradiative mode which oscillates at optical frequencies. However, as opposed to the radiowave regime, the finite width of the wire and the reduced wavelength of the propagating mode need to be taken into account. Both types of ``driving circuits'' have to deal with the problem that the antenna impedance and impedance matching of the antenna to the driving circuit are difficult to define. On the one hand, this is because of the obvious experimental difficulty of direct probing nanoscale currents and voltages. On the other hand, there is the general problem that the RF-inspired lumped-circuit approach \cite{Engheta05} only holds {\em strictly} in the quasistatic limit in which all constituents of the driving circuit are supposed to be much smaller than the respective free-space wavelength, which is not easy to achieve in plasmonics.

\subsection{Nanoantennas driven by quantum emitters}
\label{nanoantenna+quantumemitter}
\begin{figure}[htbp]
        \centering
        \includegraphics[width=0.5\textwidth]{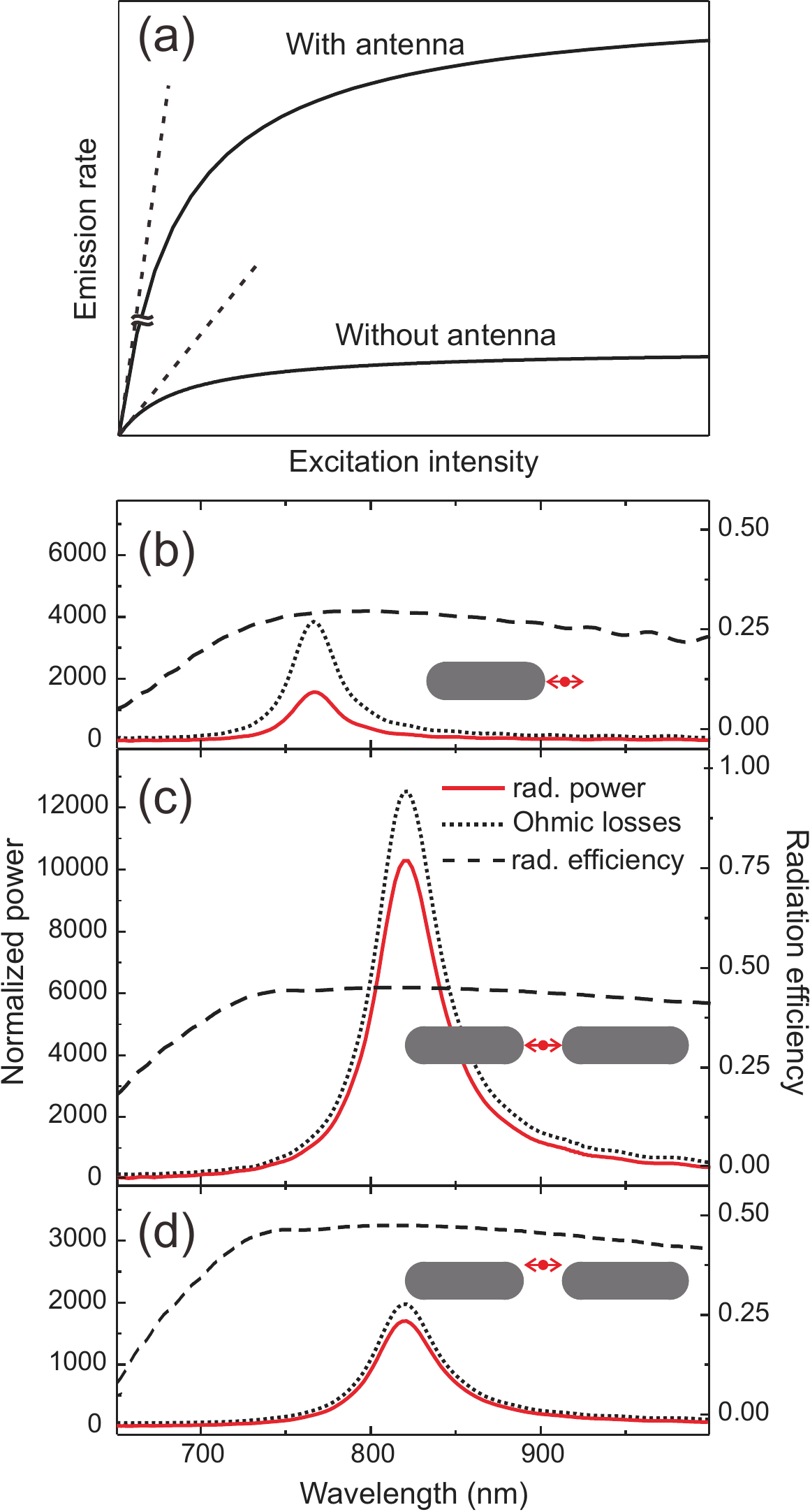}
        \caption{Antenna-coupled quantum emitter: (a) sketch of the emission behavior with and without the antenna; (b) power dissipated to heat (black dotted line), radiated power (red solid line), and radiation efficiency (black dashed line) for a quantum emitter coupled to a single-wire antenna; (c) same for a quantum emitter coupled to the feed-gap of a two-wire antenna; (d) same for a quantum emitter displaced by 10~nm from the center of the feed-gap of the two-wire antenna. All powers are normalized to the power radiated by the emitter in vacuum.}
        \label{emission}
\end{figure}

When placing a quantum emitter in close proximity to a nanoantenna, due to the modified local density of final states available for the decay of the system compared to the case of homogeneous space, the decay rate $k$ of the emitter is also modified. The description of such effects in principle requires a quantum analysis of the problem in which the so-called partial local density of states needs to be determined. It can be shown \cite{novotny}, however, that the partial local density of final states can be expressed via the imaginary part of Green's dyadic of the system evaluated at the position ${\bf r}_0$ of the emitter, i.e. purely based on electromagnetic theory,
\begin{equation}
k = \frac{\pi\omega_0}{3\hbar\varepsilon_0} \left| {\bf p} \right|^2\rho_{\rm p}({\bf r}_0,\omega_0) \textrm{,}
\end{equation}
with
\begin{equation}
\rho_{\rm p}({\bf r}_0,\omega_0) = \frac{\omega_0}{\pi c^2}\left[{\bf n}_{\rm p}\!\cdot\! {\rm Im}\! \left\{ {\bf G}\!({\bf r}_0,{\bf r}_0;\omega_0)\right\} \!\cdot\! {\bf n}_{\rm p}\right]\textrm{,}
\end{equation}
where $\rho_{\rm p}({\bf r}_0,\omega_0)$ is the partial local density of states \cite{novotny} and  ${\bf n}_{\rm p}$ is the unit vector pointing along the direction of the dipole moment ${\bf p}$. The Green's dyadic $ {\bf G}\!({\bf r},{\bf r}')$ of the system determines the electric field at a point ${\bf r}$ that is generated by a dipole at point  ${\bf r}'$ for the three fundamental dipole orientations as
\begin{equation}
{\bf E}(\bf r) = \omega^2 \mu \mu_0 {\bf G}\!({\bf r},{\bf r}')\cdot{\bf p}\textrm{.}
\end{equation}
Once the Green's dyadic for a given system is known, all relevant electromagnetic properties of the system may be derived.
The total decay rate $k$ of the quantum system coupled to the antenna is the sum of the radiative $k_{\rm r}$ and the nonradiative decay rate $k_{\rm nr}$.
When a quantum emitter is coupled to the local fields of an optical antenna, the coupled system has a very large absorption cross section compared to that of the isolated emitter. Qualitatively, this is explained by the high local fields that are created upon external far-field illumination. Using the reciprocity theorem one can argue that an emitter placed in the feed-gap will itself produce a strongly localized field upon excitation, which then efficiently couples to an antenna mode and thus significantly increases both $k_{\rm r}$ and $k_{\rm nr}$. This is the result of a strongly reduced excited-state lifetime, with the benefit of a larger saturation rate for the generation of far-field radiation. This is shown in Fig.~\ref{emission}(a), where the emission rate for an isolated emitter and for an emitter coupled to an optical antenna is qualitatively sketched.

Figs.~\ref{emission}(b) and (c) show the results of FDTD simulations where a point dipole is coupled either to a single- or to a two-wire antenna with 10~nm gap. The distance of the emitter from the wire apex is 5~nm in both cases. By calculating the time-averaged flux of the Poynting vector through boxes surrounding the antenna arms without including the emitting dipole, the Ohmic losses in the antenna can be obtained \cite{Mohammadi08}. A significant increase in the radiated power can be observed for the case of the two-wire antenna compared to the single-wire antenna. Moreover, by comparing the radiated power with the power dissipated to heat by Ohmic losses, and by considering the resulting  radiation efficiency, the improved performance achieved by coupling the emitter to the two-wire antenna, with a radiation efficiency improved by almost a factor of 2 compared to the single wire, becomes evident. Finally, in panel~(d) we show the effect of changing the position of the emitter by displacing it 10~nm away from the feed-gap center. While the radiation efficiency is largely unaffected, the overall radiation enhancement gets significantly lower.

Therefore, a single quantum emitter coupled to an optical antenna can have strongly improved emission and absorption properties while retaining its quantum character, e.g.~its single-photon emission ability. Such coupled systems have been investigated experimentally and are often referred to as ``superemitters'' for the just mentioned reasons \cite{Farahani05,Anger06,Kuhn06,Akimov07}.

The term ``superemitter'' should not lead to the wrong assumption that the coupling of a point-like emitter to a nanoantenna can have no adverse effects. The emission rate of a point dipole coupled to an optical antenna is enhanced when the emitter efficiently couples to the fundamental dipolar antenna mode. As a general trend, it is observed that the coupling increases with decreasing emitter-to-antenna distance. However, for distances that are sufficiently small, the strong field gradients of the point dipole source can efficiently excite higher multipolar lossy modes of the antenna which are mostly dark or weakly coupled to the radiation field and therefore convert e.m. energy into heat \cite{Anger06,Carminati06}. As a result, below a certain optimal distance which depends on geometry, material and wavelength, the non-radiative emission rate of the dipole will strongly increase and pull the overall radiation efficiency close to zero. This effect is also named ``quenching''.

\begin{figure}[htbp]
\centering
\includegraphics[width=0.5\textwidth]{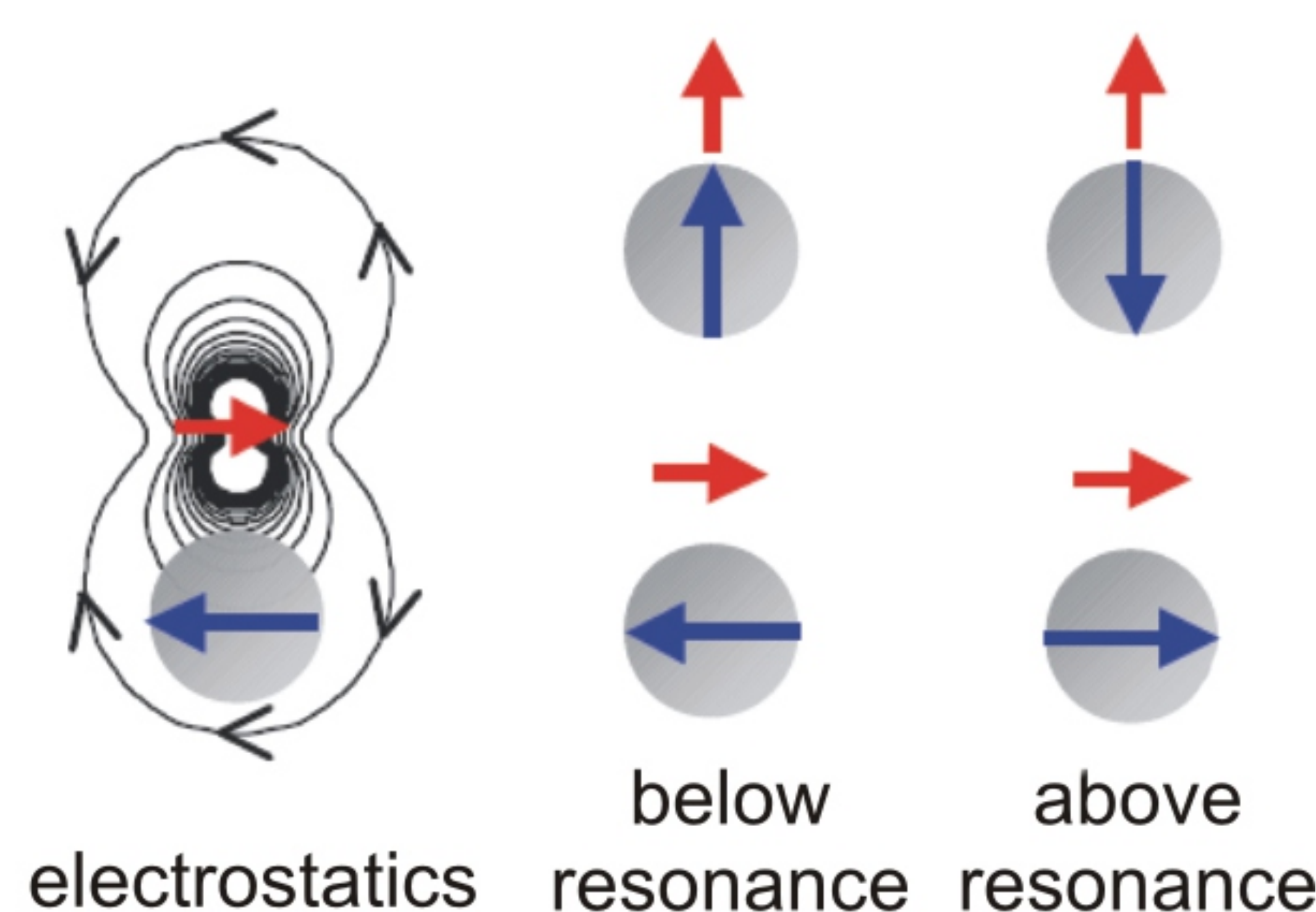}
\caption{Wavelength dependence of radiative rate enhancement. Left: Cartoon of a source dipole (red) and induced dipole (blue) in the nanoantenna in the electrostatic limit. Middle and right: Configurations of source and induced dipoles for source dipole frequencies far below and far above the dipolar resonance of the nanoantenna. Reprinted with permission from Mertens \textit{et al.} \cite{Mertens07}. Copyright 2007, American Physical Society.}
\label{quenching}
\end{figure}

We now assume an emitter at a distance to the antenna which is larger or equal than the optimum distance. In this case the emitter mainly couples to the dipolar mode of the nanoantenna. For the two fundamental orientations of the emitter, its radiative rate enhancement shows a nontrivial wavelength dependence around the antenna resonance, which depends on the relative phase of the antenna near fields and the orientation of the dipole with respect to the antenna. Since a resonance is involved, a 180$^\circ$ phase shift occurs when the resonance is crossed. As illustrated in Fig.~\ref{quenching}, depending on the source dipole orientation, the latter will oscillate in phase with the induced dipole either only above or only below the antenna resonance frequency \cite{Mertens07}. While in-phase oscillation leads to emission enhancement, anti-phase oscillation leads to a reduced emission. Such effects have been first predicted \cite{Thomas04} and then experimentally observed \cite{Anger06,Bharadwaj07a,Dulkeith05}. They are also visible in the simulation results of Fig.~\ref{emission}, where the radiation efficiency clearly decreases for photon wavelengths shorter than the resonance wavelength.

\subsection{Lumped elements at optical frequencies}
Within quasistatic limit, the concept of lumped circuit elements is a well-established method to simplify the analysis of complex electrical circuits. The concept of impedance is then introduced to describe the properties of each element and the interactions between different lumped elements in such a circuit.

\begin{figure}[htbp]
    \centering
     \includegraphics[width=0.5\textwidth]{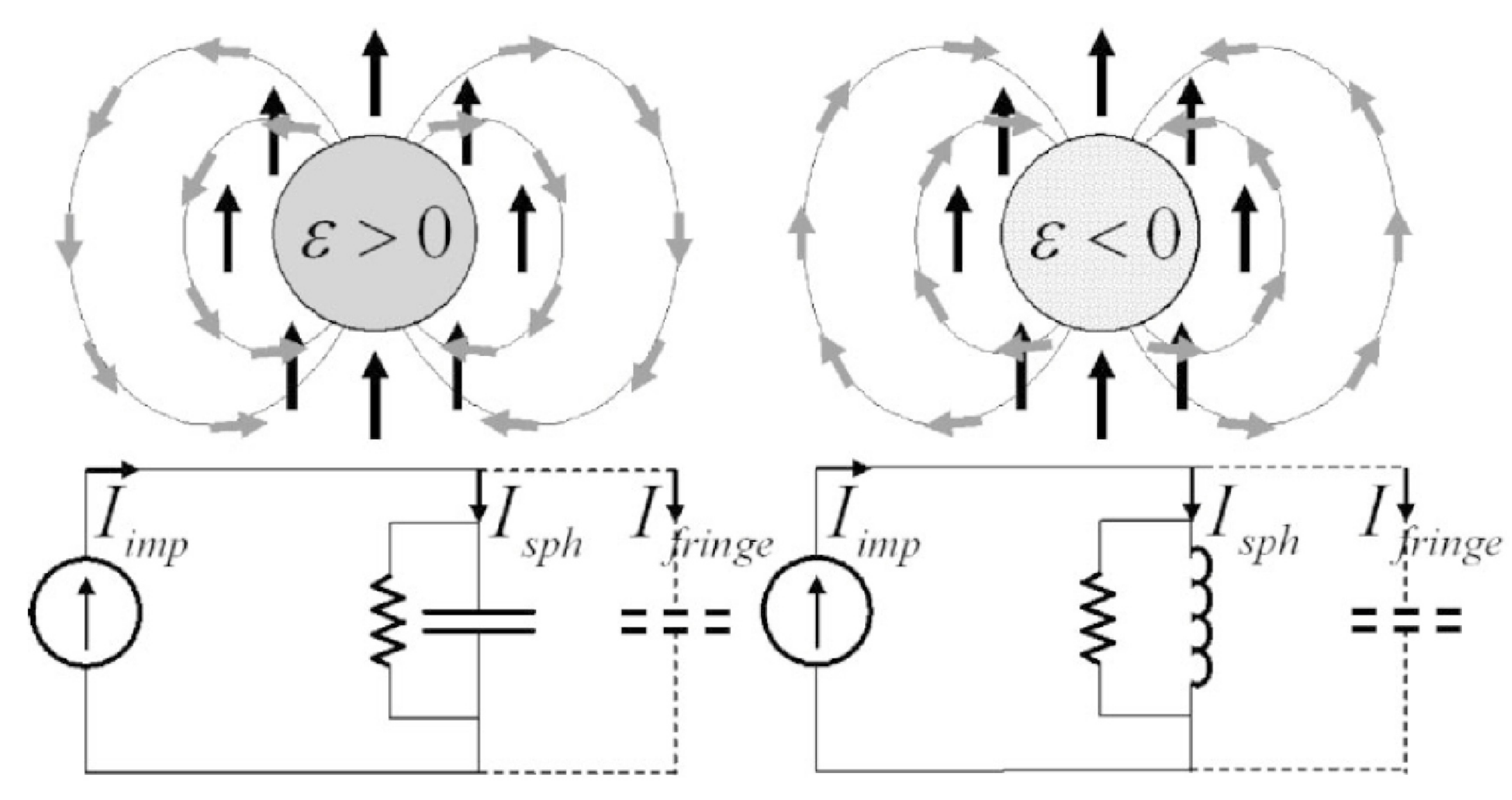}
     \caption{Sketch of the dipolar fields and equivalent circuits for a nonplasmonic (left) and a plasmonic (right) nanosphere. Black arrows represent the external field, gray arrows the induced dipolar field. Reprinted with permission from Engheta \textit{et al.} \cite{Engheta05}. Copyright 2005, American Physical Society.}
     \label{fig:engheta}
\end{figure}

The extension of these concepts to the optical realm would allow one to describe the response of a nanoparticle, for which the quasistatic approximation reasonably holds, to visible light in terms of an ``optical impedance''. Such an impedance in perspective would allow for the engineering of complex optical circuits by providing a rationale for the interconnection between nanoelements. N.~Engheta and coworkers \cite{Engheta05} have proposed a model to calculate the impedance of a spherical nanoparticle in the quasistatic limit, and demonstrated that its plasmonic resonance can effectively be described as that of an equivalent RLC circuit constituted of two parallel impedances
\begin{equation}
\label{imp}
Z_{\rm{sph}}=(-i\omega\varepsilon\pi R)^{-1}\; \rm{, }\;\;Z_{\rm{fringe}}=(-i\omega2\pi R\varepsilon_0)^{-1}\rm{,}
\end{equation}
representing the ratio between the average potential difference and the displacement currents inside the sphere and in the surrounding vacuum, respectively. In particular, a plasmonic sphere ($\rm{Re}[\varepsilon]<0$) can be described by the parallel connection of an inductor and a resistor, while a non-plasmonic sphere ($\rm{Re}[\varepsilon]>0$) is equivalent to the parallel connection of a capacitor and a resistor, as shown in Fig.~\ref{fig:engheta}. The surrounding vacuum behaves like a further capacitor connected in parallel. Although the concept has been pushed forward over the years \cite{Salandrino07,Alu07,Silveirinha08}, it is not yet widely applied in the practical design of optical antennas and plasmonic circuits.

The challenge for the practical application of the concept lies in the assumptions under which it is developed. First of all, the lumped circuit theory requires that the mutual influence of particles on each other's local fields can be neglected. Since this is not generally the case for near-field coupling between nano-objects, a dependent current source should be added to the equivalent circuit of each element \cite{Engheta05}, which potentially constitutes an obstacle towards a straightforward application of these concepts. Moreover, the quasistatic approximation is at the basis of the applicability of  Kirchhoff's voltage law, which locally requires $\nabla \times \textbf{E}\simeq 0$. When the object size is increased and becomes comparable with the operating wavelength this might not be valid anymore.

Although antennas are clearly not sub-wavelength elements, their feed points are usually very close to each other, and in the specific region of the feed-gap - where the quasistatic-like fields closely resemble those of a parallel plate capacitor - the condition $\nabla \times \textbf{E}\simeq 0$ is fulfilled, thus justifying the introduction of an input impedance which relates the voltage and the currents at the feed points. For a linear dipole antenna, one can therefore calculate an input impedance by taking the voltage to (displacement) current ratio across the feed-gap. In this way, numeric calculations provide impedance values which overall compare well with those of standard RF antennas \cite{Alu08b,Locatelli09}. Moreover, the extension of  standard radiowave concepts to optical antennas allows predicting the effect of filling the antenna gap with an arbitrary material, thus loading the antenna with an extra impedance in order to tune the overall input impedance (e.g. for impedance matching to a transmission line) \cite{Alu08b}. While such a treatment nicely illustrates the fact that nanoantennas can be described by equivalent resonant circuits as much as in the RF realm, one always needs to keep in mind that an impedance determined in this way might have limited meaning in terms of experimentally relevant impedances, since the waveguide wires that will be attached to the antenna will generally possess a finite width which makes the concept of feed points critical and might require particular care in its application.

The calculation of the antenna impedance by straight application of its definition relies on the full knowledge of field distributions. Since analytical approaches usually fail to accurately describe the antenna system, calculations based on numeric simulations are widely implemented. As an alternative and possibly more practical approach to determine a nano-optical impedance, one can rely on the relation between impedance mismatch and reflection coefficient when a transmission line is used to feed the antenna \cite{Huang09b}. The complex reflection coefficient $\Gamma$ [see Eq.~(\ref{reflection})] allows one to extract the antenna impedance once the characteristic impedance of the transmission line is known. Experimentally, $\Gamma$ can be obtained by imaging the field distribution along the transmission line by means of near-field \cite{Dorfmuller09,Weeber01,Krenz10,Schnell11,Fang11} or photoemission electron microscopy \cite{Douillard08,Cinchetti05} techniques. Once the characteristic impedance of the line is calculated, this approach allows one to measure the impedance of any optical circuit element connected to the line as a load.

\begin{figure}[htbp]
    \centering
     \includegraphics[width=1\textwidth]{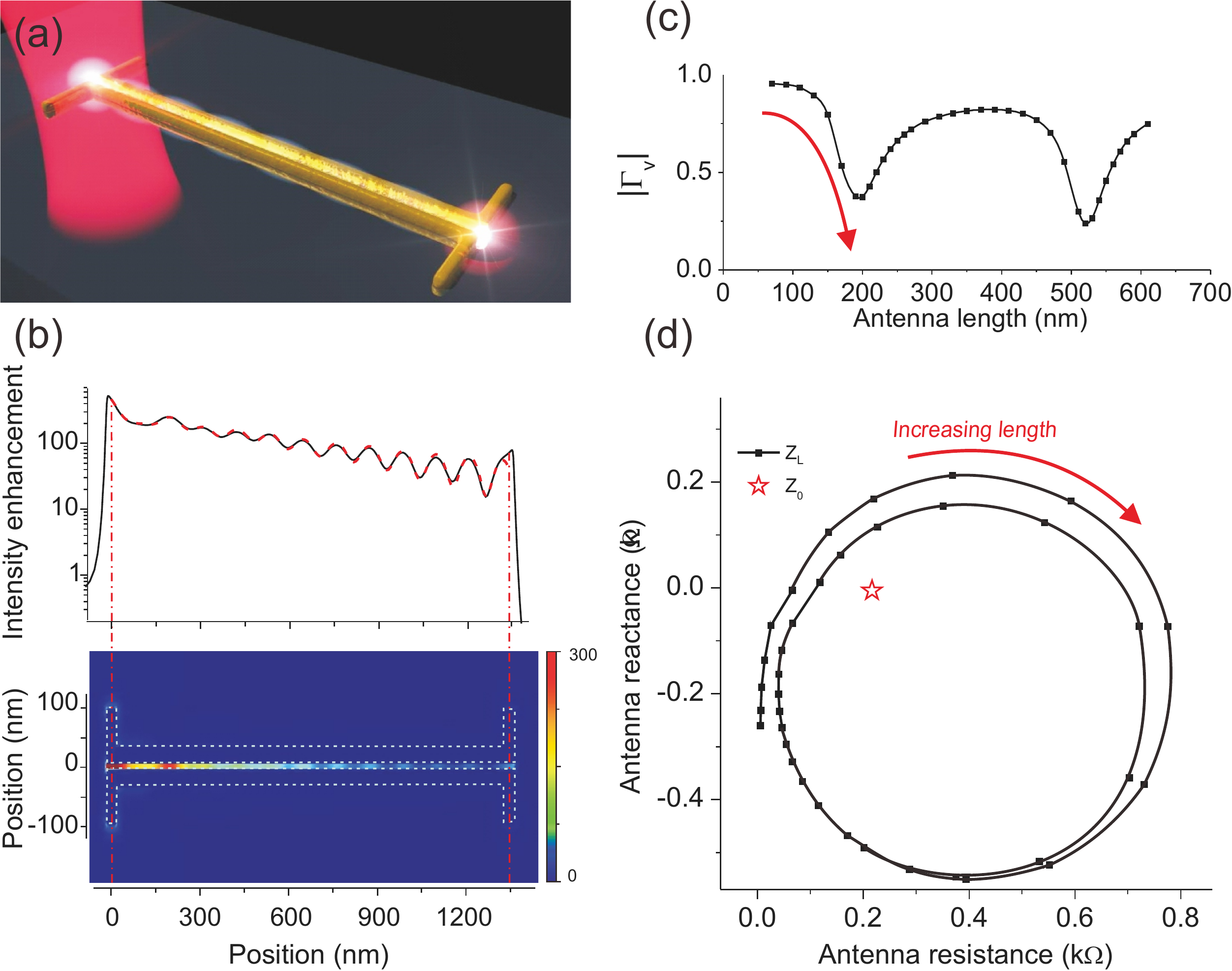}
     \caption{Impedances and impedance matching in an antenna circuit: (a) Sketch of a prototype circuit, constituted of a receiving antenna (left), a two-wire transmission line, and a transmitting antenna (right) on a glass substrate; (b) representative simulated standing-wave intensity pattern along the transmission line at 830~nm free-space wavelength; (c) calculated voltage reflection coefficient (absolute value) as a function of the transmitting antenna length; (d) calculated complex impedance of the transmitting antenna. The star designates the characteristic impedance of the feeding transmission line. Best impedance matching is related to the closest distance in the complex impedance plane. Adapted with permission from Huang \textit{et al.} \cite{Huang09b}. Copyright 2009, American Chemical Society.}
     \label{fig:nanolett}
\end{figure}

As an illustration of such concepts, Fig.~\ref{fig:nanolett} shows simulation results for a prototypical gold antenna circuit constituted by a receiving and an emitting two-wire nanoantenna connected by a two-wire transmission line \cite{Huang09b}, as depicted in panel (a). By illuminating the receiving antenna, a propagating mode along the transmission line can be launched, which eventually undergoes an impedance mismatch at the emitting antenna and builds up a standing-wave pattern [Fig.~\ref{fig:nanolett} (b)] that can be fitted in order to extract the value of the reflection coefficient as a function of the antenna length [Fig.~\ref{fig:nanolett} (c)]. Finally, from $\Gamma$ one can calculate the antenna impedance by applying Eq.~(\ref{reflection}). It is worth noticing that the impedance plot in Fig.~\ref{fig:nanolett}(d), while providing values that compare well with their RF counterparts, tends towards negative antenna reactances. This is a clear indication that parasitic reactances are present in the antenna equivalent circuit, due to the combined capacitive contribution of the substrate and the antenna gap.

Notably, a novel and generalized concept of optical impedance (termed ``specific impedance'') has recently been proposed \cite{Greffet10}, which is able to elegantly provide a unified treatment of the interaction of single emitters with nano- and microstructures such as optical antennas and dielectric cavities in terms of generalized impedance matching concepts. By the formal analogy between the power radiated by a current element in a piece of a conductor and that of an optical dipole, the specific impedance can be introduced via the imaginary part of Green's tensor. This generalized definition of impedance, which has the dimension of $\Omega\cdot \rm{m}^{-2}$, has been applied to dipole emission in vacuum,  in a microcavity, and in the vicinity of a spherical nanoparticle antenna, demonstrating that the Green tensor point of view is indeed capable of reconciling concepts from largely separated fields of research.

\subsection{What optical antenna engineers may be concerned with}
What we have learned in the previous sections is that optical antenna engineers in many cases will have to deal with isolated antennas that may be coupled to (or fed by) single emitters in their vicinity. The resonances of isolated antennas are well understood. There are generally applicable rules that can be used to qualitatively predict the resonances of moderately complex assemblies of nanoparticles which are based on effective wavelength concepts and resonance splittings due to strong coupling (plasmon hybridization). For sharp corners and in nanogaps strong near fields may appear for eigenmodes that create opposing surface charge densities at both sides of the gap accompanied by strong electromagnetic coupling.

Optimization of the coupling between a localized emitter and a nanoantenna is a field of intense research which has no counterpart in RF antenna design. However, although significant steps have been made \cite{Anger06,Greffet10,Taminiau08b,Vandenbem10} it is still an open problem how to optimize the transfer of energy between a single emitter and a nanoantenna, which represents its most important and practically relevant feeding mechanism. Even if a perfect impedance match could be achieved, this still would only optimize the energy transferred to the antenna while its radiation efficiency may still be small.
Conditions for minimal Ohmic losses and therefore optimal radiation efficiency might lead to different sets of optimal antenna parameters. For example, when complex optical antennas are build, each additional passive element lowers the overall radiation efficiency of the antenna, even though it might strongly improve certain properties.
It is therefore advantageous for optical antenna engineers to look for antenna designs that realize a certain function in the simplest possible way.

When integrating nanoantennas into complex plasmonic circuits the different behavior of RF and optical antennas should also be carefully taken into account. While, as extensively discussed, many efforts are made to apply lumped-circuit approaches and impedance-matching arguments to the optical realm, these concepts at optical frequencies still need to be interpreted and validated against experiments.

\section{On the defining properties of optical antennas}
\label{definition}
So far, we emphasized both the close similarities of optical antennas and standard RF antennas as well as differences that arise in the optical realm because of the non-ideal conductive properties of metals and the occurrence of plasmonic resonances.
It can be fairly said that all the significant differences between the behavior of radiowave and optical antennas stem from the fact that: (i) optical antennas are characterized by large Ohmic losses and a finite skin depth and (ii) the wavelengths of plasmonic wire modes and of free space waves are largely mismatched. These observations hold many consequences for the behavior of isolated optical antennas as compared to the RF regime, among which the most important can be summarized as follows: (i) reduced radiation efficiency, (ii) lower quality factors of the resonances (larger bandwidth), (iii) a peculiar scaling law for the antenna resonance length, (iv) deviating radiation patterns, and finally (v) a deviating current distribution.

Let us now consider in a bit more detail the question of what are the characteristic properties of an optical antenna as it has been discussed so far. Since any interaction of light with matter leads to the formation of localized near fields \cite{Wolf85} one may ask the question what is the significance of the particular metal nanostructures we have introduced. Indeed the term ``optical antenna'' is used in very broad sense in the literature. When categorized according to materials involved, besides metal antennas there are also ``antennas''  made of semiconducting \cite{Cao09,Cao10b} and dielectric materials \cite{Lee11}.

In the following we will formulate some criteria to define a subset of nanostructures from all possible shapes of matter that we call ``optical antenna'' for the purpose of this Report. We would like to emphasize that we do not intend to provide a universal definition of ``optical antennas'', although, the very unspecific use of the term in the literature can cause some confusion. For the remainder of this Report we will stick to these criteria in order to focus and limit our discussion.

The first criterion for a structure to qualify as an optical antenna in the sense of this Report is that it should be able to concentrate (i.e. enhance and localize) propagating fields of plane waves within a certain spectral bandwidth for a range of directions of incidence, thereby creating enhanced and localized and therefore non-propagating near fields in absence of inelastic processes. Plane waves may be considered here without loss of generality since any propagating field can be decomposed into plane waves propagating in different directions \cite{novotny}. This first criterion does not yet exclude any specific structure according to Ref. \cite{Wolf85}.

The second criterion is that the reciprocity theorem should be applicable to the involved fields and currents (see section \ref{reciprocitytheorem}). The validity of the reciprocity theorem automatically implies also the reverse way of operation of an antenna: the structure can convert - again within a certain bandwidth - local fields into a superposition of plane waves with an emission pattern that equals the receiving pattern. Here it is reasonable to require that the amplitude of at least some of the plane waves emitted in a range of directions must be larger in presence of the nanoantenna than without it. This second criterion together with the first one excludes structures from the following discussion that show Stokes-shifted fluorescence, as well as optical processes in photosynthetic and other supramolecular complexes, where irreversible energy funneling towards a reaction center occurs.

A third criterion is that the light confinement principle does not only rely on free-space propagating waves. If some part of a structure under consideration is based on such a description this part will not be considered a part of the actual antenna but rather a part of the illumination/detection optics. Obviously this excludes mirrors and lenses of all kinds including Fresnel lenses in 3D and in 2D (plasmonic) realizations.

Furthermore, to increase the field enhancement it is usually favorable if a plasmon resonance is involved which enlarges the absorption cross-section of the optical antenna beyond its projected geometrical area and increases the local intensity enhancement beyond a pure lightening rod effect. While based on RF antenna theory a resonance would be considered as an effect that limits the bandwidth \cite{balanis}, at optical frequencies the involved plasmon resonances are typically very broad and the quality factor is rather small. Therefore bandwidth limitations due to a plasmonic resonance are usually not severe.

Finally, we note that the use of an optical antenna according to the criteria above may not always be the best solution for every experimental setting. There may be situations in which certain parameters of a system are improved while others get worse when trying to optimize an optical antenna.

As a conclusion of these considerations, we limit the types of optical antennas that are discussed in the following to finite metal nanostructures that exhibit plasmon resonances in the visible to near-IR range. We will therefore not consider semi-infinite structures here, such as sharp tips, although they play an important role in tip-enhanced spectroscopies \cite{Novotny06,Brehm08}.

\section{Fabrication of nanoantennas}
\label{fabrication}
Since the resonances of optical antennas strongly depend on the exact geometry and dimensions, fabrication of nanoantennas requires reliable and reproducible structuring techniques with a typical resolution below 10~nm in order to accurately define critical dimensions, such as feed-gap size or antenna arm length. This pushes state-of-the-art nanostructuring techniques to their limit and can be considered one of the main challenges in the realization of optical antennas and plasmonic devices.

Various top-down and bottom-up nanofabrication approaches have been applied to experimentally realize optical antennas. Top-down approaches, e.g.~electron-beam lithography (EBL) and focused-ion beam (FIB) milling, typically start from a thin multi-crystalline metal film on top of an optically transparent but electrically conductive substrate (often indium tin oxide, ITO), which is needed to avoid charging effects. In general, top-down approaches are capable of fabricating large arrays of nearly identical nanostructures with well defined orientations and distances. Bottom-up approaches, on the other hand, take advantage of chemical synthesis and self-assembly of metal nanoparticles in solution with nearly perfect symmetry and crystallinity that can be put on any substrate. However, to be effective, bottom-up fabrication techniques often require precise size selection and nanopositioning as well as assembly strategies to create non-trivial structures.

There are  ongoing efforts to improve the resolution and reliability of nanostructuring and to better understand the role of crystallinity in the achievement of improved optical responses for plasmonic structures. Rough surfaces and multi-crystalline materials can be detrimental both in terms of increased scattering of plasmons and ill-defined geometric parameters that arise due to the anisotropic response of multi-crystalline materials during nanofabrication.
Recently, a combination of both approaches has shown potential to fabricate high-definition nanostructures with fine details over large areas \cite{Huang10b}.

\subsection{Electron-beam lithography}
\begin{figure}[htbp]
    \centering
     \includegraphics[width=0.5\textwidth]{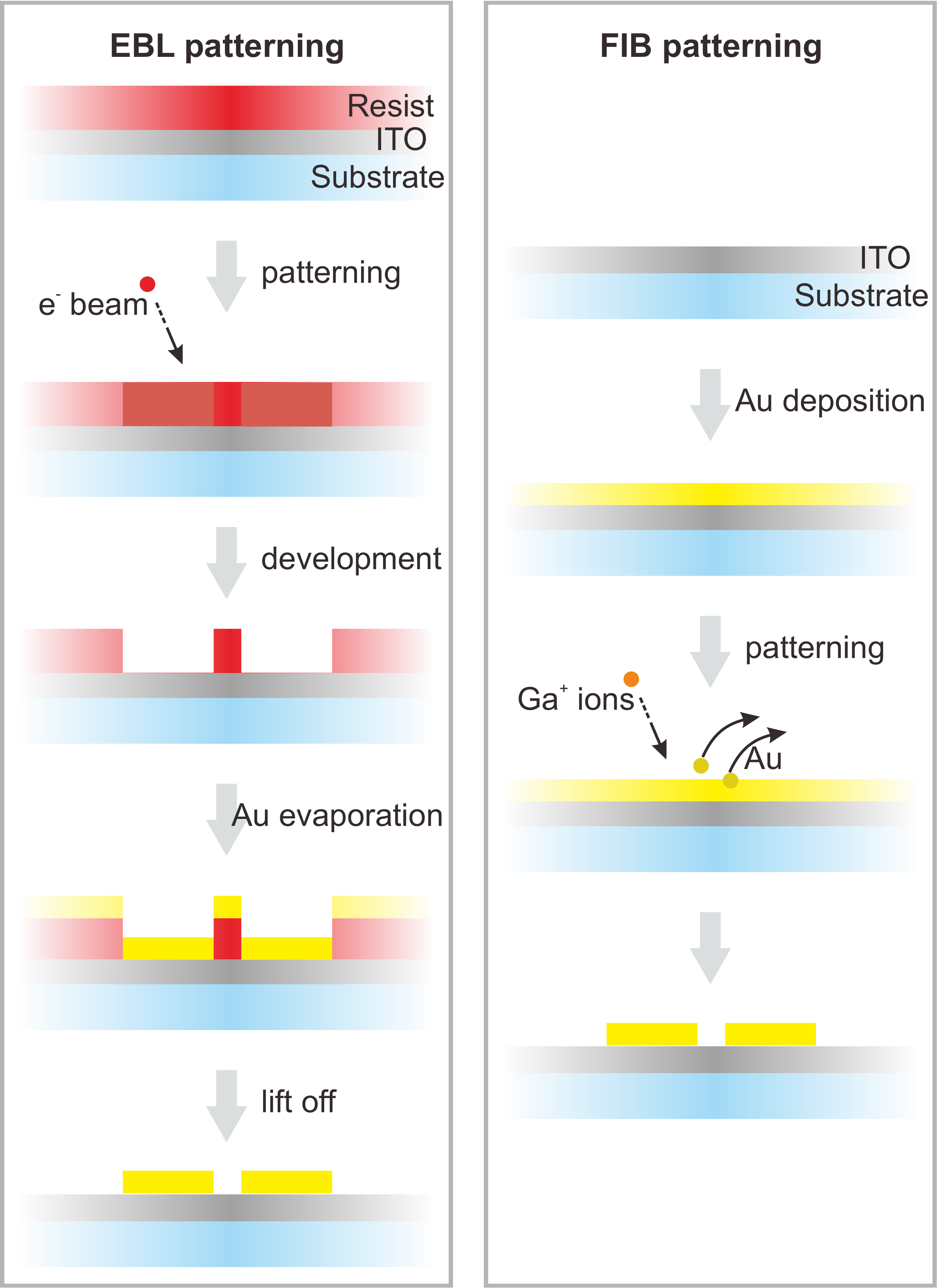}
     \caption{Sketch of the main steps for standard EBL and FIB nanostructuring of nanoantennas.}
     \label{SEMFIB}
\end{figure}

One of the most popular techniques to fabricate nanoantennas on a flat substrate is EBL \cite{Ghenuche08,Schuck05,Muskens07a,Schnell09}. In the typical implementation of EBL (see Fig.~\ref{SEMFIB}) a high-resolution electron-sensitive resist, e.g. PMMA, is patterned by means of a focused electron beam \cite{Rai-Choudhury97}. The patterns are then developed and selectively removed. A thin layer of metal with the desired thickness is then evaporated covering both the voids and the remaining resist. Finally, the sample is subjected to a solvent which removes the remaining resist and leaves the metal structures in the voids unaffected (lift-off). Since the patterning is done by an electron beam, the spatial resolution of the pattern is usually below 5~nm. However, due to the multicrystallinity of the deposited metal layer, the final structural resolution is usually not as good. A state-of-the-art nanoantenna produced by EBL is shown in Fig.~\ref{fabric}(a). In order to increase the stability of the fabricated nanostructures during lift-off, a thin layer of titanium or chromium (typically $<$ 5~nm) is often used as adhesion layer. Such layers can, however, significantly increase the damping of the surface plasmon \cite{Huang10b,Jiao09}. Only larger patches of metal survive lift-off without such precautions. FIB fabrication applied to such gold patches has been used to fabricate nanoantennas without adhesion layer \cite{Muhlschlegel05}. Recently, other EBL-based techinques without adhesion layer have been proposed and developed \cite{Grigorescu09,Chen10b}. In addition to lithography, electron-beam induced deposition has been applied to build complex nanostructures \cite{vanDorp08,Graells10} or to produce masks for further dry etching \cite{Weber10}. Electron-beam induced deposition also holds promise to be applied to engineering of the dielectric properties of the antenna environment (e.g. for dielectric loading of the antenna feed-gap) or for the deposition of free-standing conductive wires for combination of optical and electrical processes in plasmonic circuits \cite{Frabboni06}.

\subsection{Focused-ion beam milling}
Another efficient machining technique for the realization of optical antennas is FIB milling. FIB structuring is based on the localized sputtering of material using accelerated Ga ions extracted from a liquid metal ion source \cite{Orloff}. The emitted ions are accelerated, focused into a beam with a few nanometer spot, and scanned over a conductive substrate to produce a desired pattern \cite{Melngailis87}. Ion collisions generate a cascade inside the solid, with atoms being knocked off their equilibrium position, giving rise to local surface erosion (see the typical fabrication steps in Fig.~\ref{SEMFIB}). In the field of optical antennas, FIB nanofabrication has been successfully applied in a number of cases, because it couples nanoscale resolution with the high versatility of the direct patterning approach \cite{Muhlschlegel05,Farahani05,Kim08,Cubukcu06,Taminiau07a}.

\begin{figure}[htbp]
        \centering
        \includegraphics[width=0.5\textwidth]{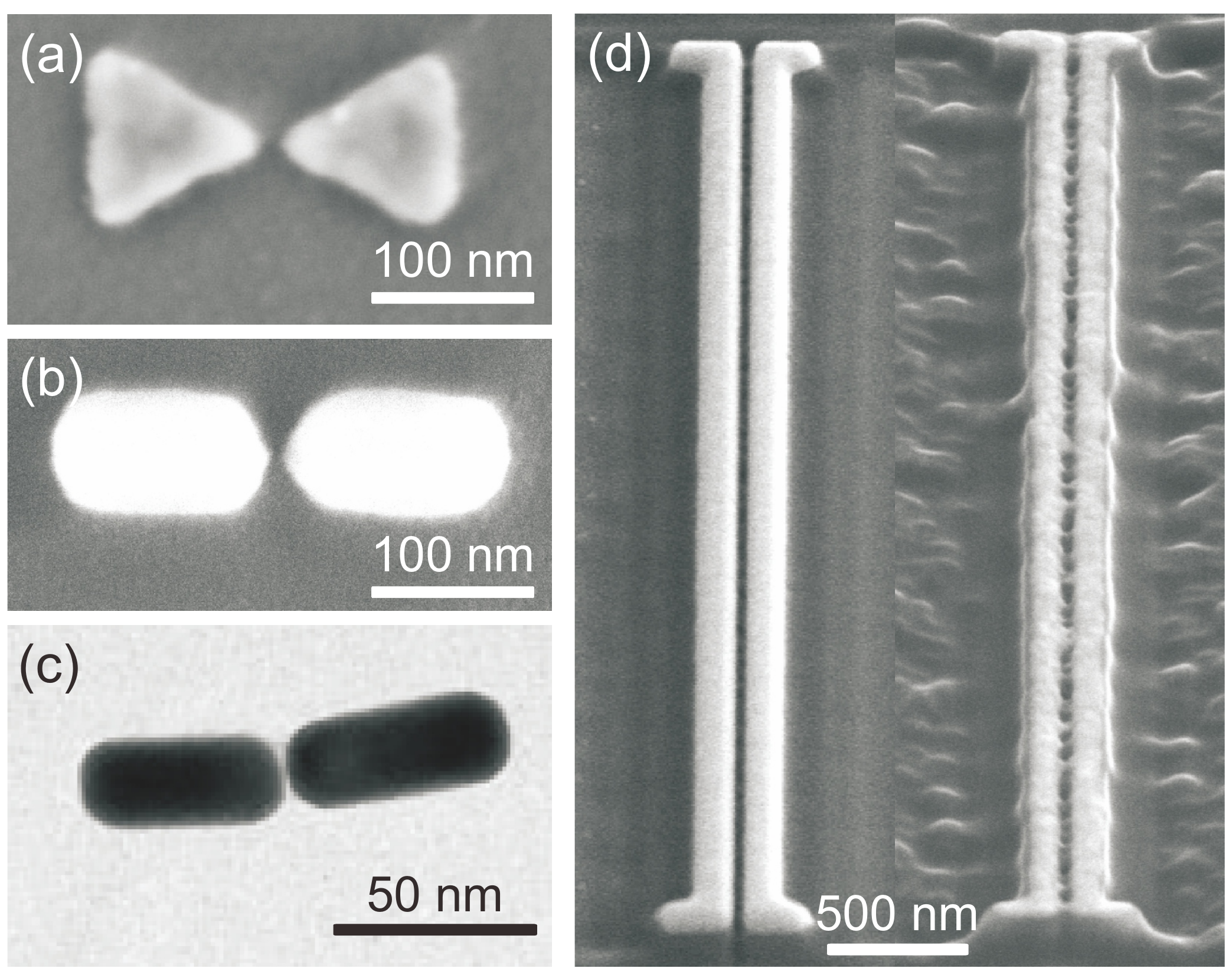}
        \caption{Fabricated nanoantennas: (a) SEM image of a bow-tie nanoantenna produced by EBL \cite{Schuck} with a sub-10~nm gap; (b) SEM image of a 8~nm gap in a dipole antenna realized by FIB starting from a single-crystalline Au microflake \cite{Huang10b}; (c) transmission electron microscopy (TEM) image of a covalently-bonded self-assembled two-wire antenna \cite{Pramod08}; (d) effect of multi-crystalline grains on FIB structuring of a plasmonic circuit (left: single-crystalline Au film; right: multi-crystalline Au film) \cite{Huang10b}. Panel (c) from Pramod \textit{et al.} \cite{Pramod08}. Copyright Wiley-VCH Verlag GmbH \& Co. KGaA. Reproduced with permission. }
        \label{fabric}
\end{figure}

The main advantages of FIB milling are the broad applicability to almost any type of (sufficiently conductive) material and the very good resolution, allowing for the fabrication of prototype surface nanostructures and gaps in the 10-20~nm range. When applied to chemically grown single-crystalline metal flakes \cite{Huang10b,Huang10}, highly reproducible antenna gaps below 10~nm can be obtained [see Fig.~\ref{fabric}(b)]. FIB is a unique structuring tool whenever the use of resist-based lithographies is difficult, for example with non-flat sample topographies such as optical antennas on atomic-force microscopy (AFM) tips \cite{Farahani05,Taminiau07a,Farahani07} or when a resist must be avoided to allow for epitaxial growth of single-crystal metal substrates. On the other hand, since FIB is a sputtering process, part of the sputtered material can be redeposited back to the surface and contaminate already fabricated structures. Therefore, careful design of the FIB pattern and the writing sequence is important. Furthermore, gaps cut by FIB usually show a tapered shape so that the gap width is not uniform across the section, an effect which might be less pronounced for EBL because of the absence of direct patterning and which can anyway be kept under control with careful procedure optimization. Although FIB sputters away the material, the accelerated ion beam can cause Ga ion implantation into the target metal film and substrate. Recently, an energy dispersive x-ray study on a FIB structured gold edge \cite{Huang10b} has shown undetectable Ga implantation, which corresponds to a Ga concentration of $< 1~\%$. However, since even a weak ion implantation might influence the dielectric properties of the materials, the effects involved need further study. As an example, the influence of Ga residues in the FIB-cut nanogaps of Au plasmonic
antennas on the energy and quality factor of their resonances has clearly been demonstrated \cite{Han11}.

\subsection{Nano-imprint lithography}
A possible low-cost, high-throughput alternative to both FIB and EBL patterning is nano-imprint lithography (NIL). As opposed to serial beam-based lithographies, where photons, electrons, or ions are used to define nanopatterns, the NIL process utilizes a hard mold that contains all the surface topographic features to be transferred onto the sample and is pressed under controlled temperature and pressure into a thin polymer film, thus creating a thickness contrast \cite{Guo07}. Resolutions on the order of 10~nm have been demonstrated more than a decade ago. A promising variation of this technique is UV-NIL, where a transparent mold, such as glass or quartz, is pressed at room temperature into a liquid precursor which is then cured by UV radiation. In order to reduce the cost of mold fabrication and to achieve patterning over larger areas at lower pressures, soft nano-imprinting techniques, based on polymeric flexible stamps replicated from a single master mold, have also been developed. In future, NIL might become the ideal technique for low-cost, highly-reproducible realization of antenna arrays covering large areas, e.g.~for the realization of bio-optical sensors on substrates or on fiber facets \cite{Boltasseva09}.

\subsection{Self- and AFM-based assembly of nanoantennas}
In addition to top-down nanofabrication techniques, bottom-up approaches have also been widely used to obtain single-crystalline metallic nanostructures. Chemically grown colloidal nanoparticles provide controlled shape, high purity, and a well-defined crystallinity. Nanoparticles made of different materials in various shapes have been synthesized by using surfactants in the redox process \cite{Yu97,Sun02,Gole04,Liu05,Chu06,Millstone09,Zhang09b,Wiley06,Kim02}. Isotropic and anisotropic core-shell nanoparticles are also obtained chemically \cite{Pham02,Wang06}. Use of chemically grown nanoparticles as nanoantennas typically requires to arrange them on a surface in well defined patterns in order to achieve the desired optical properties. Such a controlled assembly can be achieved for example by nanomanipulation \cite{Yang10}, electrophoresis \cite{Raychaudhuri09}, fluidic alignment \cite{Huang01}, or micro-contact printing \cite{Kraus07}. Recently, polymer spacer layers have also been introduced to control the interparticle distances in nanoparticle clusters \cite{Fan10}.

The fabrication of very small gaps in optical antennas is an important issue since it will allow studying effects that are expected to act against purely electromagnetic field enhancement effects, such as nonlocality \cite{McMahon09} and quantum tunneling \cite{Zuloaga09}. Gaps required to reach the regime where such effects may occur are very small, on the order of 1~nm or less \cite{Hadeed07,Strachan08}, and are difficult to achieve by standard nanofabrication. Such gaps can, however, be readily achieved by means of chemical self-assembly of gold nanorods in solution [see Fig.~\ref{fabric}(c)], where the gap is determined by the thickness of the surfactant layer that covers the particle surfaces \cite{Pramod08,Jain09}.

Another approach that can produce very small gaps is AFM nanomanipulation. Nanostructures can be manipulated on a surface by pushing them in a controlled way using the tip of an AFM. To perform the manipulation, the structures of interest are first localized using tapping-mode imaging. Then the set point of the feedback loop is increased to the manipulation threshold and pushing is performed by parking the tip ``behind'' a particle and performing a piezo movement in a desired direction. In this way, the gap of a bow-tie antenna has been reduced until contact occurred \cite{Merlein08}. In other experiments, nanoantennas containing diamond nanoparticles were assembled by AFM nanomanipulation \cite{Schietinger09}.

\subsection{Nanoantennas on tips}
By exploiting the possibilities afforded by scanning probe technology, optical antennas fabricated at the apex of scanning probe tips can be positioned freely to probe different parts of a surface with nanometer precision. In this way it is possible to place a single emitter in close proximity to the antenna feed-gap while probing its optical response with an inverted confocal microscope. The fabrication of optical antennas at the apex of a scanning probe tip, however, poses a challenge to any microfabrication tool and FIB milling is frequently used in this context. As an application example, emission enhancement and a related decrease in excited-state lifetime have been demonstrated for a bow-tie antenna on an AFM tip which was coupled to a semiconductor nanocrystal \cite{Farahani05}. In order to obtain a reliable procedure for producing tip-based antennas that are parallel to the sample plane, flat plateaux at the tip apex are desirable as a starting point before metal evaporation and nanostructuring \cite{Biagioni08}. However, antennas can also be realized perpendicular to the sample plane, following the already-established line of tip-on-aperture microscopy \cite{Frey02,Frey04}. The realization of $\lambda /4$ monopole antennas at the apex of tapered optical fibers by FIB \cite{Taminiau07a} and their interaction with single molecules \cite{Taminiau08a,Taminiau08b} have thus been investigated.

An alternative approach to the fabrication of antenna probes consists in the chemical grafting of individual gold nanoparticles by means of a dielectric probe tip for which efficient techniques have been developed \cite{Gan07}. These concepts have been applied to study the emission properties of molecules coupled to single resonant Au nanoparticles attached to tapered optical fibers \cite{Anger06,Kuhn06}. The technique has also been applied to image the distribution of fluorescently labeled proteins in biological membranes with very high resolution \cite{Hoppener08b,Hoppener08}. As grafting of nonspherical particles is not trivial, an alternative approach suggests to place nanorods inside quartz nanopipettes \cite{Novotny09}.

\subsection{Fundamental material issues}
The role played by the material quality in determining the quality factor and the resonance frequency of antenna oscillations, for a given geometry, is twofold: it can affect the scattering of individual electrons involved in the oscillation or it can increase scattering of plasmons into propagating waves.

Grain boundaries, voids, or surface contamination can result in changes of both the real and the imaginary part of the dielectric constant, thus modifying the conditions for plasmon propagation and antenna resonances \cite{Huang10b,Aspnes80,Kuttge08,Chen10}. It has been shown for the case of Au that changes in the above-bandgap dielectric properties can mainly be attributed to voids in multi-crystalline films, because of the fact that less polarizable material is available, while changes in the below-bandgap dielectric response can be strongly influenced by grain boundaries. This happens since at these energies the electron mean free path becomes quite large and comparable with the grain size \cite{Aspnes80}.

Surface plasmons and antenna oscillations are interface e.m.~waves, and as such they can be scattered by surface roughness. In optical antennas, this will result in additional dissipation channels in the form of background radiation losses (i.e.~radiation with unspecific or unwanted pattern), thus lowering the quality factor of the resonance. As a matter of fact, it has been clearly shown experimentally that the role of both limiting factors (changes in the dielectric constant and/or wave scattering due to surface roughness) can severely affect surface waves in noble metals \cite{Kuttge08,Chen10,Mills75,Nagpal09}.

Recently, novel procedures have been proposed to overcome these limitations. A combination of template stripping  with precisely defined silicon substrates has been demonstrated, which, because of the very small roughness, is able to achieve plasmon propagation lengths comparable to those of perfectly flat films \cite{Nagpal09}. Moreover, the influence of crystallinity on plasmon resonances in single Au nanorods has also been specifically discussed \cite{Laroche07} and improved optical properties have been observed in single-crystalline gold nanostructures \cite{Tang07,Vesseur08,Wiley08}. Recently, single-crystal chemically grown Au microflakes have been introduced as a means to achieve reproducible, high-resolution nanopatterning of plasmonic elements with superior optical properties \cite{Huang10b}. The use of single-crystal substrates as a starting point for FIB milling and other top-down microfabrication processes is very promising since it avoids structure imperfections due to the anisotropic response of different grains in multi-crystalline materials which are clearly visible in Fig.~\ref{fabric}(d), where a comparison is made between nanoantenna circuits realized starting from a single- and a multi-crystalline Au patch.

\section{Experimentally studied geometries of metal optical antennas}
So far we have tried to convey a general understanding of optical antennas including possibilities to fabricate them. Now we are going to introduce nanoantennas that have been proposed and investigated experimentally. We will start from very simple single-particle structures, like nanospheres and nanorods, then move to more complicated ones, like nanoparticle dimers, cross nanoantennas, and Yagi-Uda nanoantennas. For each structure we will discuss how the respective antenna properties benefit their experimental application.

\subsection{Single nanospheres and nanorods}
Single gold nanospheres and nanorods exhibit a resonance spectrum that shifts with the particle's aspect ratio, as discussed before. Their use is motivated mainly by the easy and well-established chemical synthesis methods that allow producing single-crystalline nanoparticles in solution. Because of their very simple geometry, they also represent the ideal benchmark to test theoretical predictions for plasmon-coupled quantum emitters. They have been successfully applied to enhance the sensitivity of fluorescence and Raman spectroscopy at a single-molecular level \cite{Mohammadi08,Anger06,Kuhn06,Rogobete07}.
Elongated particles as opposed to spheres also show sensitivity to the polarization of the fields. Typically, the field enhancement and confinement near the ends of nanorods is much larger than for a sphere, which can be attributed to the combined contribution of both a more favorable spectral location (in terms of associated losses) of the fundamental resonance and  of enhanced lightning-rod effects.

\subsection{Nanosphere and nanorod dimers}
Coupling of two nanoparticles results in increased near-field intensity enhancement and confinement in the gap, as discussed in Sections \ref{section:coupling} and \ref{somesimulations}. It also goes along with a further degree of freedom in tuning the resonance frequency and a larger radiation efficiency, as shown by simulations in Section \ref{nanoantenna+quantumemitter}.
While the nanorod dimers \cite{Ghenuche08,Muhlschlegel05,Pramod08} represent the simplest analogue to RF linear dipole gap antennas, the nanosphere dimers are proposed as the analogue of RF Hertzian dipole antennas. Although being physically small, such nanosphere dimers are expected to have a high radiation efficiency due to the reduced current density inside the spheres \cite{Alu08c}.

\subsection{Bow-tie nanoantennas}
\begin{figure}[htbp]
    \centering
     \includegraphics[width=0.5\textwidth]{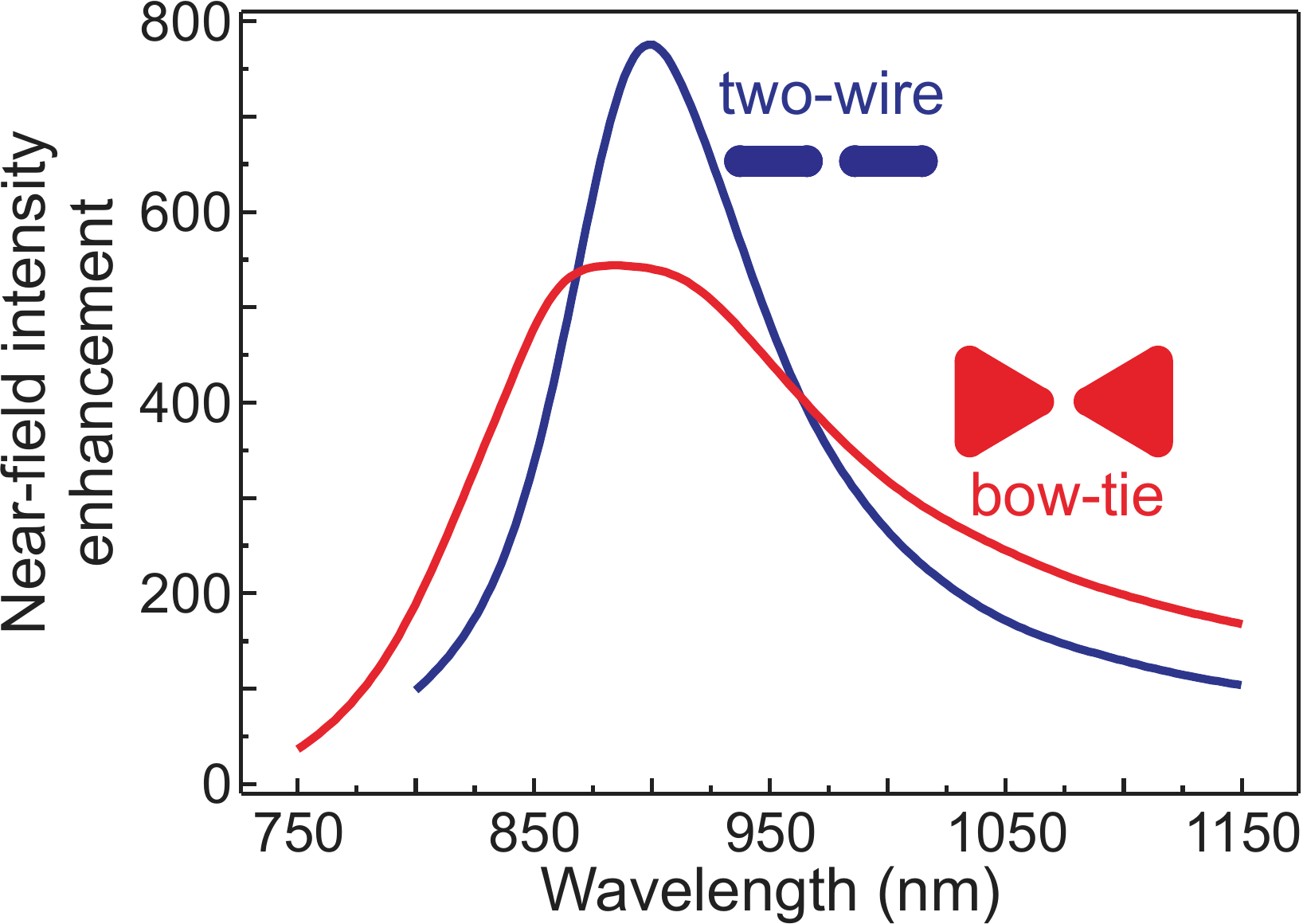}
     \caption{FDTD simulated near-field spectra in the middle of the gap for a two-wire and a bow-tie antenna on a glass substrate with the same total length, same gap size, and same radius of curvature at the apex. The reduced quality factor for the bow-tie structure is clearly observable and accordingly the field enhancement on resonance is also reduced.}
     \label{bowtie}
\end{figure}

Bow-tie dimer antennas are constituted of two triangles facing each other tip-to-tip \cite{Schuck05,Fromm04}. They are being applied to enhance molecular fluorescence \cite{Farahani05,Kinkhabwala09}, Raman scattering \cite{Fromm06,Lim10}, and for high-harmonic generation \cite{Kim08}. Bow-tie antennas are expected to possess a rather broad bandwidth since they represent the two-dimensional analogue of a biconical antenna \cite{balanis} and are also considered to have higher field enhancement in the gap compared to two-wire antennas because of larger lightning-rod effect at the apex. However, taking fabrication constraints into account, it turns out that the latter effect is practically limited by the radius of curvature at the apex. Because of this, if a two-wire and a bow-tie antenna with the same total length (300~nm), gap (30~nm), and radius of curvature (10~nm) are compared, the largest resonant field enhancement is obtained for the two-wire antenna, since for the bow-tie structure losses are increased due to the larger volume, which decreases the quality factor (see Fig.~\ref{bowtie}). On the other side, compared to nanorod antennas, bow-tie antennas suppress more efficiently near-field intensity enhancement at their outer ends \cite{Yu07a}.

\subsection{Yagi-Uda nanoantennas}
\label{yagiuda}
\begin{figure}[htbp]
    \centering
     \includegraphics[width=0.5\textwidth]{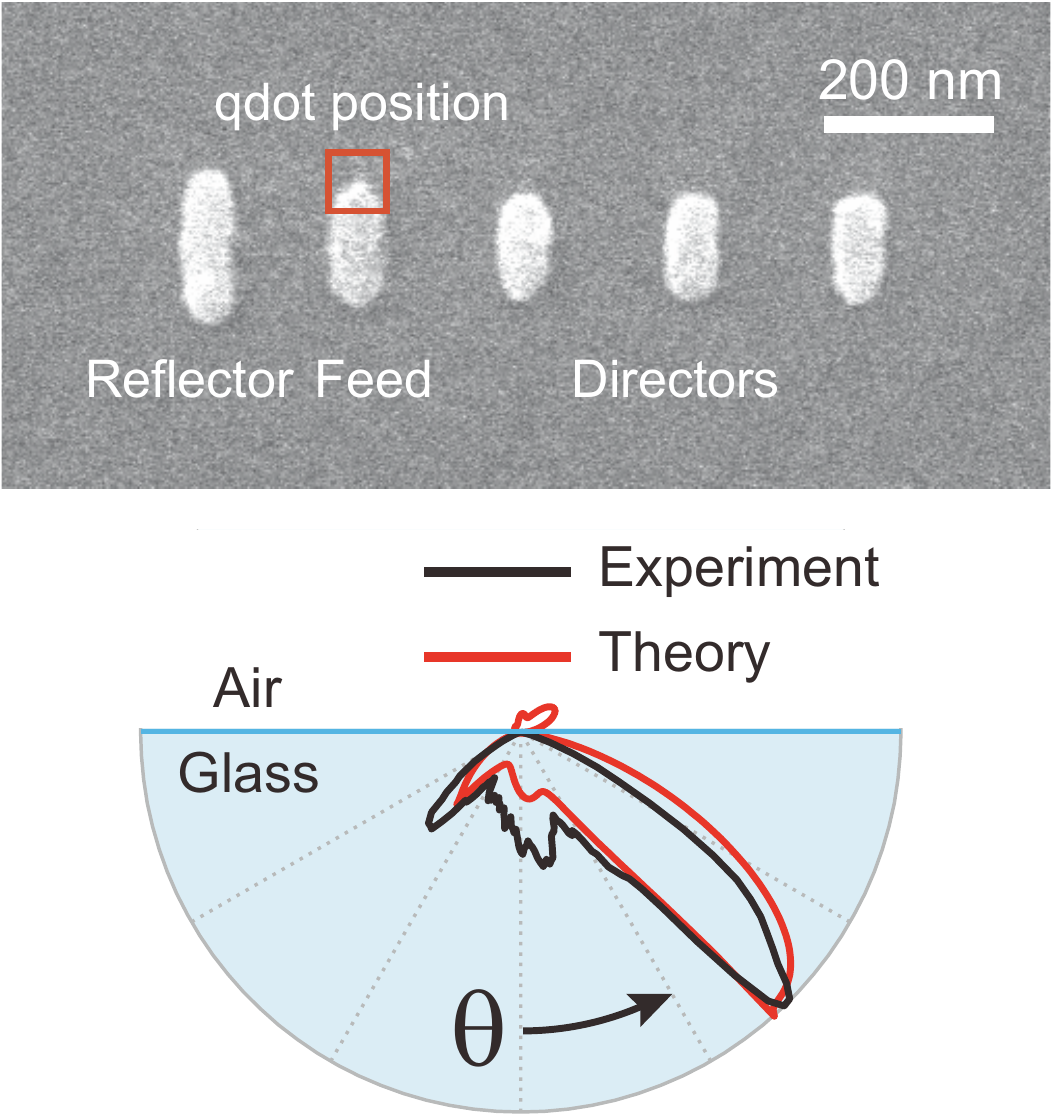}
     \caption{Yagi-Uda nanoantenna: (a) SEM image of the investigated system; (b) comparison between simulated and experimental angular radiation patterns for the resonant Yagi-Uda antenna. From Curto \textit{et al.} \cite{Curto10}. Reprinted with permission from AAAS.}
     \label{fig:YagiUda}
\end{figure}

Yagi-Uda nanoantennas for light have been theoretically proposed and experimentally demonstrated \cite{Curto10,Taminiau08c,Li07,Kosako10,Dorfmuller11} to take advantage of their very good directivity. Their unidirectional response is appealing in terms of enhanced sensitivity for detection, possible use for improved single-photon sources, and concurrently modified excitation patterns. Similarly to their RF counterparts, nanometer-sized Yagi-Uda antennas consist of a resonant single-wire antenna ($\pi/2$ phase shift between the driving field and the induced charge oscillations) arranged between a reflector (phase shift $>\pi/2$) and a set of directors (phase shift $<\pi/2$). To this purpose, the inter-element distance is important to achieve the desired interference between direct and reflected radiation. Since the existence of neighboring elements may increase Ohmic losses and modify the resonance frequency of a nanoresonator via near-field coupling, the design of an optimal Yagi-Uda nanoantenna is demanding. Very recently, the coupling between a single quantum dot and a Yagi-Uda nanoantenna has been demonstrated using multi-step e-beam lithography to fabricate the hybrid system \cite{Curto10}. Directionality of the emission pattern has been observed at the back focal plane of the objective, with a clear directional emission of single-quantum dot luminescence for optimized antenna dimensions (Fig.~\ref{fig:YagiUda}).

\subsection{Other nanoantenna geometries}
\label{other_geometries}

\begin{figure}[htbp]
    \centering
     \includegraphics[width=0.5\textwidth]{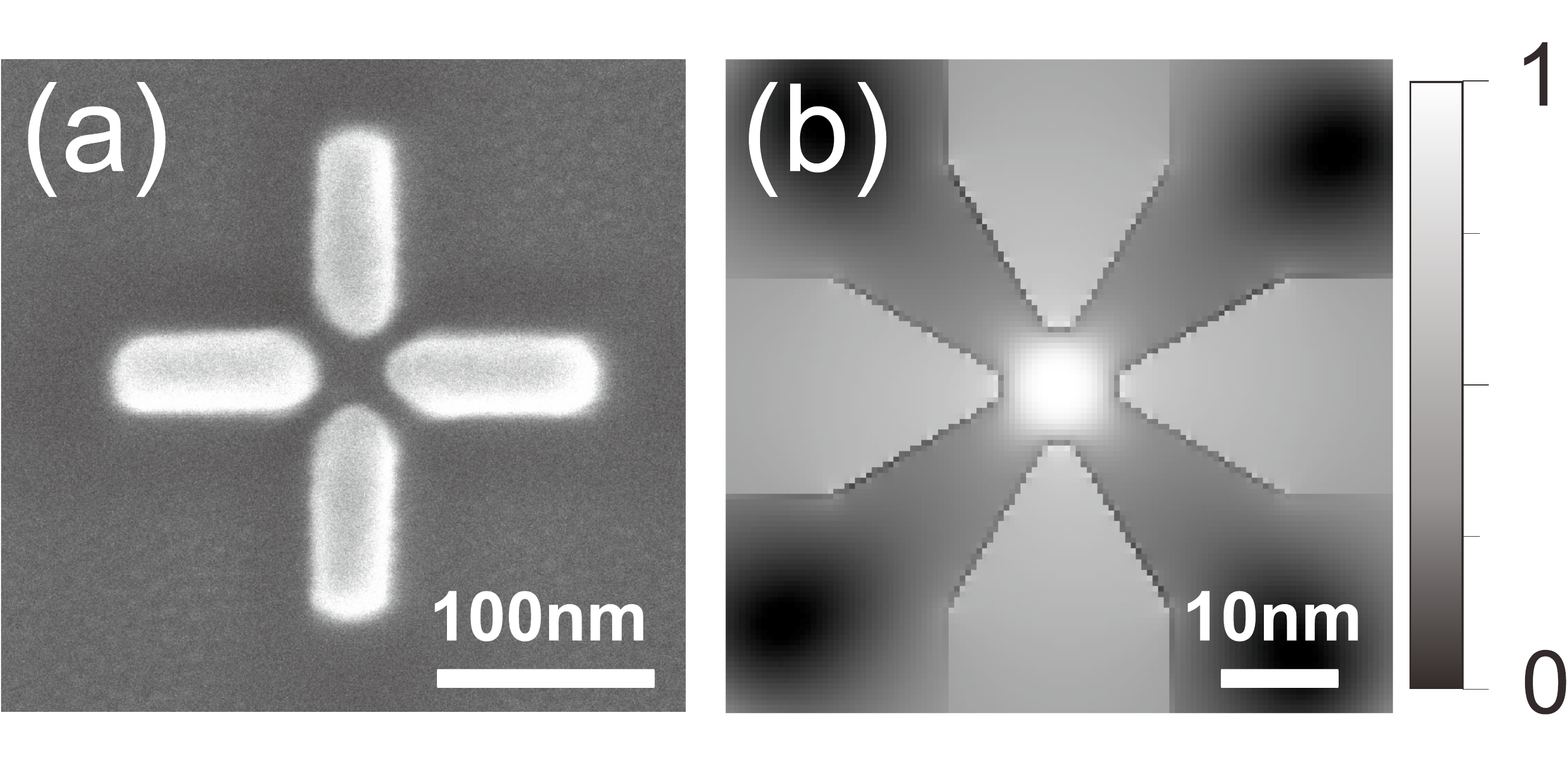}
     \caption{Cross antennas: (a) SEM image of a cross nanoantenna prototype, realized by FIB milling starting from a single-crystalline Au microflake \cite{Huang10b}. (b) simulated map of the degree of circular polarization in the gap region after circularly-polarized far-field illumination. Panel (b) adapted with permission from Biagioni \textit{et al.} \cite{Biagioni09b}. Copyright 2009, American Physical Society.}
     \label{fig:cross}
\end{figure}

Since the fields in the gap of a two-wire nanoantenna are highly polarized along the antenna axis \cite{Muhlschlegel05,Biagioni09b}, linear nanoantennas enhance only such field component and thus can be used to control the polarization of the emission from single emitters \cite{Taminiau08a,Moerland08}. However, for some applications such as polarimetry, chiral molecule mapping or optical data storage and retrieval it is necessary to enhance the field without perturbing its polarization state. To this purpose, cross nanoantennas, consisting of two identical but perpendicular linear dipole antennas sharing a common gap [see Fig.~\ref{fig:cross}(a) for a representative SEM image], have been proposed \cite{Biagioni09b}. Each linear antenna picks up and enhances the field component along its own axis. The two perpendicular field components then coherently add up in the gap region and build up a localized field with the same polarization properties as the excitation source. Upon circularly polarized excitation, a highly confined spot with an almost unitary degree of circular polarization in the plane of the antenna can thus be obtained [Fig.~\ref{fig:cross}(b)]. Based on the concept of tuning the phase of antenna oscillations with respect to that of the external field, cross antennas can be further modified not only to preserve but also to shape and control the near-field polarization. For example, a nanosized quarter waveplate has been proposed, by which a confined and enhanced circularly-polarized light field can be produced inside the common gap starting from a linearly-polarized excitation \cite{Biagioni09c}.
Along the same line, asymmetric bow-tie cross antennas have also been introduced as ``nano-colorsorters'', which are able to control field localization by achieving spectrally distinct resonances on spatially separate hot spots, within distances of only a few tens of nm \cite{Zhang09,McLeod11}.

Based on the application of Babinet's principle at optical frequencies \cite{Zentgraf07} and on the phenomenon of extraordinary transmission \cite{Ebbesen98}, resonant hole antennas have also been proposed as the optical counterparts of standard RF slot antennas, in order to achieve large field confinement on subwavelength regions \cite{Lee10}. To this purpose, bow-tie \cite{Wang07,Guo08}, c-shaped \cite{Shi03}, and crescent-shaped \cite{Wu09} hole antennas have been proposed. Also, hole arrays with specific symmetry properties have been demonstrated in order to control their spectral response and radiation directionality \cite{Alaverdyan07,Zhang08b}.

Patch antennas are largely used, especially in microwave wireless connections, because of their simplicity. They consist of a single metal patch, which radiates through the discontinuities at the edges. Esteban \textit{et al.} have shown by simulations that an optical patch antenna is a very promising system for coupling to single emitters, concurrently providing a large Purcell factor and a larger spectral width than standard linear antennas \cite{Esteban10}.

In addition to nanoantennas for electric fields, antennas for magnetic field detection have also been realized by fabricating a split ring at the facet of a coated fiber \cite{Burresi09}. For this antenna probe, the magnetic field generates a circulating current in the ring - a magnetic dipole - which then couples to a fiber mode thanks to the antenna gap which breaks the cylindrical symmetry of the system. 	A bow-tie nanoaperture antenna has also been designed to generate a confined and enhanced hot spot of magnetic fields at optical frequencies \cite{Grosjean11}.

Very recently, planar stacked geometries have also been proposed for optical antennas, where the planar feed-gap is constituted by the separation layer between two stacked antenna arms \cite{Pohl11}. Such a geometry holds promises because of its easier integration in planar device geometries and precise control of the gap width.

As the reader may have noticed, it seems that the designs for optical antennas that have been used so far were strongly inspired by the large pool of antenna structures that have been developed by RF antenna engineers over the years. While this has indeed proven to serve its purpose for a while, it is expected that with improved understanding of the clear differences between optical and RF antennas as pointed out in Section \ref{definition}, and of the particular ways of using them as discussed in Section \ref{antenna+circuits}, soon new designs for optical antennas are going to be developed that have no direct RF counterpart but represent optimized realizations for the very high frequencies of visible and near-infrared light.

\subsection{Substrate effects}
It is important to note that, in view of practical applications, nanoantennas will generally be supported by a substrate. It is well-known from RF antenna theory that, when an antenna is positioned over ``ground'', its impedance is modified by the presence of the image dipole which is generated by the interface nearby. Similar arguments are true for optical antennas, where the substrate refractive index acts as a parasitic impedance and causes a red shift of the resonance frequency, as is well-established for plasmonic nanoparticles \cite{Tamaru02}. For the special case of a linear antenna on and perpendicular to a conducting substrate at optical frequencies, the image dipole is ideally as strong as the antenna dipole, resulting in a resonance length which is only half the free-space length for a given working frequency. The optical counterpart of such a configuration has been demonstrated by realizing a tip placed on the metalized facet of an optical fiber used for near-field imaging \cite{Taminiau07a,Frey04}.

Furthermore, the larger refractive index of the substrate compared to air strongly influences the emission pattern of the nanoantenna \cite{Taminiau08c}, an occurrence that is also well-known for RF antennas \cite{balanis} and for quantum emitters on a substrate \cite{novotny}.

\section{Characterization of nanoantennas}

Once antenna structures have been successfully fabricated, their performance needs to be characterized. This entails analysis of their geometry, position with respect to other structures, near-field intensity distributions, emission pattern, as well as their spectral properties. While SEM, TEM, or AFM are typically used for the first two purposes, optical and spectroscopic characterization can be achieved by various techniques.  As to be expected from the hybrid photonic and electronic character of plasmonic resonances, these techniques are based on interactions with photons, electrons, or combinations of both. Characterization techniques involving electrons have been covered in a number of publications and reviews \cite{Douillard08,Cinchetti05,Swiech97,Aeschlimann07,Vogelgesang10}. Such methods can provide detailed insight due to their potentially high spatial resolution but experiments require vacuum environments. Here we concentrate on purely optical (photon-in/photon-out) techniques because they are straightforward to implement in any optics laboratory.

Optical characterization of nanoantennas requires a combination of spatial resolution (possibly subwavelength, to resolve nanoscale features) and spectral resolution (to properly characterize antenna resonances). Photons can interact with an optical antenna either elastically or inelastically. Broadband elastic light scattering is the most natural way to spectrally probe antenna resonances but care needs to be taken in order to sufficiently suppress excitation light. Very high sensitivity can be reached for inelastic interactions, since the measured signals are spectrally separated from the excitation wavelengths. Inelastic interactions, e.g. photoluminescence, typically also bear signatures of the antenna response since the spectrum of the emitted photons may be strongly shaped by the antenna resonances. The overall efficiency of multiphoton-excited photoluminescence processes, as well as other nonlinear effects, can be used as a probe for the near-field intensity enhancement for a certain antenna mode, since the amplitude of such signals strongly depends on the local excitation intensity.

In this section we concentrate on techniques that are able to address the optical response of {\em individual} nanostructures. Although extended arrays of nanostructures can in principle be fabricated to provide larger signals during experiments, single-antenna measurements are important to circumvent inhomogeneous broadening of spectral features which may occur due to various structural inhomogeneities. Single-antenna experiments require sufficient detection sensitivity since the scattering cross section for off-resonant structures can become small. There are two further requirements which need to be considered: (i) the response of a resonator to an applied external field is fully characterized only once both its amplitude and phase are known. To capture both quantities, interferometric techniques need to be applied. Furthermore, (ii) control over light polarization is required to selectively probe distinct antenna modes.

\subsection{Elastic and inelastic light scattering}

Interaction of light with optical antennas can be either elastic or inelastic. Elastic interaction is encountered when photons of a certain energy excite a certain antenna mode which then re-radiates photons of the same energy (elastic scattering). Inelastic interactions lead to the emission of photons of higher or lower energy. The respective increase or loss in energy is due to nonlinear and/or dissipative effects, respectively. One-photon photoluminescence (1PPL) from noble metals is an example for a dissipative effect. It was first reported in 1969 after 488~nm wavelength excitation of Au and Cu \cite{Mooradian69}. Two-photon photoluminescence (2PPL) is an example of a nonlinear effect. 2PPL in Au relies on two sequential single-photon absorption events: a first near-IR photon excites an intra-band transition within the $sp$ conduction band, while the second photon drives an inter-band transition between the $d$ and the $sp$ bands \cite{Imura05,Biagioni09}. The resulting hole distribution in the $d$ band can eventually decay radiatively, by recombination with $sp$ electrons, thus leading to a weak photon emission in the green-red spectral region. Recently, 2PPL from Ag and Al nanorods has also been reported \cite{Imura09,Castro10}, rendering the method more universal. Besides two-photon absorption, higher-order multiphoton absorption processes can also lead to photoluminescence in Au nanoparticles. Three-photon absorption is sometimes observed \cite{Farrer05,Eichelbaum07}, while a broad emission spectrum depending on the fourth power of the excitation intensity has been reported for dipole antennas \cite{Muhlschlegel05}. Even higher-order avalanche multiphoton photoluminescence (up to 18th order) has been reported in nanowire arrays \cite{Wang07b}. However, the origin of the photoluminescence processes with order higher than 2 so far is not fully understood.

The concept of evaluating antenna performances by probing photoluminescence of the antenna material is based on the dependence of the photoluminescence signal on the strength of the local fields generated upon illumination. Both 1PPL and 2PPL can be easily observed, however, 2PPL is often employed since many of the investigated structures exhibit a resonance in the near infrared such that for resonant excitation the energy of a single photon is too small to generate 1PPL. Furthermore, the amount of recorded 2PPL emission quadratically depends on the excitation intensity, which allows one to clearly distinguish resonant structures with strong local fields from off-resonant structures that exhibit much lower near-field intensity. Similar arguments apply for other nonlinear (multiphoton) processes that may occur, such as harmonic generation or frequency mixing \cite{Danckwerts07,Hanke09}.

\subsection{Near-field intensity distribution}

\begin{figure}[htbp]
    \centering
     \includegraphics[width=0.5\textwidth]{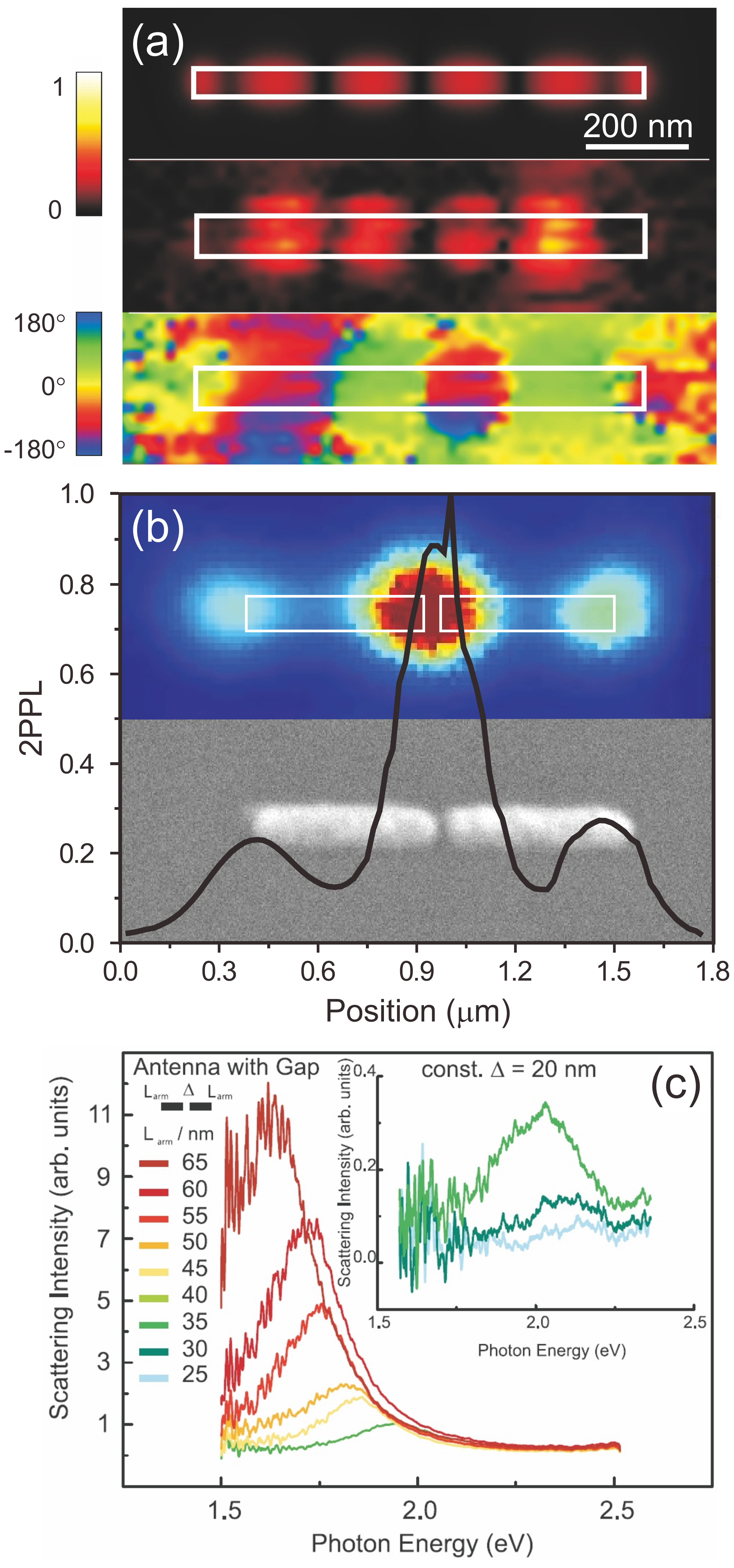}
     \caption{Characterization techniques for optical antennas: (a) interferometric near-field imaging of single-wire antennas by means of a scattering probe (upper panel
     simulated near-field intensity, mid panel experimental near-field intensity, lower panel experimental phase response); (b) 2PPL scanning confocal image of a two-wire Au nanoantenna and the corresponding SEM image; (c) scattering spectra for two-wire
     antennas for different arm lengths $L_{\rm{arm}}$, for a fixed gap size $\Delta = 20$~nm. Panel (a) reprinted  with permission from Dorfm\"{u}ller \textit{et al.} \cite{Dorfmuller09}. Copyright 2009, American Chemical Society. Panel (b) reprinted with permission from Ghenuche \textit{et al.} \cite{Ghenuche08}. Copyright 2008, American Physical Society. Panel (c) reprinted with permission from Wissert \textit{et al.} \cite{Wissert09}. Copyright 2009, IOP Publishing. }
     \label{tppl}
\end{figure}

A way to optically probe the spatial distribution of near-field intensity of an antenna mode with reasonable resolution has been the use of a near-field scattering probe oriented perpendicular to the sample surface, which is scanned in close proximity to the antenna \cite{novotny}. The basic idea behind such experiments is that a certain antenna mode is being excited by far-field illumination with a well-defined polarization (usually parallel to the antenna axis for linear antennas). At the wavelength of illumination, the near-field scattering probe is assumed to possess a non-negligible polarizability only along its main axis. This direction can be chosen to be perpendicular to the polarization of the illumination field. The probe is therefore not directly excited and its induced dipole cannot contribute to any mode hybridization. However, around the antenna structure of interest, depolarization fields build up which possess vector components along the main axis of the scattering probe. The scattering probe then becomes efficiently polarized and scatters a signal into the far field which is polarized perpendicular to the fields directly scattered by the antenna. Both signals can therefore be detected separately. In general, the scattering probe will be as small as possible, which in turn causes the scattering signal to be very small. It is therefore usually amplified by means of heterodyne and lock-in detection \cite{novotny}. It is important to note that the perpendicular orientation of scattering probe and antenna leads to a minimal back action of the probe on the antenna resonance. This explains why experimentally recorded antenna mode patterns match so well with theoretical expectations. Fig.~\ref{tppl}(a) shows an example of such sub-diffraction mapping of an antenna mode, which is able to spatially resolve 5  individual intensity maxima of a higher-order antenna mode on a Au single-wire antenna \cite{Dorfmuller09}. Since an interferometric setup with a reference beam is used, the phase of the measured fields can be determined as well. Similar experiments have been performed for a number of different plasmonic oscillators \cite{Dorfmuller10,Dorfmuller09,Schnell09,Dorfmuller11,Hillenbrand03,Schnell10,Kim09,Esteban08,Olmon08,Olmon10}.

While scattering probes allow for sub-diffraction imaging of antenna near fields after far-field illumination, aperture probes - characterized by a sub-wavelength hole at the apex - offer the possibility of local excitation of an optical antenna. The antenna then emits light into the far field which is eventually collected by a standard objective \cite{Mikhailovsky03,Celebrano09}. The measured spatial distribution of far-field photons is determined by the antenna mode and spatial resolution is roughly limited by the size of the aperture.

However, the simplest and most common way to address the spatial pattern of antenna resonances is scanning confocal optical microscopy \cite{novotny}. Although it only provides a diffraction-limited excitation spot of about 200-250~nm with blue-green light, this technique benefits from its ease of  implementation, high photon throughput and good control over light polarization and therefore has been extensively used to characterize optical antennas
\cite{Huang10b,Huang10,Ghenuche08,Muhlschlegel05,Schuck05,Hanke09,Celebrano09}. The overall lateral resolution (convolution of excitation and detection volumes) is usually sufficient to single out the response of an optical antenna within a larger ensemble, but it is usually not possible to resolve nanoscale features of the near-field intensity distribution of a single nanoantenna. This is because the excitation spot is of the order of the antenna dimension, unless the antenna dimensions are particularly large. It is, however, possible to observe pseudo-resolution in the measured excitation spot of a particular antenna in the sense that lines of zero excitation probability may occur in scan images where excitation of a probed antenna mode is forbidden due to symmetry \cite{Huang10}.

Fig.~\ref{tppl}(b) shows an example of confocal mapping of the local fields of a two-wire gold antenna. Here, since the antenna dimensions are particularly large, it is possible to resolve the individual intensity hot spots even with a diffraction-limited optical technique \cite{Ghenuche08}. The measurement is based on two-photon photoluminescence (2PPL) imaging, which over the last decade has become one of the preferred tools to probe plasmonic and antenna resonances \cite{Huang10b,Huang10,Ghenuche08,Muhlschlegel05,Schuck05,Imura05,Beversluis03,Bouhelier05}. It is here worth mentioning that a fixed-wavelength analysis of individual nanoantennas within an array of increasing length, and therefore variable near-field intensity enhancement, allows studying the antenna resonance without direct spectral analysis and with the advantage of a fixed dielectric constant \cite{Muhlschlegel05,Hanke09,Celebrano09}.

\subsection{Emission patterns}

Emission patterns of nanoantennas, i.e.~the angular distribution of the radiated photons, are very important to characterize, since they can drastically differ from their RF counterparts, as discussed in section~\ref{radiation_section}. Knowledge of the emission pattern of a particular antenna mode allows optimizing its coupling to light by way of the reciprocity theorem (see section~\ref{reciprocitytheorem}) and distinguishing between different antenna modes \cite{Dorfmuller10}. Typically, emission patterns of individual antennas are measured by first localizing a particular antenna in a confocal microscope and subsequent imaging of the back aperture of the microscope objective onto a CCD chip with a well-chosen magnification. The measured pattern corresponds to the intensity distribution in the Fourier plane of the objective lens \cite{Curto10,Huang08} and therefore represents a two-dimensional projection of the emission pattern. An example of a Yagi-Uda angular emission pattern has already been discussed in Fig.~\ref{fig:YagiUda}.

\subsection{Spectral properties}

The goal of spectrally-resolved experiments on optical antennas is to verify their performance over the frequency range of interest. The most straightforward way to spectrally characterize the behavior of a nanoantenna is to make use of elastic light scattering. Elastic light scattering reveals characteristic resonances in the scattering cross section of nanoantennas. A broad-band, ``white-light'' source excites a nanoantenna simultaneously at a wide range of illumination wavelengths. The spectrum of the elastically scattered light, after proper normalization, then directly reveals the scattering cross section
\cite{Sonnichsen,Funston09,Merlein08,Fromm04,Wissert09,Sonnichsen00,Muskens07b,Zhang10}. A representative set of antenna scattering spectra,
acquired for antennas of different lengths, is shown in Fig.~\ref{tppl}(c).

For elastic light scattering techniques, it is crucial to achieve sufficient suppression of the background caused by direct reflection of the excitation light. This can be achieved by decoupling the illumination and collection paths via dark-field microscopy techniques \cite{Prescott06,Funston09,Sonnichsen02}. As we have already mentioned, it is also possible to study the elastic scattering properties to observe resonant behavior by studying the antenna scattering as a function of its length for a fixed illumination wavelength \cite{Celebrano09}.

Scattering spectra directly reveal the peak position and quality factor of observable (``bright'', symmetry-allowed) resonances that efficiently radiate into the far field. Note that many symmetric structures support dark modes, which, due to their symmetry, cannot be excited and therefore do not appear in scattering spectra. However, even with symmetric excitation, dark modes can become visible when they couple to a bright mode. Such coupled resonances are also so-called Fano resonances which recently gained much attention \cite{Lukyanchuk10,Miroshnichenko10}. In a particular type of dark-field setup the sample is illuminated via an evanescent field created by total internal reflection. This approach has the advantage that, by breaking the cylindrical symmetry of the illumination, dark modes can be excited and their weak emission can be recorded \cite{Yang10}.

A complete characterization of an antenna resonance requires both phase and amplitude of the local near fields to be investigated with respect to the driving field. In the far field, a combination of coherent white-light illumination and differential interference contrast microscopy is able to address the complex polarizability of single nanoparticles \cite{Stoller06}. Near-field techniques, in addition to providing subwavelength resolution \cite{Klar98}, combine a weak propagating field with a very strong local field, which makes the far-field interference between the illumination dipole and the particle's dipole very strong. Therefore, the whole complex response (amplitude and phase) of the nanoantenna is encoded in the detected signal. Spectrally-resolved near-field extinction is therefore a powerful means to address the response of plasmonic oscillators \cite{Mikhailovsky03}.

When studying inelastic interactions of light with optical antennas, particularly with luminescence processes, not only the integrated emission efficiency provides valuable insight into the antenna response, but also its spectral shape. While Imura \textit{et al.} reported that the 2PPL spectrum is dominated by two features related to the bulk Au band structure \cite{Imura05}, other groups showed that a plasmon resonance is able to shape the luminescence spectrum \cite{Bouhelier05,Mohamed00,Dulkeith04,Wissert10}. The spectral properties of noble-metal nanostructure luminescence therefore seem to depend on a subtle interplay between the electronic density of states of the antenna material and the occurrence of a plasmon resonance and deserve further investigation. It is important to note that, in order to exploit the spectral shape of photoluminescence to study antenna resonances, one needs to illuminate the nanoantenna with an off-resonant excitation, so that the excitation wavelength does not overlap with the resonance of interest, which would then have to be cut off by the filters that block the excitation light. In addition to radiative modes, the local coupling of luminescence dipoles to the antenna resonance breaks the symmetry of the system and allows dark modes to become populated and observable, an advantage compared to elastic scattering with symmetric illumination.

\section{Applications and perspectives of nanoantennas}

The topic of optical antennas is a young and aggressively progressing field of research. As outlined in the introduction, in particular in Section \ref{potentialofopticalantennas}, it holds promises for a variety of possible applications that take advantage of enhanced light-matter interaction. In the present Section we review different fields in which nanoantennas are already applied today and in which they might play a significant role in the near future. Since the number of publications using the key word ``optical antenna'' is increasing exponentially, this overview must be incomplete and is influenced by the personal taste of the authors.

\subsection{Scanning near-field optical microscopy, spectroscopy, and lithography}
In order to make use of optical antennas as imaging and spectroscopic probes, a viable solution is to scan the antenna in close proximity over some area of a sample. To this end, nanoantennas need to be fabricated at the apex of a scanning probe. Super-resolved imaging can then be achieved either by inverted confocal illumination of the antenna-sample system, or by realizing antenna-on-aperture probes to drive the antenna resonance by the localized field of a small aperture. Single-molecule imaging with bow-tie \cite{Farahani07} and $\lambda$/4 antennas \cite{Taminiau07a}, respectively, has been achieved in this way, with resolutions down to 25~nm in the latter case. In related experiments single gold spheres have been attached to dielectric tips in order to create well-controlled antenna probes. These probes were used to image single molecules at a resolution smaller than the sphere diameter but compatible with the expected extent of the localized field \cite{Anger06,Kuhn06}. Using  these single- and also multiple-sphere nanoantennas on optical fiber near-field probes, single proteins have been imaged at very high resolution in their native cell membranes \cite{Novotny11,Hoppener08b,Hoppener08}.

Highly-sensitive spectroscopy represents another key area for nanoantenna applications, where the antenna serves the purpose of both providing enhanced excitation as well as enhanced emission of the nanoobject under investigation. In particular, Raman signals can be largely enhanced following the already well-established route of surface-enhanced \cite{Nie97} and tip-enhanced \cite{Anderson07,Bailo08} Raman scattering. Along this line, enhanced Raman signals have been measured for EBL-fabricated bow-tie antennas covered with a layer of adsorbed molecules \cite{Fromm06}, showing signatures that are typical for single-molecule Raman experiments. Since the signal enhancement is known to strongly depend on the gap size and geometry, control over these parameters is crucial. In this respect, the gap in a plasmonic nanodimer for Raman spectroscopy has been deterministically varied both by scanning an Au nanoparticle attached to a fiber probe \cite{Olk07} or by AFM nanomanipulation of Au-Ag nanoshell particle pairs on the sample substrate \cite{Lim10}. In view of practical sensing applications, an array of e-beam fabricated antennas has also been transferred to the facet of an optical fiber, used for illumination and collection \cite{Smythe09}. Not only Raman scattering, but also vibrational spectroscopies based on infrared absorption can largely benefit from suitably engineered resonances in plasmonic antennas \cite{Neubrech08}. Finally, following the original demonstration of fluorescence correlation spectroscopy (FCS) at high analyte concentrations using reduced sensing volumes in subwavelength hole arrays \cite{Levene03}, recently a 10000 times reduction of the FCS sensing volume has been demonstrated by exploiting the field confinement in optical antennas \cite{Estrada08}.

The highly localized near field of a nanoantenna also finds a natural application in optical lithography, where the fabrication of nanostructures has been accomplished via nonlinear photopolymerization of a photoresist by exploiting both the largely confined and enhanced fields in the gap of a bow-tie antenna and the nonlinearity of resist response \cite{Sundaramurthy06}.

\subsection{Nanoantenna-based single-photon superemitters}
Experimental studies of single emitters coupled to optical antennas address another key application of resonant optical antennas. If the emitter is placed in a ``hot spot'' of a resonant antenna most of the single emitter decay processes will not generate a free propagating photon but will rather create a single plasmon in the resonant mode of the antenna. Upon radiative decay of these plasmons, single propagating photons are created that bear the properties of the antenna resonance, e.g.~its resonance spectrum, polarization and emission pattern. One can therefore envisage the possibility to build single-photon sources \cite{Lounis05} with well-defined polarization, optimized radiation patterns (see e.g.~section \ref{yagiuda}) and several thousand times enhanced emission rates as discussed in section \ref{nanoantenna+quantumemitter} by carefully adjusting the position of the emitter to avoid quenching.

A simple way to achieve the positioning of single emitters in the feed-gap is to directly cover the antenna with a stochastically distributed ensemble of emitters. However, in such a configuration, depending on their density, many emitters are excited simultaneously and may interact via energy transfer with each other. This renders quantitative analysis of assumed single-emitter effects challenging. Nevertheless, spin coating of a thin polymer films containing fluorescent molecules in a small concentration on top of  antenna structures has been exploited to gain insight into the emission enhancement that can be achieved in the vicinity of an optical antenna \cite{Muskens07a,Kinkhabwala09,Bakker08}. In this way, emission enhancements up to three orders of magnitude have been observed compared to the case without antenna \cite{Kinkhabwala09}, resulting from the combined enhancements in the excitation efficiency and quantum efficiency.

Deterministic positioning of single emitters in an antenna hot spot, e.g. the feed-gap of a two-wire antenna, would of course be a much better approach. This constitutes a technical challenge and so far has been achieved mainly in three ways: (i) by positioning of nanoantennas on scanning tips on top of emitters, (ii) by AFM nanopositioning of the emitter and/or the antenna arms with respect to each others, and (iii) by spontaneous or directed self-assembly of antenna and emitter.
Using the first method, experimental investigations demonstrated that coupling between the antenna and a two-level system can lead to a significant reduction of the excited-state lifetime \cite{Farahani05} and to a re-direction of the emission dipole of the molecule along the antenna axis \cite{Taminiau08a}.  The second method has been proven to be very effective for precise positioning of nanoparticles \cite{Hansen98,vanderSar09,Custance09}. In this way, antennas can be realized on a suitable substrate, onto which quantum emitters are co-deposited. An AFM tip is used both for precise imaging of the position of the nanoparticles and to push them in a controlled way inside the antenna feed-gap. Diamond nanocrystals between two Au nanoparticles have been produced and studied in this way, obtaining an enhancement in the radiative decay rate of about one order of magnitude \cite{Schietinger09}. As for the third method, quantum dots have been coupled to Ag nanowires \cite{Akimov07} which support a propagating plasmonic mode (see section \ref{fabryperot}). When a single-photon emitter, such as a quantum dot, is coupled to the Ag nanowire, it can excite plasmon quanta that propagate along it and are re-converted to single photons upon emission from the wire end. Notably, non-classical photon correlation has been demonstrated between the emission from the quantum dot and single photons resulting from the decay of the plasmon quanta launched in the Ag nanowire. Very recently, even unidirectional emission of a quantum dot which was deterministically coupled to a Yagi-Uda nanoantenna by means of two-step electron-beam lithography has been experimentally demonstrated \cite{Curto10}, as discussed in Section \ref{yagiuda}.

\subsection{Optical tweezing with nanoantennas}
Optical tweezing based on field gradients in tightly-focused beams is nowadays a quite established tool in atomic physics and biology \cite{novotny}, where it is usually implemented by exploiting microscope objectives with high numerical aperture and properly shaped beam profiles. Since highly localized optical near fields naturally possess strong gradients, their use as localized trapping spots is very appealing and has been proposed already more than a decade ago \cite{Novotny97}. More recently, the first experimental demonstrations of optical trapping with well-controlled hot spots in the gap of nanoantennas have been published, where the large antenna field enhancement allows trapping with lower excitation power and higher efficiency and stability \cite{Zhang10,Grigorenko08,Righini09}.

\subsection{Antenna-based photovoltaics and infrared detection}
In the field of optical detectors and solar cells, plasmonic structures have been introduced as a means to achieve substantial absorption of incident light within small active volumes and thin layers. Indeed, the earliest interest for resonant antennas in the sub-mm wavelength regime was mostly driven by the need for efficient IR detectors and, with first realization for 10 to 100 $\mu$m wavelengths \cite{Twu75,Neikirk82,Grossman91}, dates back to the Eighties and even before. Dipole, spiral, cat-whisker or bow-tie nanoantennas were produced by means of EBL and the IR detected light was measured by means of a microbolometer or with a micrometer-sized metal-oxide-metal diode. More recently, an antenna-coupled Ge nanodetector for near-IR wavelengths has been demonstrated \cite{Tang08}.

Absorption enhancement is particularly crucial for modern solar cell technologies \cite{Catchpole08,Atwater10}, especially for those realizations that move towards thin single-crystal films and intermediate-band nanocrystal insertions. In this context, efficient coupling of light must be achieved both with broad collection angles and wide spectral response \cite{Ferry08}. Noticeably, semiconductor-based resonant nanostructures, also often called antennas, have been proposed to improve the performances of detectors and solar cells \cite{Cao09,Cao10b}.

In the field of optical detection and electrical conversion with nanoantennas, it was recently shown that nonlinear tunnelling conduction between gold electrodes separated by a subnanometer gap can lead to optical rectification, producing a dc photocurrent when the gap is illuminated \cite{Ward10}.

\subsection{Optical antenna sensors}
Plasmonic sensors based on the adsorption of molecules onto functionalized noble-metal systems have nowadays become a reality and find many applications. Gold is the preferred material in this field because of its biocompatibility and easy functionalization mainly by means of sulfur bonds between gold atoms and molecules. Commercially applied plasmonic sensing is based on attenuated total internal reflection and exploits the fact that the wavevector of surface plasmon polaritons (SPP) propagating at the interface between a thin gold film and an adjacent dielectric is very sensitive to changes in the refractive index of the dielectric \cite{Homola99}. High sensitivity is thus achieved thanks to the strong confinement of fields in the close proximity of the interface. On the other side, such SPP sensors are based on a relatively complicated setup, which does not always meet the criteria required for commercialization, mass production and use, especially for portable apparatuses and point-of-care testing. Moreover, such systems require a rather large surface area and are therefore not suitable for parallelization and integration in view of lab-on-a-chip platforms.

More recently, however, the demonstration of sensors based on localized particle resonances opened new perspectives for plasmonic sensing, where in principle much simpler transmission, reflection, or scattering measurements can be performed because the local refractive index change upon ligand binding directly translates into a shift of the particle's resonance frequency. Sensing based on particle arrays on a fiber facet \cite{Smythe09,Mitsui04} or on a substrate \cite{Liu10b} has thus been demonstrated, ultimately with sensitivities down to the single-particle level \cite{Raschke03,Raschke04,Eah05,Anker08,Liu11}.
While the improved field localization provides sensitivities comparable to commercial SPP devices \cite{Svedendahl09}, there is also a potential for parallelization and miniaturization. Along this road, since plasmonic nanoantenna systems display localized resonances that strongly depend on the dielectric properties of the environment, they are very good candidates for sensing applications down to extremely low concentrations. Narrow resonances with large quality factors are advantageous in this context since they provide increased sensitivity. Therefore the use of Fano-like resonances \cite{Lukyanchuk10,Hao08} and of dark antibonding antenna modes \cite{Huang10} has been proposed in this context.

\subsection{Ultrafast and nonlinear optics with nanoantennas}
So far, we mainly focused our attention on the field enhancement and spatial confinement in optical antennas as well as on their resonances in the frequency-domain. However, one may also consider the behavior of nanoantennas in the time-domain and discuss the temporal characteristics of the confined fields as well as their coherent control. It has been shown theoretically by M.~I.~Stockman that, upon ultrafast excitation, random metal surfaces with nanoscale geometrical features can exhibit ultrafast nanoscale localization and enhancement in their near fields due to multiple interference \cite{Stockman2000}. By controlling the temporal profile of the excitation pulse, field localization, i.e.~the position of hot spots and the exact moment of time when they become active, can be coherently controlled \cite{Stockman2002}. The temporal profile of the excitation pulse can be shaped at will by manipulating the spectral phase and amplitude of the laser pulses used for excitation \cite{Wollenhaupt}. In this way coherent spatio-temporal control of localized near fields has been realized experimentally using antenna-like gold nanostructures \cite{Aeschlimann07,Aeschlimann10}. Since the behavior of metallic nanostructures, including nanoantennas, is time-invariant, once the impulse response of the system under certain excitation conditions is known, one may apply shaped excitation pulses and coherently control the antenna near fields. Using this concept, Huang \textit{et al.} theoretically demonstrated that the temporal profile of the local fields in the gap of a nanoantenna or at a given position on an antenna nanocircuit  can be optimized \cite {Huang09c}.
Similar techniques might facilitate signal processing in future plasmonic optical nanocircuitry. Also recently, Utikal \textit{et al.} have shown that the near-field in a hybrid nanoplasmonic-photonic system induced by an ultrashort pulse can be enhanced or turned off by a well controlled pulse following the excitation pulse \cite{Utikal2010}. Such a switching control based on near-field interference should in principle be applicable to plasmonic nanoantennas as well. The detection and analysis of all such phenomena, however, experimentally relies on the generation of non-linear signals that propagate to the far field, such as e.g.~third harmonic generation \cite{Utikal2010}.

As a general remark, it is worth recalling that noble metals are known to possess extremely large optical nonlinearities, but have generally not been considered as nonlinear materials because of their high reflectivity. In this context, nanostructured systems offer the possibility to engineer the field penetration inside the material and fully exploit their nonlinear response \cite{Lepeshkin04}. It is also important to note that, when e.g.~second-harmonic generation is considered for nanoparticles illuminated with a localized field, not only the symmetry of the bulk crystal structure, but also that of the field and of the nanoparticles needs to be taken into account in order to derive the relevant selection rules that can be significantly different from those of plane-wave illumination of bulk systems \cite{Dadap99,Finazzi07}.

Second-harmonic enhancement at correspondence with a plasmonic resonance has clearly been shown \cite{Hubert07,Zavelani08}. Taking advantage of the field concentration ability of optical antennas, practical applications using nanoantennas to generate higher harmonic emission signals have been proposed and demonstrated experimentally. Third-harmonic photons have been efficiently generated in isolated gold dipole antennas \cite{Hanke09} and a clear connection between third-harmonic efficiency and antenna resonances has been demonstrated. Kim \textit{et al.} exploited the enhanced and confined field of nanosized bow-tie antennas on a sapphire substrate to reduce the excitation pulse energy required to generate extreme ultraviolet light by high-harmonic generation \cite{Kim08,Husakou11}. Nanoantennas can also be applied to enhance the efficiency of wave mixing of infrared laser beams. Danckwerts \textit{et al.}~have shown that the four-wave mixing signal can be enhanced by 4 orders of magnitude as the gap of a nanoparticle pair shrinks \cite{Danckwerts07}. The four-wave mixing signal drastically changes when the two particles are in contact because the gap is shortened and the charge can redistribute via the conductive bridging. The use of the frequency mixing signal as a precise and sensitive indication of touching contact between two plasmonic nanoparticles has therefore been proposed \cite{Grady10}.

In perspective, the combination of laser pulse shaping and well-designed nanoantenna geometries might lead to a number of novel applications since it opens the possibility to manipulate the temporal behavior of the antenna near fields. Also here, the rather broad antenna resonances are advantageous to achieve a very high temporal resolution.

\subsection{Perspectives for lasing in nanoantennas}
A very intriguing perspective  is given by the possibility of using a plasmonic resonator as a localized cavity for surface plasmon amplification by stimulated emission of radiation (SPASER), as originally proposed by Bergman and Stockmann \cite{Bergman03}. A first step along this line has been achieved with the demonstration of stimulated emission of SPPs propagating at the interface with a gain medium \cite{Noginov08,Ambati08}. In one case, also laser-like emission accompanied by a clear threshold behavior and moderate line narrowing was observed \cite{Noginov08}, although no clear feedback mechanism could be identified.

In this context, metal nanoparticles acting as optical antennas can play a role because of their large local field enhancement and extremely reduced interaction volume. Very recently, subwavelength plasmon lasers based on the original SPASER proposal have been demonstrated \cite{Noginov09,Oulton09,Ma10,Hill10}. Larger enhancement and confinement, compared to single particles, can especially be encountered in the gap of two-wire or bow-tie nanoantennas. Along this line, laser operation for bow-tie antennas coupled to semiconducting quantum dots or multiple quantum wells has been theoretically addressed \cite{Chang08}.

\subsection{Nanoantennas and plasmonic circuits}
Sub-diffraction propagation in plasmonic waveguides offers the possibility of distributing and processing e.m.~signals on a very small footprint, thus offering the perspective of combining the speed of photonics with the high degree of integration of modern microelectronics \cite{Ozbay06}. In this frame, optical antennas - working as receiving and transmitting devices - represent an interface between subwavelength localized modes that propagate along transmission lines and free-space propagating waves \cite{Huang09b,Wen09,Fang11,Wen11}. Wireless optical interconnects based on matched optical antennas have also been suggested \cite{Alu10}, which could be used as high-speed links in chip architectures.

In terms of interfacing electronics and plasmonics, it is of course mandatory to efficiently convert electrical signals into plasmon waves and vice-versa. Recently, electrical sources \cite{Walters09,Denisyuk10,Bharadwaj11} and electrical detection \cite{Neutens09,Falk09} of plasmons have been realized. Also, electrical, mechanical, or optical tuning of nanoantennas by means of anisotropic load materials \cite{Barthelot09,Deangelis10}, stretchable elastomeric films \cite{Huang10c}, or photoconductive gap loads \cite{Large10,Abb11} has been proposed and demonstrated.

\subsection{Nanoantennas and thermal fields}
Resonant metal nanoparticles at optical frequencies are very efficient absorbers mostly due to considerable Ohmic losses. However, there are fields in which this supposed drawback turns into a dramatic advantage, since plasmonic nanoparticles can be viewed as nanolocalized heat sources that can be turned on and off by standard optical means at relatively low powers. This idea finds its most significant application in photothermal therapies, where functionalized metal nanoparticles and nanoshells have been used to demonstrate selective targeting of tumor tissues \cite{Huang09d,Lal08} and drug delivery \cite{Ghosh08}. Plasmon-assisted photoacoustic imaging has been proposed as well \cite{Eghtedari07,Mallidi09}. Thermal hot spots are also at the basis of heat-assisted magnetic recording, in which writing heads equipped with nanoantenna systems are used to confine heating within a mesoscopic volume, thus achieving very small bit sizes \cite{Stipe10}. In the context of plasmonic nanostructures, localized thermal fields have also been successfully exploited to achieve thermal imaging of single Au nanoparticle down to 2.5~nm \cite{Boyer02} and to develop hybrid heat-assisted nano-patterning techniques \cite{Fedoruk11}.
Many efforts have been devoted and are still needed to gain a correct understanding of heat flows at the nanoscale, where evanescent thermal fields play a significant role in radiative heat transfer \cite{Rousseau09,Shen09}.

Remarkably, when plasmonic nanostructures are used to generate localized thermal fields, care needs to be taken since optimization for local heating can lead to different requirements than for enhanced scattering of photons \cite{Baffou09}. Consequently, also the localization of thermal hot spots can differ significantly form that of optical hot spots, as demonstrated experimentally by Baffou \textit{et al.} for a Au dipole antenna \cite{Baffou10}.

In addition to using the heat itself, the heat-induced morphology transformation of nanorods has also been utilized to achieve five-dimensional high density recording \cite{Zijlstra09}. While three dimensions are provided by controlling the position of the laser focal spot, two additional dimensions can be achieved utilizing the wavelength and polarization sensitive resonances of the nanorods, which work as optical antennas.

Finally, an intriguing possibility is provided by the use of nanoantennas that are specifically designed to achieve efficient radiation of thermal fields \cite{bohren}, as recently demonstrated with dielectric nanostructures \cite{Greffet02,Schuller09}.

\section{Conclusions}
We hope that this {\it tour de force} on optical antennas has been able to show how lively and flowering this area of research is. The first ideas can be traced back to the efforts of bringing RF antennas down to the THz domain in the Eighties and to the development of optimized optical near-field probes \cite{pohl99,Pohl91}. The regime of optical frequencies has been fully addressed in 2004-2005, which triggered an exponentially increasing interest in nanoantennas accompanied by numerous major breakthroughs which seems to continue with unwaning vigour. Today the possibility of using optical antennas to manipulate light at the nanometer scale has become a reality and an enormous number of applications is at hand.

During the preparation of this Review, we obviously had to make choices about what to include and what to exclude from our discussion, as explicitly stated in Section 6. This unavoidably makes our Report fragmentary but, as we hope, still complete within its subset of arguments.

As we have seen, proof of principles which take advantage of the enhanced light-matter interaction afforded by optical antennas have already been demonstrated for many of their potential applications. This includes nanoscale microscopy, spectroscopy, lithography, quantum optics, sensing, trapping, photovoltaics, and many others. While all these results clearly show the potential of nanoantennas, we can probably safely state that a widespread use of all these possibilities in everyday research - not even to mention everyday life - is still far ahead. Many big steps are still needed which so far could not be reached because of the difficulty in achieving the required levels of nanostructuring precision and reproducibility towards ``mass production'' of nanoantennas over extended areas. A second open key point, which is under intensive study both theoretically and experimentally, is how to achieve the best coupling between a quantum emitter and an optical antenna system. Finally, in view of true plasmonic circuitry, the concepts of impedance matching and lumped plasmonic elements are still calling for a deeper understanding when it comes to the application of standard RF strategies to the plasmonic realm.
Hopefully, this Review will help many researchers and students in particular to enter this exciting field of research. For people already working in the field, we hope that it will spark new discussions between colleagues and foster collaborations to soon make ``future'' applications a reality.

\section*{Acknowledgments}
The authors warmly acknowledge A.~Cattoni, C.~De~Angelis, J.~Dorfm\"{u}ller, L.~Du\`{o}, T.~Feichtner, M.~Finazzi, J.~Kern, D.~W.~Pohl, J.~Prangsma, and M.~Savoini for stimulating discussions and for their help during the preparation of this manuscript.

\end{document}